\documentclass[10pt,a4paper,reqno]{amsart}
\usepackage{acronym}
\usepackage{amsfonts}
\usepackage{amsmath}
\usepackage{amssymb}
\usepackage{amsthm}
\usepackage{bm}
\usepackage{braket}
\usepackage{cite}
\usepackage{color}
\usepackage{geometry}
\usepackage{graphicx}
\usepackage{hyperref}
\usepackage{mathrsfs}
\usepackage{mathtools}
\usepackage{setspace}
\usepackage{stmaryrd}
\usepackage{subcaption}
\usepackage{tikz}
\usepackage{CJKutf8} 
\usepackage{epsfig}
\usepackage{setspace}
\usepackage{booktabs}
\usepackage{threeparttable}
\usepackage{diagbox}
\usepackage{float}
\usepackage{cancel}

\newgeometry{top=2cm,bottom=2cm,outer=1.5cm,inner=1.5cm}
\newcommand{\bse}{\begin{subequations}}
\newcommand{\ese}{\end{subequations}}

\numberwithin{equation}{section}

\title[VC-PINN: variable coefficient physical information neural network for forward and inverse problems of PDEs with variable coefficient]{VC-PINN: variable coefficient physical information neural network for forward and inverse PDEs problems with variable coefficient}
\author{Zhengwu Miao,}
\address[MZ]{School of Mathematical Sciences, Shanghai Key Laboratory of Pure Mathematics and Mathematical Practice, and Shanghai Key Laboratory of Trustworthy Computing \\
East China Normal University \\ Shanghai 200241 \\ People's Republic of China}
\author{Yong Chen$^*$}
\address[YC]{School of Mathematical Sciences, Shanghai Key Laboratory of Pure Mathematics and Mathematical Practice, and Shanghai Key Laboratory of Trustworthy Computing \\
East China Normal University \\ Shanghai 200241 \\ People's Republic of China}
\address[YC]{College of Mathematics and Systems Science \\ Shandong University of Science and Technology \\ Qingdao 266590 \\ People's Republic of China}
\email{ychen@sei.ecnu.edu.cn}

\begin{document}

\begin{abstract}
The paper proposes a deep learning method specifically dealing with the forward and inverse problem of variable coefficient partial differential equations -- Variable Coefficient Physical Information Neural Network (VC-PINN). The shortcut connections (ResNet structure) introduced into the network alleviates the ``vanishing gradient" and unify the linear and nonlinear coefficients. The developed method was applied to four equations including the variable coefficient Sine-Gordon (vSG), the generalized variable coefficient Kadomtsev–Petviashvili equation (gvKP), the variable coefficient Korteweg-de Vries equation (vKdV), the variable coefficient Sawada-Kotera equation (vSK). Numerical results show that VC-PINN is successful in the case of high dimensionality, various variable coefficients (polynomials, trigonometric functions, fractions, oscillation attenuation coefficients), and the coexistence of multiple variable coefficients. We also conducted an in-depth analysis of VC-PINN in a combination of theory and numerical experiments, including four aspects: the necessity of ResNet; the relationship between the convexity of variable coefficients and learning; anti-noise analysis; the unity of forward and inverse problems/relationship with standard PINN.
\end{abstract}

\maketitle
\section{Introduction}

Deep learning has achieved success in applications including image recognition \cite{krizhevsky2017imagenet}, natural language processing \cite{lecun2015deep}, and more. The research on using neural networks to solve partial differential equations (PDEs) can be traced back to the work of \cite{dissanayake1994neural} in 1994. However, limited by the computing power of computers at that time, they stopped after trying on shallow networks. Until 2019, the emergence of a deep learning framework based on physical constraints -- Physical Information Neural Network (PINN) provided a new idea for the numerical solution of PDEs \cite{raissi2019physics}. Under the theoretical support of the universal approximation theorem \cite{cybenko1989approximation}, PINN implements a mesh-free numerical algorithm by embedding PDEs into the loss of the neural network by using the automatic differentiation technique (AD) \cite{baydin2018automatic}. PINN can easily incorporate various mechanism-based constraints and data measurements into the loss function. Its strong scalability and generality make it more flexible than the finite difference method (FDM) and the finite element method (FEM). The above capabilities of PINN and its advantages in high-dimensional and inverse problems make it quickly become a powerful tool for solving the forward and inverse problems of PDEs through deep neural networks (DNNs).

Although the success that has been achieved is encouraging, PINN has to face the challenges brought about by more complex problems, the need for higher accuracy and efficiency, and the requirement for stronger robustness. To improve the performance of the network, researchers have extended the standard PINN method in various aspects. For example, the locally adaptive activation function is proposed to accelerate network training \cite{jagtap2020adaptive}, and various adaptive weight methods \cite{xiang2022self,wang2021understanding,wang2022and} and point-weighted methods \cite{mcclenny2020self, li2022revisiting} are developed to balance each loss item. The form of the loss function is optimized by using Meta-learning PINNs \cite{psaros2022meta} and adding gradient information of PDEs residuals (gPINNs) \cite{yu2022gradient}. The proposed adaptive sampling method (RAR, RAD) \cite{wu2023comprehensive} and the combination of PINN with numerical discrete format (CAN-PINN) \cite{chiu2022can} can effectively improve accuracy. Facing the problem of large space-time domain, cPINN \cite{jagtap2020conservative}, XPINNs \cite{jagtap2021extended}, Parallel PINNs \cite{shukla2021parallel}, etc. based on the regional decomposition strategy have been developed. DeepM\&Mnet \cite{mao2021deepm} and Multi-Head PINNs (MH-PINNs) \cite{zou2023hydra} are proposed to deal with multi-scale/multi-physics problems and multi-task collaboration problems respectively. Two mainstream operator networks, DeepONet \cite {lu2021learning} and FNO \cite{li2020fourier}, are proposed to learn the mapping from infinite dimension to infinite dimension, which avoids the problem of repeated training. In addition, there are various variants of PINNs including B-PINNs \cite{yang2021b}, fPINN \cite{pang2019fpinns}, hPINN \cite{lu2021physics}, hp-VPINNs \cite{kharazmi2021hp}, gwPINNs \cite{xiong2022gradient}, etc., which have injected new vitality into this field. PINN and its extensions have been widely applied to many scientific and engineering fields including fluid mechanics \cite{raissi2020hidden}, nano-optics \cite{chen2020physics}, biological systems \cite{yazdani2020systems}, thermodynamics \cite{cai2021physics}, to deal with various forms of equations (standard PDEs, integro-differential equation \cite{lu2021deepxde}, stochastic differential equation \cite{zhang2020learning}, fractional differential equation \cite{pang2019fpinns}, etc.). Of course, other frameworks based on deep learning to solve the forward and inverse problem of PDE are also worthy of attention, such as Deep Ritz Method (DRM) \cite{yu2018deep}, Deep Galerkin Method (DGM) \cite{sirignano2018dgm}, Weak Adversarial Networks (WAN) \cite{zang2020weak}, Sparse Identification of Nonlinear Dynamics (SINDy) \cite{brunton2016discovering}, PINN with sparse regression \cite{chen2021physics}, Local Extreme Learning Machines (loc-ELM) \cite{dong2021local}. In short, the proposal of the PINN method is revolutionary and has a milestone significance, which greatly promotes the development of scientific computing and related fields.

The coefficients of a constant coefficient differential equation are fixed constants, and it is usually a model described in a homogeneous medium and constant physical quantities. In the variable coefficient differential equation, the coefficient is a variable quantity, which is a function of the space-time variables ${\bm x}$ and $t$. The variable coefficient equation is more complex than the constant coefficient equation, but it can more realistically simulate the abundant physical phenomena in the real world, especially when we consider boundary effects, non-uniform media, and variable external forces. For example, the generalized variable coefficient Kadomtsev–Petviashvili equation \cite{li2010painleve} describing water waves in channels with variable width, depth, and density; the heat equation with variable coefficient \cite{costin2012borel} describing heat diffusion in heterogeneous media, and so on. Variable coefficient differential equations are not only more meaningful in practical applications but also more mathematically challenging.


In view of the importance of the variable coefficient model, the idea of applying the powerful PINN method to the variable coefficient equation is very natural. However, to the best of our knowledge research in this area is still lacking \cite{zhou2022data, song2023deep}. The standard PINN is difficult to deal with the case where the independent variable dimension of the coefficient function is different from the equation dimension. For example, in $(1+1)$-dimensional equations, variable coefficients related only to the time variable $t$ are common. But PINN can't do anything about this situation except to use the unsatisfactory strategy of adding soft constraints to the loss function. Therefore, this paper aims to propose a deep learning method that specifically deals with variable coefficient forward and inverse problems -- Variable Coefficient-Physical Information Neural Network (VC-PINN). It adds branch networks responsible for approximating variable coefficients on the basis of standard PINN, which adds hard constraints to variable coefficients, avoiding the problem of different dimensions between coefficients and equations. In addition, shortcut connections (ResNet structure) are introduced into the feed-forward neural network. This structure has solved an important problem (``network degradation") that hinders the learning of deep neural networks in the field of image processing \cite{he2016identity,he2016deep}. However, in the context of variable coefficient problems, ResNet not only alleviates gradient vanishing but also serves a new purpose. It has the function of unifying linear and nonlinear coefficients (to solve the network degradation problem of linear coefficients).


For the proposed new method, it is necessary to conduct extensive numerical experiments to test its performance. Therefore, the accuracy of the standard results required in the comparison of performance tests is extremely important. However, it is not easy to obtain the exact solution of the nonlinear partial differential equation, and it is even more difficult to obtain the exact solution of the variable coefficient problem. But there is a special kind of PDEs with good properties -- Integrable System. Its abundant exact solutions (multi-soliton, lump, breather, rogue wave) have provided sufficient sample space for PINN with constant coefficients \cite{li2020solving, peng2022n, pu2022data, pu2021solving, miao2022physics, wang2021data, fang2021data}. In addition, special integrable structures and integrable properties including conserved quantities \cite{lin2022two}, symmetries \cite{zhang2022enforcing} and Mirua transformations \cite{lin2023physics} have been successfully incorporated into PINN. And it is surprising that in the field of integrable systems, the exact solution of the variable coefficient problem can be obtained by generalizing the classical integrable method. This provides an exact sample rather than a high-precision numerical sample for testing the performance of the developed VC-PINN. It is worth mentioning that the combination of deep learning and integrable methods such as Lax pair, Darboux transform, inverse scattering transform, and Hirota bilinear is also expected in the future.


This paper is organized as follows. In Section \ref{method}, VC-PINN is introduced from four aspects: problem setting, ResNet structure, forward problem, and inverse problem. Section \ref{Forward_ex} and Section \ref{Inverse_ex} test the performance of the proposed method on forward and inverse problems on four equations (the variable coefficient Sine-Gordon, the generalized variable coefficient Kadomtsev–Petviashvili equation, the variable coefficient Korteweg-de Vries equation, the variable coefficient Sawada-Kotera equation), including the case of high dimensionality and coexistence of multiple variable coefficients. Several different forms of variable coefficients are involved (polynomials, trigonometric functions, fractions, oscillation damping coefficients). In Section \ref{analy_Dis}, an in-depth analysis of VC-PINN is made by combining theory and numerical experiments, including the necessity of ResNet; the relationship between the convexity of variable coefficients and learning; anti-noise analysis; the unity of forward and inverse problems/relationship with standard PINN. Finally, a summary of the article and some empirical conclusions are given in Section \ref{conclusion}.

\section{Variable coefficient Physics-informed neural networks}\label{method}
Based on the classic PINN, this section will propose a new deep learning framework to specifically deal with the forward and inverse problems of variable coefficient PDE, which is called VC-PINN. This section introduces this framework from four aspects: problem setting, ResNet structure, forward problem, and inverse problem.
\subsection{Problem setup}\label{setup}
\quad

Consider a class of evolution equations with time-varying coefficients in real space, as follows:
\begin{equation}\label{vceq1}
	 u_{t}={\bm {N}}[u]\cdot {\bm C}[t]^{T},\ {\bm x}\in \Omega,\ t\in [T_0,T_1],
\end{equation}
where $u=u({\bm {x},t})$ represents the real-valued solution of equation \eqref{vceq1}, $\Omega$ is a subset of $\mathbb{R}^N$, and the $N$-dimensional space vector ${\bm x}$ is recorded as ${\bm x} = (x_1,x_2, \cdots,x_N)$, so it is actually a $(N+1)$-dimensional evolution equation. ${\bm {N}}[\cdot]$ represents an operator vector, expressed in component form as ${\bm {N}}[u] = (N_1[u],N_2[u],\cdots)$. Each component $N_i$ is an operator, which usually includes but is not limited to linear or nonlinear differential operators. However, ${\bm C}[t]=(c_1(t),c_2(t),\cdots)$ is a coefficient vector whose component $c_i(t)$ is an analytical function of the time variable $t$, and ${\bm C}[t]$ has the same dimension as ${\bm N}[u]$. Furthermore, the product of the corresponding components of ${\bm N}[u]$ and ${\bm C}[t]$ represents an operator with time-varying coefficients ($c_i[t]N_i[u]$), usually interpreted as dispersion terms, higher-order terms of the equation. In particular, the variable coefficient considered here is only a function of the time variable $t$. The case where the variable coefficients are also correlated with the spatial variable ${\bm x}$ will be discussed in Section \ref{analy_Dis}.




For forward problems in the continuous sense, the expressions for the variable coefficients are fully known and written into the equations. In the treatment of such problems, it is sufficient to use the standard PINN method like the constant coefficient problem. But in engineering applications, the requirement of fully knowing the expressions of variable coefficients is harsh. Therefore, subsequent discussions based on forward and inverse problems are carried out in a discrete sense. Specifically, we use whether the variable coefficients are known or not in the discrete sense as the basis for distinguishing forward and inverse problems. Although this division is not too strict, it can be seen from the discussion and analysis in Section \ref{analy_Dis} that there are indeed essential differences in the performance of forward and inverse problems in this sense. Therefore, it is reasonable and meaningful to make such a distinction.

In view of the difference between the variable coefficient equation and the constant coefficient equation in the description of the forward and inverse problem, it is necessary to give a formal definition of the forward and inverse problem of PDEs under the variable coefficient version. As follows (Fig. \ref{foward_inverse_fig}):


\begin{figure}[htpb]
\centering
\includegraphics[width=18cm,height=10.16cm]{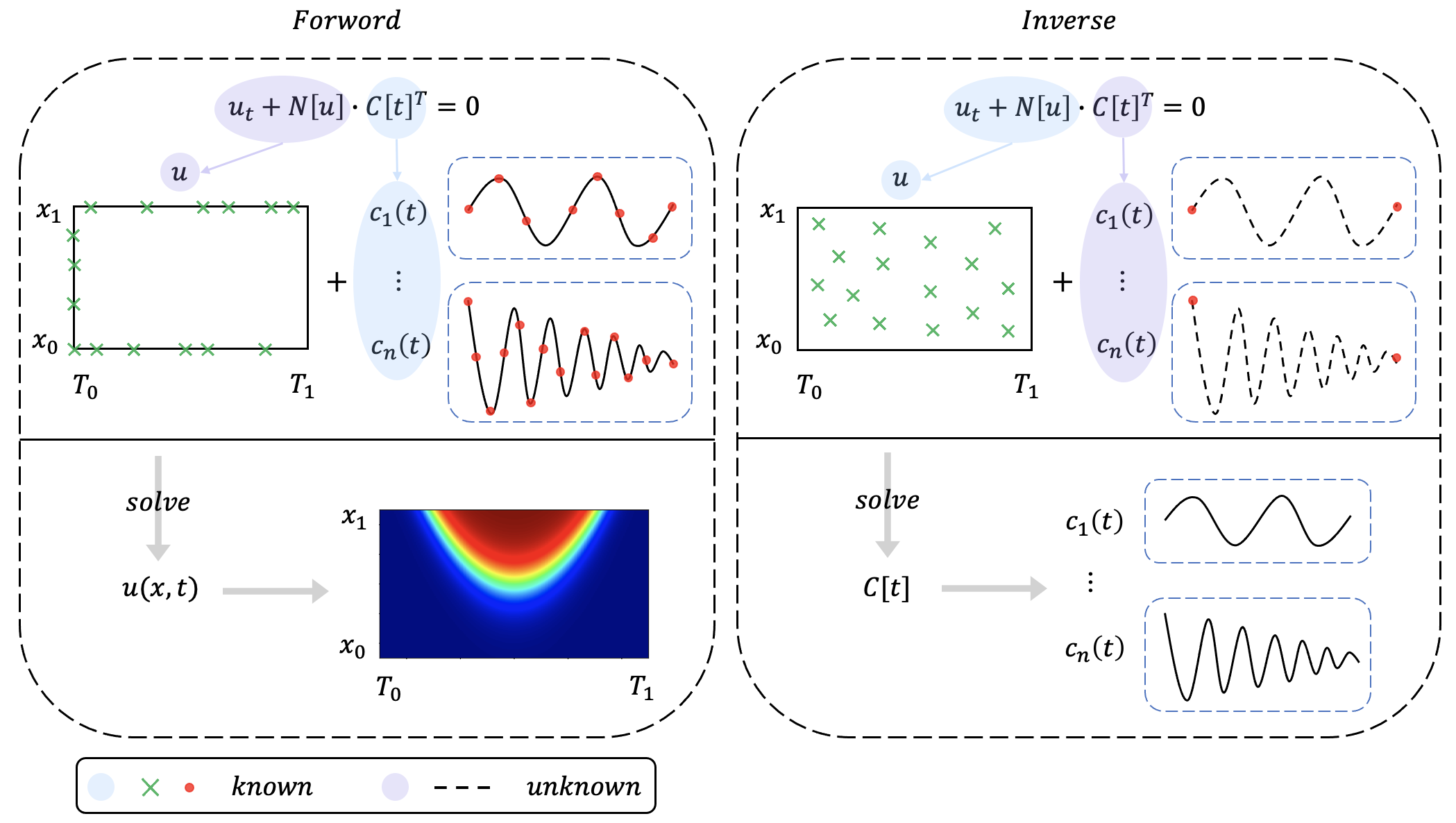}

\caption{(Color online) Schematic diagram of the forward and inverse problem of the variable coefficient equation in the discrete sense. (for simplicity, consider the space variables to be one-dimensional)}
\label{foward_inverse_fig}
\end{figure}

\begin{itemize}
	\item In the forward problem, variable coefficients $c_i(t)$ are known in a discrete sense. Specifically, the values of the coefficients $c_i(t)$ at a finite number of discrete points on $[T_0, T_1]$ are known information. These coefficient values correspond to some observable physical quantities (with varying degrees of noise) in practical applications. The forward problem with variable coefficients is thus formulated as solving $u$ over the region using the initial boundary value conditions for $u$ (consistent with constant coefficients) and the above discrete coefficient values.
	
	
	\item In the inverse problem, the coefficients to be determined are no longer several fixed constants, but a set of functions related to the time variable $t$. Consistent with the constant coefficient problem, the value of $u$ at finite discrete points in the region is known information. Besides, due to the multi-solution nature of the inverse problem, it is necessary to provide boundary values ($c_i(T_0)$ and $c_i(T_1)$) of the variable coefficients $c_i(t)$. \footnote{This is not a mandatory condition, and the requirements for boundary values can be appropriately relaxed according to the difficulty of the problem and the limitations of observation conditions.} This boundary information corresponds to the initial and terminal values of the observed quantities in the experiment. In summary, the inverse problem is described as using the above two known information to obtain the complete coefficient variation in the time domain $[T_0,T_1]$, that is, the discrete value of $c_i(t)$ at any time.

\end{itemize}

Because of the difference between the description of forward and inverse problems in the time-varying coefficient equation and the constant coefficient equation, it is necessary to propose a new PINN framework to deal with this type of specific equation. In addition, variable coefficients also bring new problems. It is well known that most complex physical phenomena are described by nonlinear models, but the nonlinearity of the equation does not mean that the function coefficients must also be nonlinear. In fact, many familiar physical quantities are linear as a function of an independent variable (not necessarily the time variable $t$), and of course, they may become function coefficients in nonlinear models. Therefore, how to unify linearity and nonlinearity in neural network methods will be a challenge brought by variable coefficients.

\subsection{ResNet structure.}
\quad

In the variable coefficient problem, not only $u$ needs to be represented by a neural network, but the variable coefficient $c_i(t)$ with different numbers of independent variables also needs to be approximated by a new network. Without loss of generality, in the method description, it is assumed that equation \eqref{vceq1} only involves a single variable coefficient, i.e. ${\bm C}[t]=c_1(t)$ and the corresponding operator vector ${\bm N}[u]$ is also abbreviated as ${\mathcal N}[u]$. The method in this paper is also applicable to the case of multiple variable coefficients. This simplification is only for a clearer description. In fact, examples of multiple variable coefficients are also shown in numerical experiments. 

In 2015, He et al. discovered an important problem that hinders the learning of deep neural networks--the problem of network degradation \cite{he2016deep}. That is, when using a deep network directly stacked by a shallow network, it is not only difficult to use the powerful feature extraction capabilities of the deep network, but even the accuracy is reduced. At the same time, they proposed a network structure (residual network: ResNet) that adds shortcut connections between network layers, which not only alleviates the disappearance of gradients but also solves the problem of network degradation. This structure is applied to image processing problems based on convolutional neural networks, and the accuracy has been improved unprecedentedly. This simple yet effective design is widely used and has developed many variants including DenseNet \cite{huang2017densely}. However, ResNet also appears in the known research on solving PDEs using deep learning frameworks. Our team introduces residual blocks in PINN to handle sine-Gordon with highly nonlinear terms that make classical PINN difficult to solve \cite{li2020physics}. Cheng et al. used PINN with ResNet blocks to achieve better results than classic PINN in fluid flow problems such as the Burgers equation and the Navier-Stokes equation \cite{cheng2021deep}. Niu's team respectively proposed adaptive learning rate residual network \cite{chen2022adaptive} and adaptive multi-scale neural network with resnet blocks \cite{chen2022adaptive2} based on the idea of shortcut connection to alleviate the gradient imbalance and multi-frequency oscillation encountered in the process of solving PDE.

In the variable coefficient problem, in addition to the above-mentioned known advantages, ResNet can better unify linearity and nonlinearity in a network to adapt to different variable coefficients. However, the unification of the above two is the problem of how the deep network approaches identity mapping, which is what He et al. mentioned in \cite{he2016deep}. Therefore, this paper also hopes to introduce the design of shortcut connections in the network structure of VC-PINN. Moreover, in Section \ref{nece_ResNet}, the necessity of using the ResNet structure in variable coefficient problems will be discussed more deeply in combination with the results of numerical experiments.

The ResNet used in this paper is not the original ResNet, but a ResNet with a new residual unit \cite{he2016identity}, which was also proposed by He et al. shortly after \cite{he2016deep} was published. The difference between the new ResNet and the original ResNet lies in the relative position of the activation function and shortcut connections. The activation function of the new ResNet is moved before the shortcut connection, and this connection mode is called ``pre-activation", which makes the prediction accuracy of the network improved again. (Unless otherwise specified, the ResNet mentioned later refers to the new ResNet.)

First, consider the most common feed-forward neural network (FNN) with a depth of $D$. The $0^{th}$ layer and the $D^{th}$ layer are respectively called the input layer and the output layer, and naturally there are $D-1$ hidden layers. A special requirement that must be considered before introducing shortcut connections is that the two vectors to be connected need to have the same dimension. A common approach is to use linear projections to match dimensions to satisfy the above conditions. In line with the principle of not introducing new network parameters as much as possible, it may be assumed that the number of nodes in each hidden layer is $N_d$. Then, a ResNet structure diagram with $N_B$ residual blocks and each residual block containing $N_h$ hidden layers is as follows:

\begin{figure}[htpb]
\centering
\includegraphics[width=17cm,height=6cm]{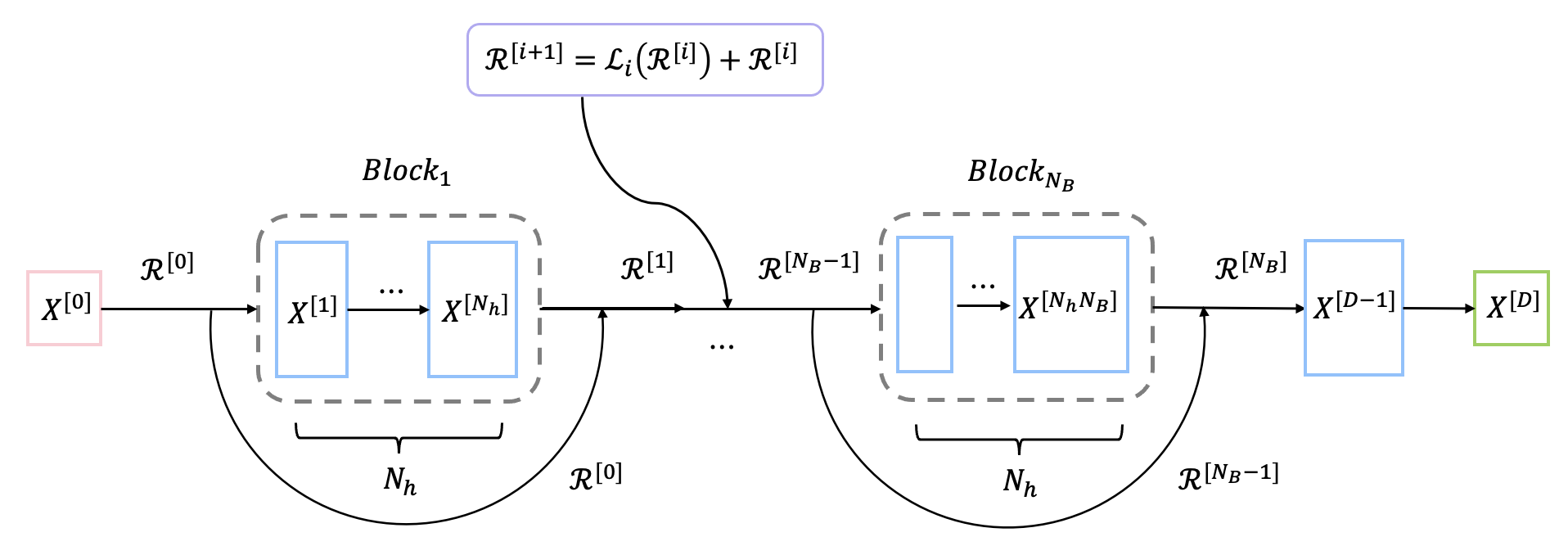}

\caption{(Color online) ResNet network structure diagram. It includes $N_B$ residual blocks, and each residual block contains $N_h$ hidden layers.}
\label{ResNet}
\end{figure}

As shown in Fig. \ref{ResNet}, $X^{[d]}$ represents the state vector of the $d^{th}$ layer node, and $R^{[i]}$ is both the output of the $i^{th}$ residual block and the input of the $(i+1)^{th}$ residual block. The $i^{th}$ residual block consists of layers $[(i-1)N_h + 1]^{th}$ to $(iN_h)^{th}$ of the network. So $R^{[i]}$ is equal to the state vector $X^{[iN_h+1]}$ of the $(iN_h+1)^{th}$ layer node. In order to show the structure of the network more clearly, and to express ResNet and ordinary FNN in a unified way, a more mathematical expression is used here to replace the simplified representation of ResNet mentioned in \cite{he2016identity}. The transformation relationship between the input and output of the residual block is expressed as:
\begin{equation}\label{eq2}
	\begin{split}
	X^{[0]}_{L}&=W^{[0]}X^{[0]}+b^{[0]},\\
	   {X}^{[1]}=\mathcal{R}^{[0]}&=\mathcal{K}X^{[0]}_{L}+(1-\mathcal{K})\mathcal{F}(X^{[0]}_{L}),\\
	{X}^{[iN_h+1]}=\mathcal{R}^{[i]}&=\mathcal{L}_{i}(\mathcal{R}^{[i-1]})+\mathcal{K}\mathcal{R}^{[i-1]},\ i=1,2,...,N_B,\\
	X^{[D]}&=W^{[D-1]}X^{[D-1]}+b^{[D-1]},\\
	&=W^{[D-1]}\mathcal{R}^{[N_B]}+b^{[D-1]},
	\end{split}
\end{equation}
where $X^{[0]}_{L}$ is an intermediate variable, and the coefficient $\mathcal{K}\in \{0, 1\}$ is mainly controls whether shortcut connections are included. Specifically, when $\mathcal{K}=1$, it is the ResNet structure, and when $\mathcal{K}=0$ is the ordinary FNN structure. The nonlinear map $\mathcal{L}_i$ is the nonlinear part between the input and output of the $i^{th}$ residual block, defined as follows:
\begin{equation}\label{eq3}
	\begin{split}
			&\mathcal{L}_i\triangleq\mathcal{T}_{iN_h}\circ \mathcal{T}_{iN_h-1}\circ\cdots\circ\mathcal{T}_{(i-1)N_h+1},\ i=1,2,...,N_B,\\
	&\mathcal{T}_i(X)\triangleq\mathcal{F}(W^{[i]}X+b^{[i]}),\ i=1,2,...,N_BN_h,
	\end{split}
\end{equation}
The above $W^{[i]}\in \mathbb{R}^{N_i\times N_{i+1}}$ and $b^{[i]}\in \mathbb{R}^{N_{i+1}}$ respectively represent the weight matrix and bias vector between the $i^{th}$ layer and the $(i+1)^{th}$ layer network, where $N_i$ is the number of nodes in the $i^{th}$ layer network, and $N_i=N_d,i=1,2,...,D-1$, which is the previous assumption. However, ``$\circ$" represents a function composite operator, and $\mathcal{F}$ is a nonlinear activation function, usually chosen as $Sigmoid$ function, $Tanh$ function, or $RelU$ function, etc. In addition, $\mathcal{T}_i$ is a nonlinear transformation, which is composed of a nonlinear activation function $\mathcal{F}$ and an affine transformation. There are a few key points to note about this ResNet structure:
\begin{itemize}
	\item In order to make the function represented by the above ResNet structure more directly approximate any linear function from a mathematical point of view, not only the connection mode of ``pre-activation" is adopted here, but also the linear projection without activation function is used to match the dimensions of the input layer and the first hidden layer. In this design, as long as the appropriate weight and bias are found to make $\mathcal{F}(W^{[i]}X+b^{[i]})=0$, the input will be truncated in the nonlinear layer, and only rely on the shortcut connections to propagate in the network, so it is easy for the linear output layer to approach any linear function (Fig. \ref{ResNet_2}). In particular, for activation functions that satisfy $\mathcal{F}(0)=0$ (such as $tanh$, etc.), only $W^{[i]}=0,\ b^{[i]}=0$ is required. This requirement is also very suitable for initialization methods with zero mean characteristics (such as $Glorot$, $Kaiming$, etc.).
\begin{figure}[htpb]
\centering
\includegraphics[width=15cm,height=6.88cm]{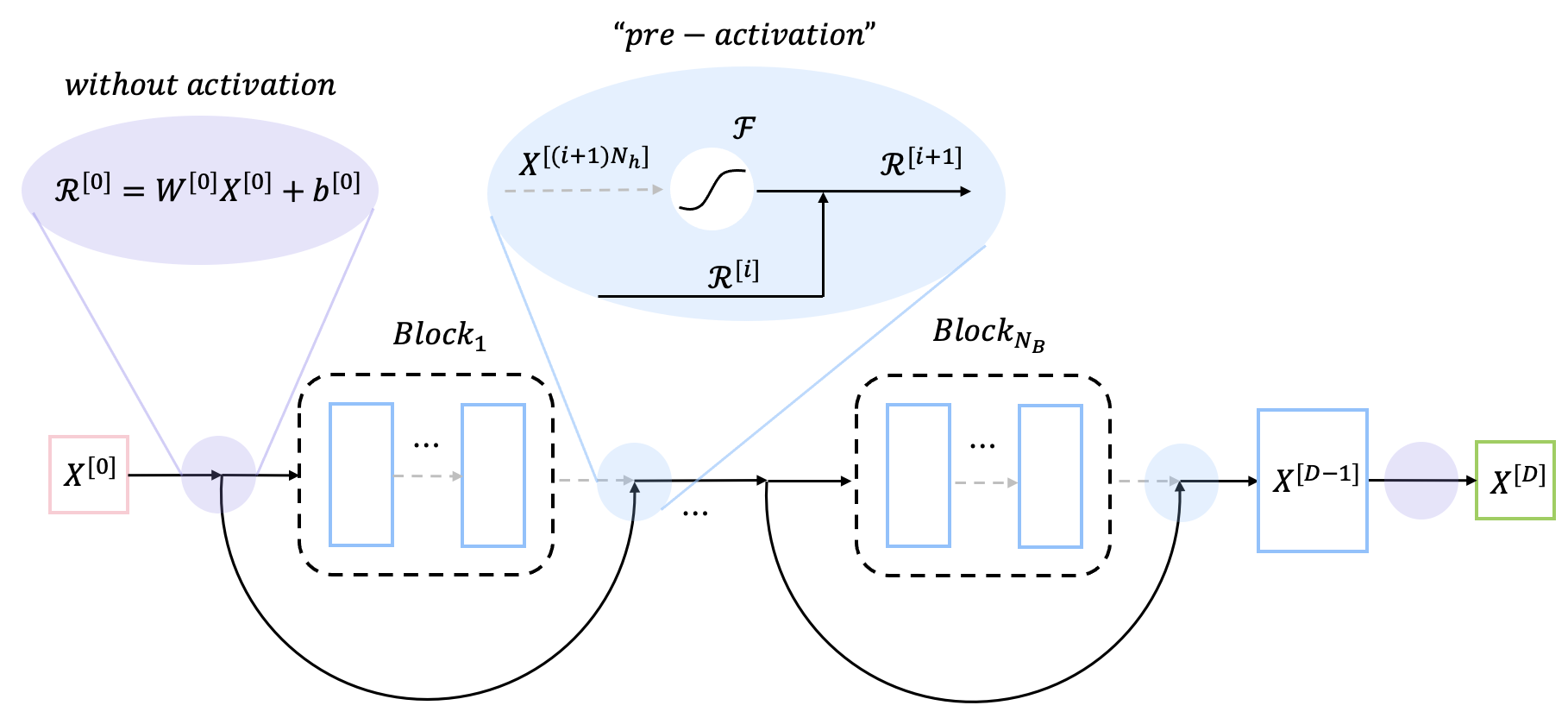}

\caption{(Color online) The ResNet network information flow diagram when approximating a linear map. The dotted line indicates that the information flow is 0 or a small amount. The figure also shows the ``pre-activation" connection mode.}
\label{ResNet_2}
\end{figure}
	
	\item The ResNet structure represented by formulas \eqref{eq2} and \eqref{eq3} is a very special case, each residual block has the same number of internal layers, and there is no nonlinear fully connected layer independent of the residual block except for the input layer and output layer (This means there is relation $D=N_BN_h+2$). In fact, according to the needs of mathematical and physical problems, the structure of residual blocks with different numbers of internal layers or the strategy of cross-use of residual blocks and nonlinear full connection layers are feasible. Of course, from the perspective of unified linearity and nonlinearity, the idea that we do not add additional independent nonlinear fully connected layers is particularly appropriate.
\end{itemize}

In order to distinguish different networks in the subsequent introduction, a specific ResNet is expressed as a structural parameter list form $NN\{D,N_d,N_B,N_h\}_{\mathcal{K}=1}$, where the definition of $\mathcal{K}$ is equivalent to formula \eqref{eq2}. To revisit, $D$ is the depth of the neural network, $N_d$ is the number of hidden layer nodes, $N_B$ is the number of residual blocks, and $N_h$ represents the number of network layers contained in each residual block. In fact, a specific ResNet depends entirely on the above four structural parameters when the dimensions of the input and output are given. On this basis, after choosing an appropriate initialization strategy, the initial state of the network will be completely clear. In particular, when $\mathcal{K}=0$, the parameter list $NN\{D,N_d,N_B,N_h\}_{\mathcal{K}=0}$ represents a common FNN, then the network structure is only determined by $D$ and $N_h$, so it is abbreviated as $NN\{D,N_h\}_{\mathcal{K}=0}$.

The $Glorot$ initialization method is taken into consideration, which is described as the biases $b^{[i]}$ are initialized to zero vectors, and the weights obey the normal distribution with a mathematical expectation of zero, as follows:
\begin{equation}
	W^{[i]}_{j,k}\sim N(0,\sigma_i^2),\ \sigma_i^2=\frac{2}{N_i+N_{i+1}},\ i=0,2,...,D-1,
\end{equation}
where $W^{[i]}_{j,k}$ represents the elements in the weight matrix $W^{[i]}$, and each element in the same weight matrix is independent and identically distributed.

As mentioned at the beginning of this section, two networks need to be constructed to approximate the solution $u$ and the variable coefficient $c_1(t)$ respectively. Both networks will adopt the above mentioned ResNet structure, call them trunk network $NN_u\{D_u,N_d^{u},N_B^{u},N_h^{u}\}$ and branch network $NN_c\{D_c,N_d^{c},N_B^{c},N_h^{c}\}$.\footnote{\ The parameter $\mathcal{K}$ will be omitted here and in the following description, because the number of parameters in the list can determine whether the network is a ResNet structure or FNN.}
\subsection{Forward Problem}\label{Forward_pro}
\quad

This section introduces the VC-PINN method from the forward problem of the time-varying coefficient equation articulated in Section \ref{setup}. Consider an initial boundary value problem ($Dirichlet$ boundary condition) for a partial differential equation involving a single time-varying coefficient:
\begin{equation}\label{cbeq}
	\begin{split}
		&u_{t}=c_1(t)\mathcal{N}[u],\ {\bm x}\in \Omega,\ t\in [T_0,T_1],\\
		&u({\bm x}, T_0)=g_0(\bm x),\ {\bm x}\in \Omega, \\
		&u({\bm x}, t)=g_{\Gamma}(\bm x ,t),\ {\bm x}\in \partial \Omega,\ t\in [T_0,T_1],
	\end{split}
\end{equation}
where $\partial \Omega$ represents the boundary of the space domain $\Omega$, the first equation of \eqref{cbeq} is a special case of \eqref{vceq1}, and the last two equations of \eqref{cbeq} correspond to the initial value condition and the $Dirichlet$ boundary condition respectively. When $c_1(t)$ is known (known in the discrete sense), the neural network method is used to solve the initial boundary value problem \eqref{cbeq}, and the key is to construct an optimization problem. 

The function represented by the trunk network $NN_u\{D_u,N_B^{u},N_h^{u},N_d^{u}\}$ is denoted as $\tilde{u}({\bm x},t;\theta_u)$, which will approach the real solution $u({\bm x}, t)$ of the initial boundary value problem \eqref{cbeq}, and the function represented by the branch network $NN_c\{D_c,N_B^{c},N_h^{c},N_d^{c}\}$ which is used to approach the real variable coefficient $c_1(t)$ is denoted as $\tilde{c}(t;\theta_c)$. $\theta_u \in \Theta_u$ and $\theta_c \in \Theta_c$ are the parameter spaces (weight and bias spaces) of the two networks of $NN_u$ and $NN_c$, respectively. In order to introduce the loss function, define the residual of the equation at the point $(\tilde{{\bm x}},\tilde{t})$ as follows:
\begin{equation}\label{feq}
	f(\tilde{\textbf{x}},\tilde{t};u,c):=\mathcal{N}_0[u,c]\big|_{\textbf{x}=\tilde{\textbf{x}},t=\tilde{t}},\ \mathcal{N}_0[u,c]=\partial_t[u]-c\mathcal{N}[u].
\end{equation}
The residual is derived from the first equation of \eqref{cbeq}, and $\mathcal{N}_0[\cdot,\cdot]$ is a new operator composed of $\mathcal{N}[\cdot]$ and $\partial_t$. However, $f(\sim,\sim;u,c)$ is treated as a function of functions that maps a point $(u,c)$ in the function space to a function on domain $\Omega\times[T_0,T_1]$, where $u=u({\bm x}, t)$ and $c=c(t)$ are interpreted as function types parameters. So the residual defined by \eqref{feq} measures how well the equation satisfies at the point $(\tilde{{\bm x}},\tilde{t})$ given the function $u({\bm x},t)$ and the coefficient $c(t)$. In particular, if $u_0$ is the solution of the initial boundary value problem \eqref{cbeq} under the variable coefficient $c_1(t)$, then obviously $f(\tilde{\textbf{x}},\tilde{t};u_0,c_1)=0,\ \forall \tilde{\textbf{x}}\in \Omega,\tilde{t}\in [T_0,T_1]$. Each set of parameters $\theta=\{\theta_u,\theta_c\}$ in parameter space $\Theta=\{\Theta_u,\Theta_c\}$ defines a function $\tilde{u}(\textbf{x},t;\theta_u)$ and variable coefficients $\tilde{c}(t;\theta_c)$. However, finding a suitable parameter $\theta^{\star}=\{\theta_u^{\star},\theta_c^{\star}\}$ from the parameter space $\Theta$ so that the residual $f(\tilde{\textbf{x}},\tilde{t};\tilde{u}^{\star},\tilde{c}^{\star})$ is close enough to zero on the domain $\Omega\times[T_0,T_1]$ is our goal. At the same time, if $\tilde{u}^{\star}$ satisfies the initial boundary value condition, then it is close enough to the real solution of the initial boundary value problem \eqref{cbeq}.

The loss function is the key to network optimization. In order to better measure the gap between real solution $u_0(\textbf{x},t)$ and $\tilde{u}(\textbf{x},t;\theta_u)$, a loss function composed of initial value constraints, boundary constraints, coefficient constraints, and physical equation constraints is constructed:
\begin{equation}\label{loss1}
    Loss(\theta) = Loss_{I}(\theta)+Loss_b(\theta)+Loss_f(\theta)+Loss_c(\theta)	,
\end{equation}
where
\begin{align}
			Loss_I(\theta)&=\frac{1}{n_I}\sum_{i=1}^{n_I}|\tilde{u}({\bm x}_I^i,T_0;\theta_u)-g_0(\textbf{x}_I^i)|^2,\\
		Loss_b(\theta)&=\frac{1}{n_b}\sum_{i=1}^{n_b}|\tilde{u}({\bm x}_b^i,t_b^i;\theta_u)-g_{\Gamma}({\bm x}_b^i,t_b^i)|^2,\\
		Loss_f(\theta)&=\frac{1}{n_f}\sum_{i=1}^{n_f}|f({\bm x}_f^i,t_f^i;\tilde{u}({\bm x},t;\theta_u),\tilde{c}(t;\theta_c))|^2\\
		Loss_c(\theta)&=\frac{1}{n_c}\sum_{i=1}^{n_c}|\tilde{c}(t_c^i;\theta_c)-c_c^i|^2.
\end{align}

However, $\{{\bm x}_I^i, u_I^i\}_{i=1}^{n_I}$, $\{{\bm x}_b^i, t_b^i, u_b^i\}_{i=1}^{n_b}$, $\{{\bm x}_f^i, t_f^i\}_{i=1}^{n_f}$ and $\{t_c^i, c_c^i\}_{i=1}^{n_c}$ represent four different types of point sets, which may be referred to as $I$-$type$ points, $b$-$type$ points, $f$-$type$ points, and $c$-$type$ points. $I$-$type$ points are initial value discrete points, and $u_I^i=g_0({\bm x}_I^i)$ is the value of the real solution $u_0$ at the spatial position ${\bm x}_I^i$ at $T_0$. Similarly, $b$-$type$ points are boundary value discrete points, and the value of the real solution $u_0$ at the spatiotemporal position $({\bm x}_b^i, t_b^i)$ is $u_b^i=g_{\Gamma}({\bm x}_b^i, t_b^i)$. The $f$-$type$ points represent internal collocation points, which are obtained by random sampling (uniform random sampling or Latin hypercube sampling, etc.) in $\Omega\times[T_0,T_1]$, and only contain space-time position information but not function values. Finally, the $c$-$type$ points are coefficient discrete points, and $c_c^i$ represents the real coefficient value at time $t_c^i$, which is the discretization of variable coefficients in the entire time domain in the forward problem. 

In the loss function \eqref{loss1}, $Loss_I$ and $Loss_b$ are initial value constraints and boundary constraints respectively, $Loss_f$ is the physical constraint, and $Loss_c$ is a unique coefficient constraint in variable coefficient problems. As an initial attempt at the variable coefficient problem, only the simplest balanced weight loss is considered here, which is beneficial to the analysis of the effect of the ResNet structure. Nevertheless, the performance improvement of regularization technology and a series of weight adjustment-based methods such as $Self$-$Adaptive$ $Loss$, $Point$-$Weighting$ $Method$ and $Soft$ $Attention$ $Mechanism$ on traditional PINN is obvious to all. Therefore, the effect of transplanting these modular technologies into our framework in the future is also expected. 

In order to find a local minimum point with good generalization of the loss function as a substitute for the global minimum point, our training strategy is to use a combination of first-order ($Adam$) and second-order ($L$-$BFGS$) optimization algorithms. From the perspective of the landscape, Adam first makes the iteration point quickly reach a good-quality area, and L-BFGS takes advantage of its second-order accuracy to explore ideal extreme points in this area. Furthermore, finding the derivatives of the network output $\tilde{u}({\bm x},t;\theta_u)$ with respect to ${\bm x}$ and $t$ involved in $Loss_f$ is trivial for AD. The generalization error is a measure of the generalization effect of the model and is defined as:
\begin{align}\label{error}
	e_u^r=\frac{\sqrt{{\sum_{i=1}^{n_{gu}}}|\tilde{u}({\bm x}_{gu}^i,t_{gu}^i;\theta^{\star})-u_{gu}^i|^2}}{\sqrt{{\sum_{i=1}^{n_{gu}}|u_{gu}^i|^2}}},\quad
	e_c^r=\frac{\sqrt{{\sum_{i=1}^{n_{gc}}}|\tilde{c}(t_{gc}^i;\theta^{\star})-c_{gc}^i|^2}}{\sqrt{{\sum_{i=1}^{n_{gc}}|c_{gc}^i|^2}}},
\end{align}
where $e_u^r$ and $e_c^r$ are the relative errors of the solution $u$ and the variable coefficient $c_1$, respectively. And $\tilde{u}({\bm x}_{gu}^i,t_{gu}^i;\theta^{\star})$ represents the generalization result of the optimal solution of the model at $({\bm x}_{gu}^i, t_{gu}^i)$, which is similar to $\tilde{c}(t_{gc}^i;\theta^{\star})$. The points in the set $\{{\bm x}_{gu}^i,t_{gu}^i,u_{gu}^i\}_{i=1}^{n_{gu}}$ are called $g_u$-$type$ points, which are the grid points in the full space-time domain and the corresponding value of the real solution $u_0$. The equidistant discontinuities in the time domain and the corresponding real values of the variable coefficient $c_1$ constitute point set $\{t_{gc}^i,c_{gc}^i\}_{i=1}^{n_{gc}}$, and the points included in it are called $g_c$-$type$ points. The above two types of point sets are the ``Standard Ruler" to measure the errors of solutions and coefficients. However \eqref{error} is a relative error calculated based on the $L^2$ norm, which mainly measures the average level of the error. If the generalization error is considered from different levels, the error formula based on other norms can also be used. For example, the error based on the $L^\infty$ norm measures the maximum error of the model.

\subsection{Inverse Problem}\label{Inverse_pro}
\quad

As described in Section \ref{setup}, the real solution $u_0$ of the equation in the inverse problem is known in a discrete sense, and the variable coefficient $c_1$ becomes our goal. Therefore, the formulation of the question has also changed from \eqref{cbeq} to
\begin{equation}\label{ip}
	\begin{split}
		&u_{t}=c_1(t)\mathcal{N}[u],\ {\bm x}\in \Omega,\ t\in [T_0,T_1],\\
&c_1(T_0)=\mathcal{C}_0,\ c_1(T_1)=\mathcal{C}_1,
	\end{split}
\end{equation}
where the first line is the original equation, and the second line represents the two-terminal conditions of the variable coefficients. In simple problems, this condition can be relaxed to a single endpoint or even not needed, whereas in more challenging problems information on higher derivatives at both endpoints and even interior point information is required. The condition for higher-order derivatives is also given here:
\begin{equation}\label{hdc}
	\frac{\partial ^{k} c_1}{\partial t^{k}}\bigg|_{t=T_0}=\mathcal{C}_0^{(k)},\ 	\frac{\partial ^{k} c_1}{\partial t^{k}}\bigg|_{t=T_1}=\mathcal{C}_1^{(k)},k=1,2,\cdots,
\end{equation}
where $\mathcal{C}_0^{(k)}$ and $\mathcal{C}_1^{(k)}$ are the corresponding higher order derivative values at both ends. This condition is mentioned here because we used it in the multiple variable coefficient example in Section \ref{vSK}, but it is not required for all problems. Under the framework of VC-PINN, dealing with the inverse problem of variable coefficients has a certain unity with its forward problem, and the change almost only occurs in the composition of the loss function. The specific differences are as follows:
\begin{equation}
	\begin{split}
		&Loss_I(\theta)+Loss_b(\theta)\rightarrow Loss_s(\theta),\\
		&Loss_s(\theta)=\frac{1}{n_s}\sum_{i=1}^{n_s}|\tilde{u}({\bm x}_s^i,t_s^i;\theta_u)-u_s^i|^2,
	\end{split}
\end{equation}
where the sampling points of the real solution $u_0$ in the full space-time region are called $s$-$type$ points, they constitute the set $\{{\bm x}_s^i,t_s^i,u_s^i\}_{i=1}^{n_s}$, and $u_s^i$ is the value of the real solution $u_0$ at point $({\bm x}_s^i, t_s^i)$. The $s$-$type$ point represents the known solution information (considered as an observable quantity in practice), and it is regarded as obtained by random sampling in the method introduction.\footnote{\ In specific problems, because efficiency and cost need to be considered, the distribution of s-type points is usually determined after careful consideration, such as the distribution of observation buoys in the ocean.} Thus, the loss function of inverse problem \eqref{ip} is (regardless of condition \eqref{hdc}):
\begin{equation}
	Loss(\theta)=Loss_s(\theta)+Loss_f(\theta)+Loss_c(\theta).
\end{equation}
The loss function of the inverse problem replaces the $I$-$type$ point and the $b$-$type$ point with the $s$-$type$ point. In fact, they all represent the known information of the solution. The difference is that one represents the initial value and boundary information, and the other represents internal information. In addition, the change to the loss item $Loss_c$ is exactly the opposite of $Loss_s$. The form of the loss remains unchanged, but the known information changes from the inside to the boundary, so the $c$-$type$ points in the inverse problem usually only contain $2$ points.

\begin{figure}[htpb]
\centering
\includegraphics[width=16cm,height=8.5cm]{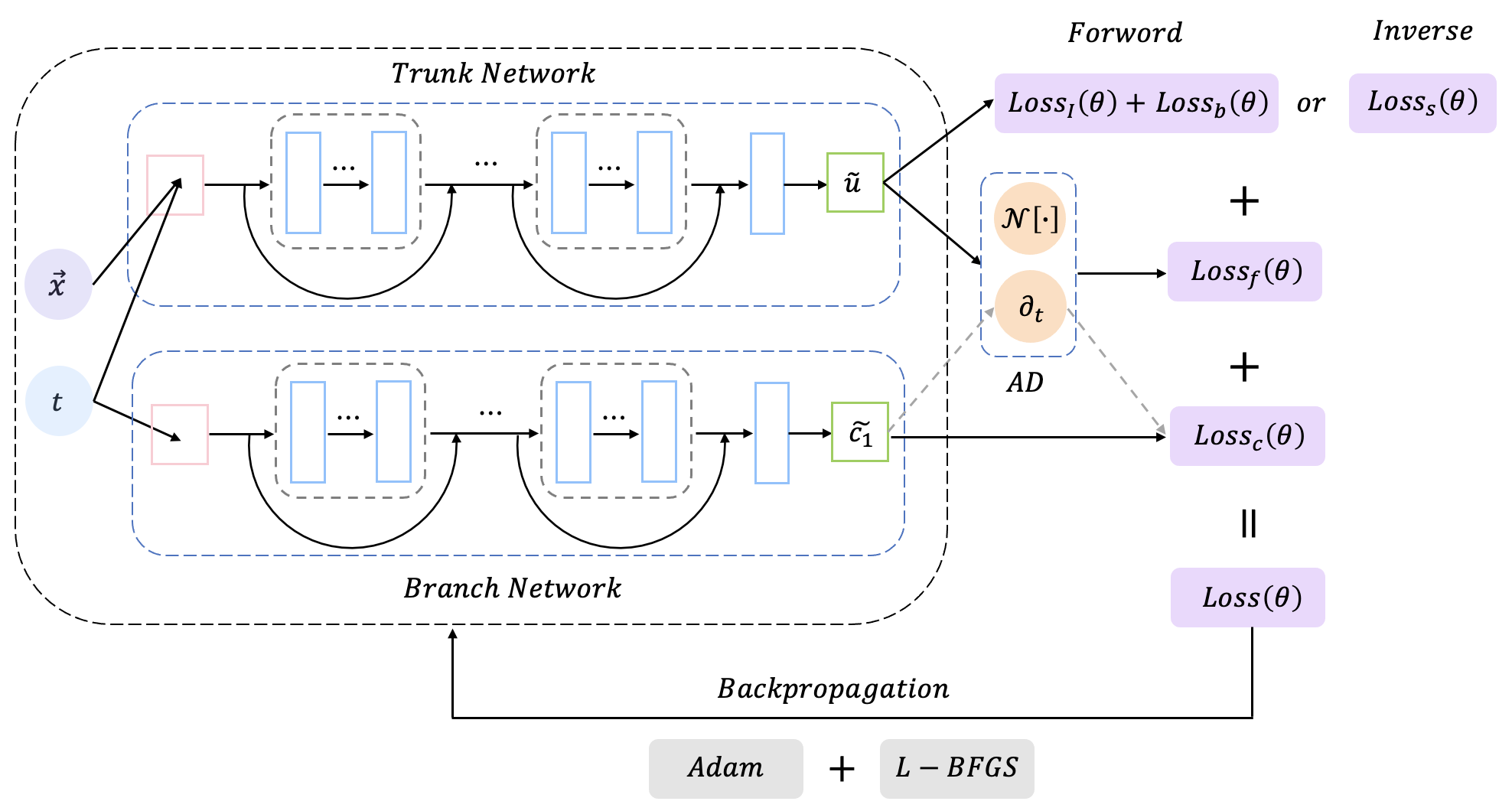}

\caption{(Color online) Flow chart of VC-PINN in forward and inverse problems.}
\label{flow_chart}
\end{figure}

In summary, Fig. \ref{flow_chart} shows the framework of the VC-PINN method, which includes forward and inverse problems. All code in this article is based on Python 3.7 and TensorFlow 1.15, and all numerical examples reported later were run on a DELL Precision 7920 Tower computer with a 2.10 GHz 8-core Xeon Silver 4110 processor, 64 GB of memory, and a GTX 1080Ti GPU.
\section{Numerical experiments on forward problems}\label{Forward_ex}
This section shows two different numerical examples of the VC-PINN method in the variable coefficient forward problem. They are the variable coefficient Sine-Gordon equation and the generalized variable coefficient Kadomtsev–Petviashvili equation. 
The two equations are different in dimension (the first equation is $(1+1)$-dimensional, while the second equation is $(2+1)$-dimensional), and this setting is for the consideration of testing the effect of the proposed method in different dimensions.


\subsection{Sine-Gordon equation with variable coefficient}\label{vSG_forward}
\quad

The Sine-Gordon (SG) equation with constant coefficients is
\begin{equation}
	u_{xt}+\text{sin}(u)=0,
\end{equation}
which is a hyperbolic partial differential equation. Edmond Bour originally proposed it in his study of surfaces with constant negative curvature\cite{bour1891theorie}, and Frenkel and Kontorova rediscovered it in their study of crystal dislocations in 1938\cite{kontorova1938theory}. In addition to differential geometry and crystal dislocation motion, the SG equations explain important nonlinear phenomena in branches of modern science including nonlinear quantum field theory, plasma physics, ultrashort optical pulse propagation, and DNA soliton dynamics \cite{gl1971analytical, sazonov2001extremely, yang2014analytical, fabian2009perturbation}. However, the non-uniformity exhibited by non-autonomous SG equations with time-varying coefficients is also worthy of attention. Give the SG equation with variable coefficients (vSG):
\begin{equation}\label{vSG}
	u_{xt}+h(t)\text{sin}(u)=0,
\end{equation}
where $h(t)$ is an analytical function that represents how the coefficients of the equation change over time. The vSG equation plays a crucial role in spin-wave propagation with variable interaction strength and in the flux dynamics of Josephson junctions with impurities \cite{braun2004frenkel}. The work of \cite{wazwaz2019integrable} proves that for any analytical function $h(t)$, equation \eqref{vSG} passes the Painlevé test (verifying the integrability) and provides its analytical solution, as follows:
\begin{equation}
	u(x,t)=4\text{arctan}\left(\frac{f(x,t)}{g(x,t)}\right).
\end{equation}
The form of this solution is consistent with the SG with constant coefficients, the difference is the constraints on the two auxiliary functions. In particular, considering the single soliton solution (multiple solitons solution is usually singular) of equation \eqref{vSG}, then we have
\begin{align}
	&f(x,t)=e^{k_1x-\omega_1(t)},\ g(x,t)=1,\\
	\label{eq1}&\omega_1(t)=\int\frac{h(t)}{k_1}dt,
\end{align}
where $k_1\in \mathbb{R}$ is a free parameter. Then the single soliton solution of \eqref{vSG} is derived as:
\begin{equation}\label{sol_vSG}
	u(x,t)=4\text{arctan}\left(e^{k_1x-\omega_1(t)}\right),
\end{equation}
where $\omega_1$ is determined by \eqref{eq1}. Although the solution \eqref{sol_vSG} represents a single soliton solution, in fact, $\omega_1$ in the formula is obtained by the indefinite integral of the coefficient function $h(t)$, so choosing different $h(t)$ will produce many solutions with rich dynamic behavior, which is not found in the constant coefficient problem. Next, we discuss the coefficient functions of the three forms (first-degree polynomial, quadratic polynomial, trigonometric function), and use the proposed VC-PINN method to obtain a data-driven solution to the corresponding initial boundary value problem. The initial boundary value data required in $Loss_I$ and $Loss_b$ and the discrete coefficient value required in $Loss_c$ are respectively obtained by the exact solution \eqref{sol_vSG} and the given coefficient function. Details are as follows:
\begin{itemize}
	\item \textbf{First-degree polynomial:} Assuming $h(t)=t$, and taking the integral constant\footnote{\ Unless otherwise specified, when it comes to integral constants, it defaults to 0.} as $0$ in the indefinite integral \eqref{eq1}, the exact solution of equation \eqref{vSG} is given as follows:
	\begin{equation}\label{vSG1}
		u^{(vSG)}_1= 4\text{arctan}\left(e^{-\frac{t^2}{2k_1}+k_1x}\right).
	\end{equation}
	
	The two data-driven solutions found by the VC-PINN approach in the scenario where the free parameter $k_1=\pm 1$ and the accompanying error outcomes are shown in Fig. \ref{SG-p1}.
	\begin{figure}[htpb]
    \centering
    \begin{subfigure}[b]{0.48\textwidth}
    \includegraphics[width=9cm,height=5.5cm]{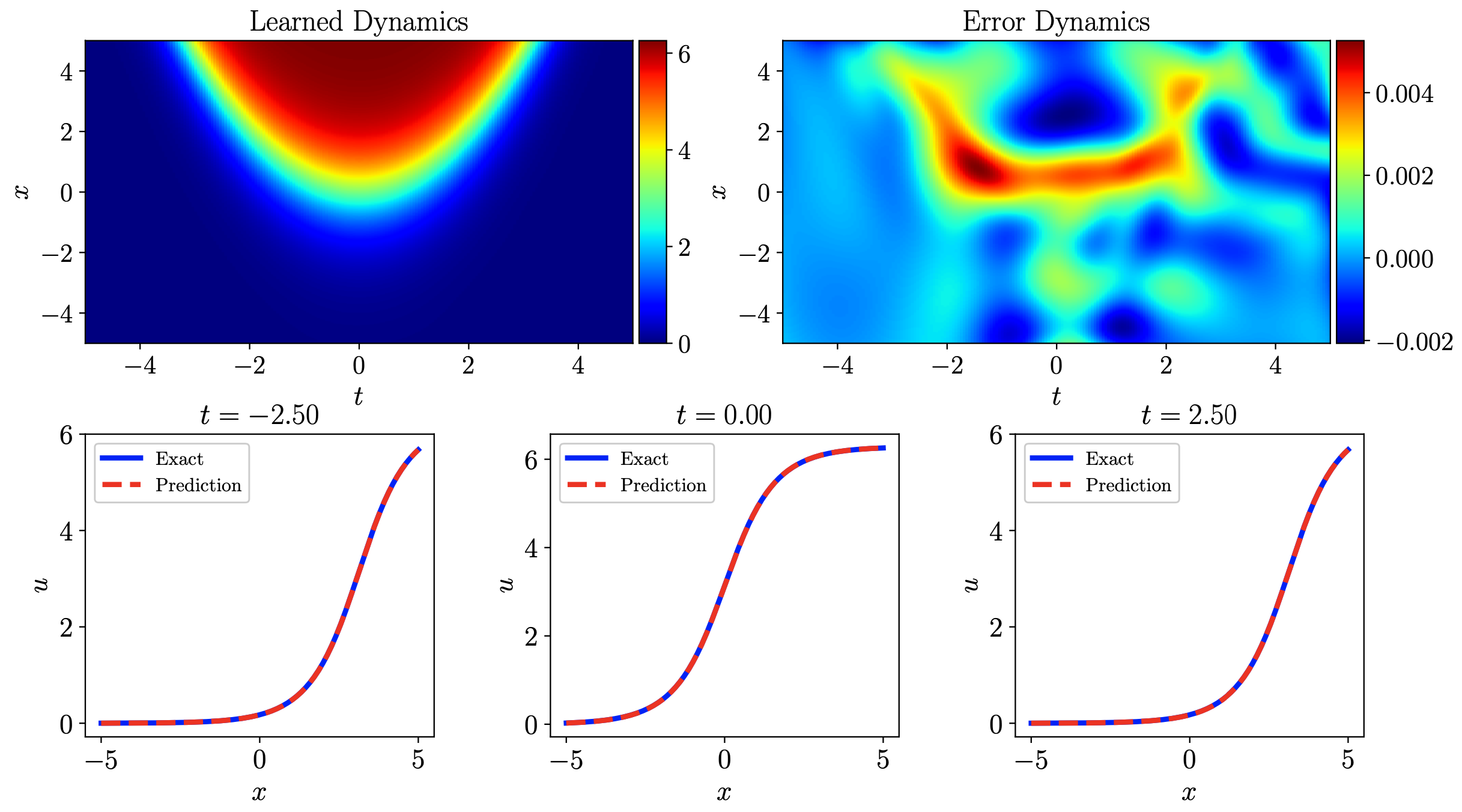}
    \caption*{(a)}
    \end{subfigure}
    \hspace{0.3cm} 
    \begin{subfigure}[b]{0.39\textwidth}
    \includegraphics[width=7cm,height=5.5cm]{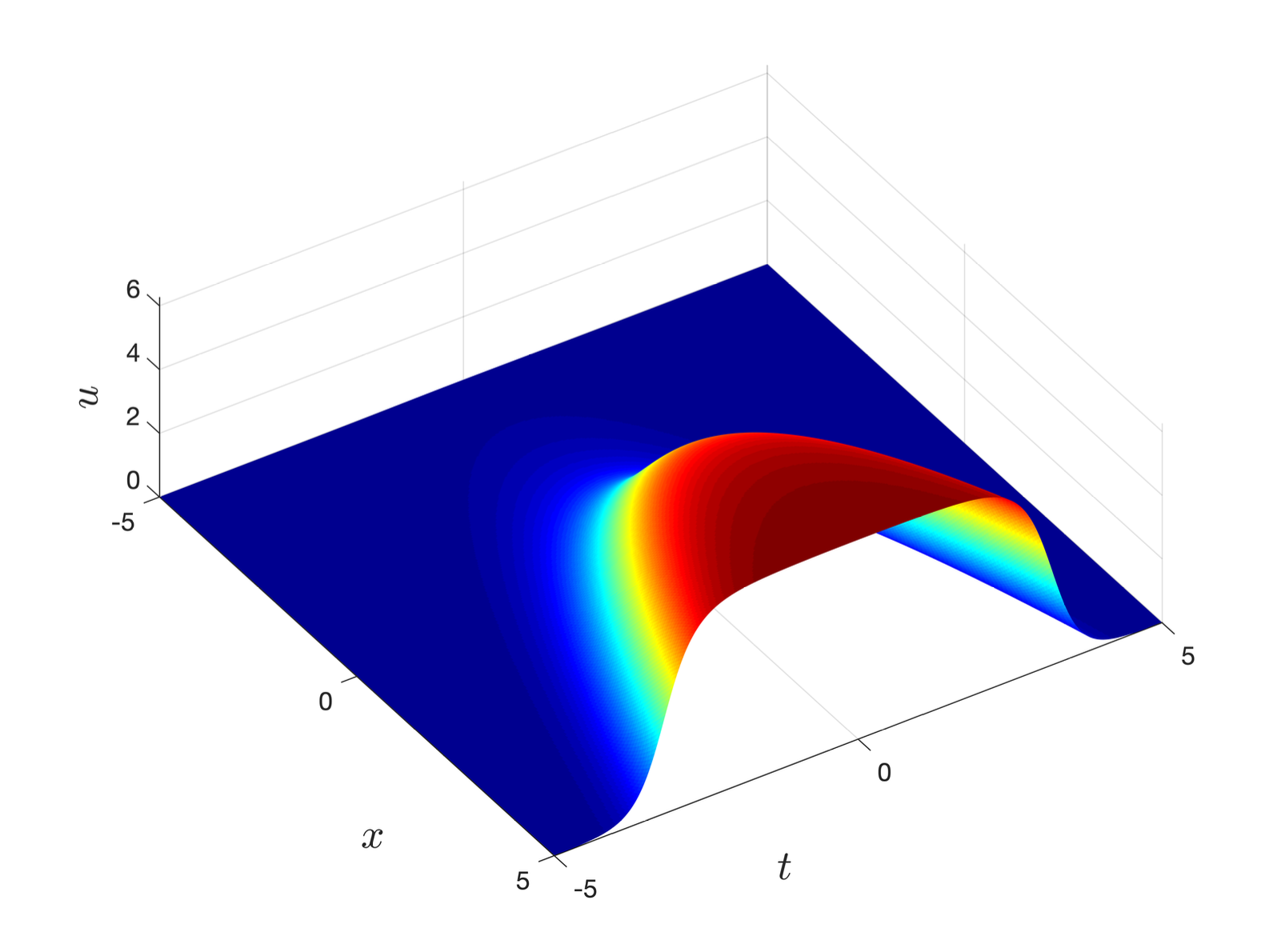}
    \caption*{(b)}
    \end{subfigure}\\
    \begin{subfigure}[b]{0.48\textwidth}
    \includegraphics[width=9cm,height=5.5cm]{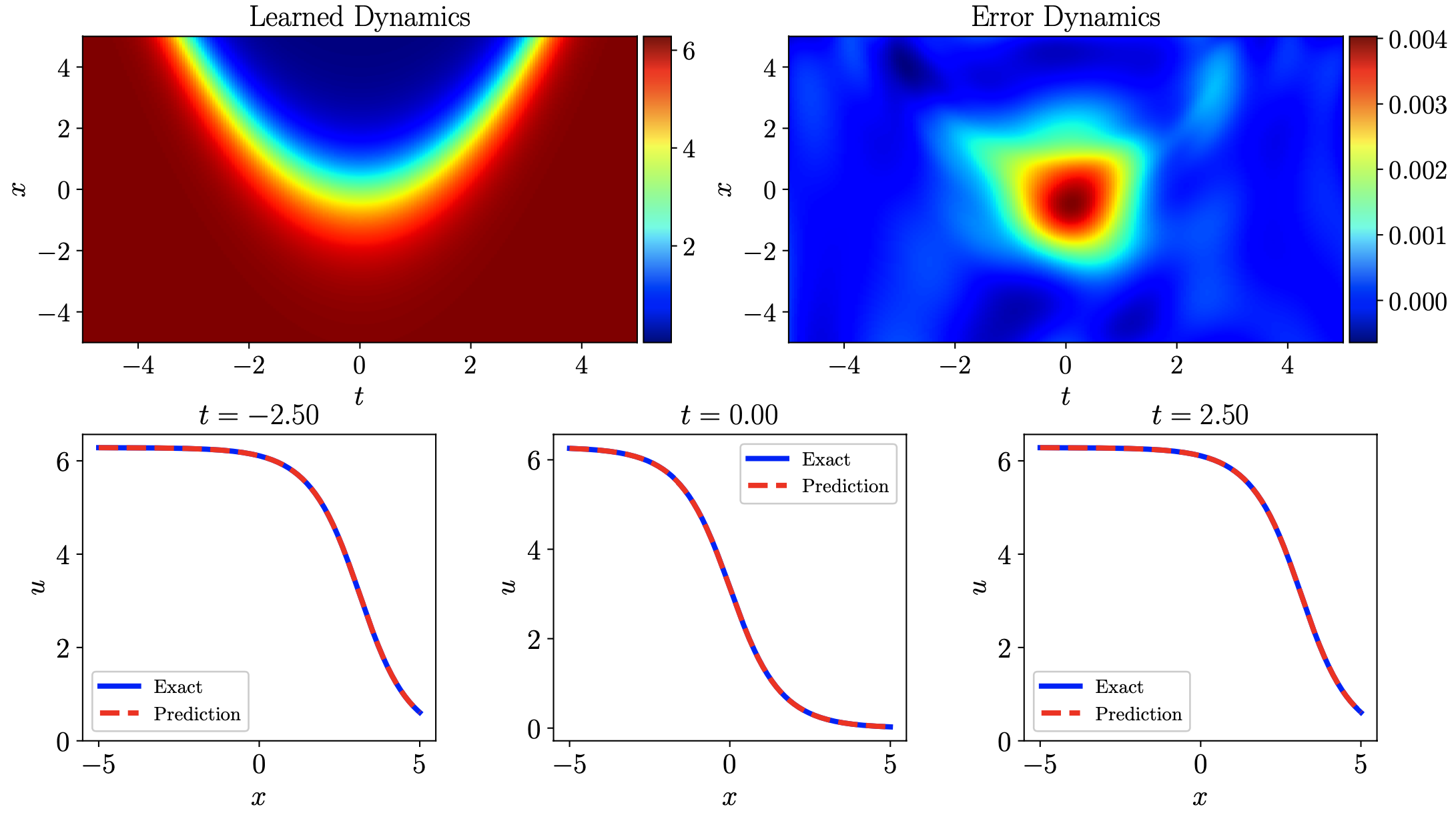}
    \caption*{(c)}
    \end{subfigure}
    \hspace{0.3cm} 
    \begin{subfigure}[b]{0.39\textwidth}
    \includegraphics[width=7cm,height=5.5cm]{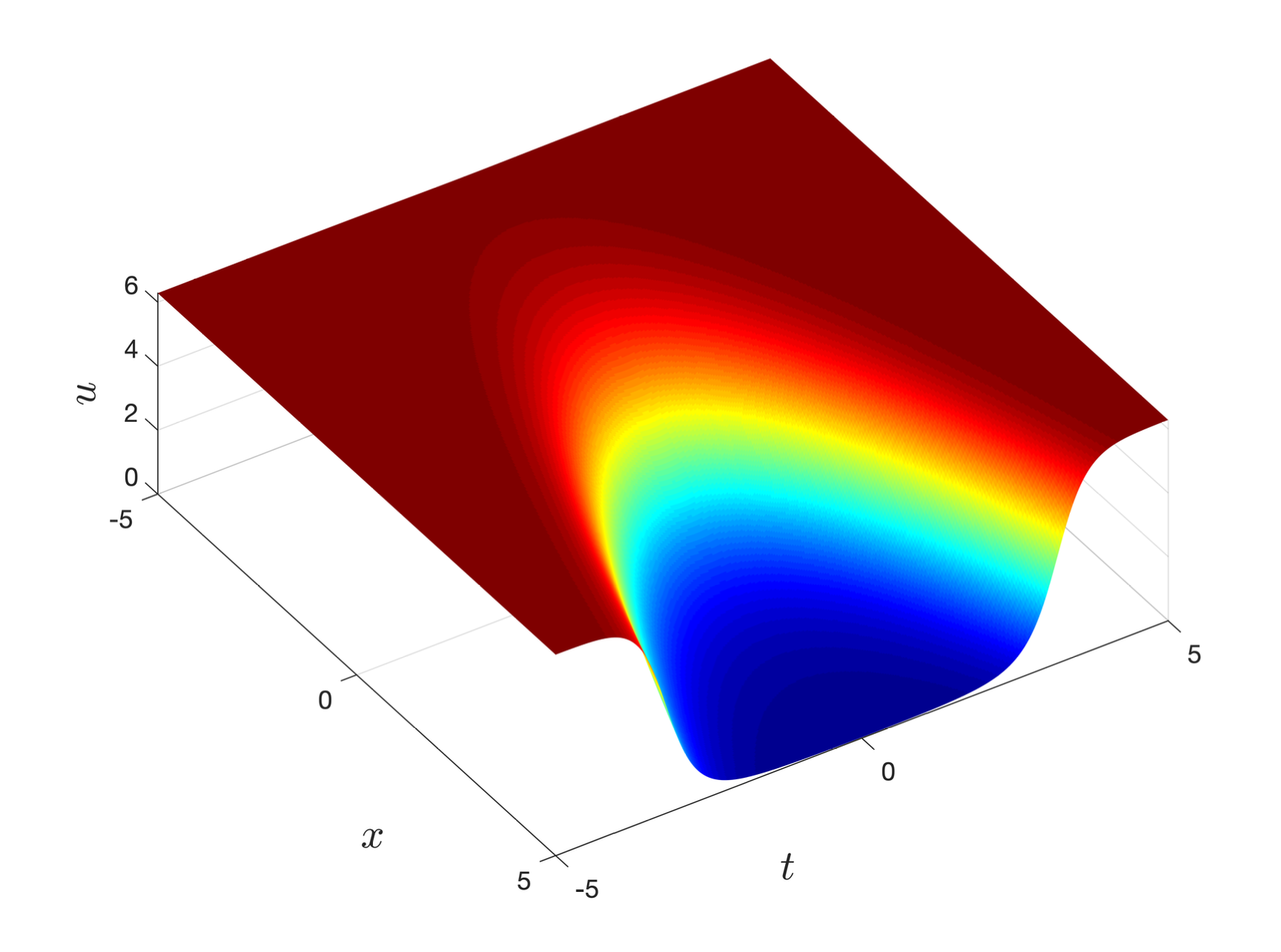}
    \caption*{(d)}
    \end{subfigure}
    \caption{(Color online) The kink-like solution $u_1^{(vSG)}$ for vSG equation with linear coefficients((a)-(b) corresponds to $k_1=1$; (c)-(d) corresponds to $k_1=-1$). (a) and (c): The density plot of the data-driven solution and corresponding error are located in the upper part. The comparison of the three time snapshot curves of the exact solution and the data-driven solution is located in the lower part. (b) and (d): $3D$ surface plots of the data-driven solution.}
    \label{SG-p1}
    \end{figure}
   
    The positive or negative of the parameter $k_1$ controls whether the solution presents a bulging convex hull or a collapsed concave hull. In fact, they all evolved from the single kink solution of the SG equation in the case of constant coefficients, and the coefficient function determines the moment and position of the kink. The linear coefficient function becomes a quadratic polynomial form through indefinite integration, which is why the kink shown in Fig. \ref{SG-p1} appears roughly as a parabola. The $L^2$ relative generalization errors $e^r_u$ of the data-driven solution for the case $k_1=\pm 1$ are $1.92\times 10^{-4}$ and $4.11 \times 10^{-4}$, respectively, which shows that the proposed method well captures the dynamical behavior of the kink-like solution of the vSG equation.
	\item \textbf{Quadratic polynomial:} Assuming that the coefficient function is $h(t)=t^2$, the exact solution of equation \eqref{vSG} is derived as:
	\begin{equation}\label{vSG2}
		u_2^{(vSG)}=4\text{arctan}\left(e^{-\frac{t^3}{3k_1}+k_1x}\right).
	\end{equation}
	
	From the expression \eqref{vSG2} of the solution $u_2^{(vSG)}$, it is observed that when the coefficient function is a quadratic polynomial, the degree of the time variable $t$ is no longer an even power as in \eqref{vSG1}, but an odd power. Therefore, the data-driven solution obtained by taking the free parameter $k_1$ as the opposite number is only a mirror symmetry of the space-time coordinates (it does not present two completely different behaviors as in Fig. \ref{SG-p1}), so only $k_1=1$ is discussed here. Similar to the linear coefficients, the locations where the kinks occur completely reveal the shape of a cubic polynomial, which is evident in Fig. \ref{SG-p2}. The $L^2$ relative generalization errors of the obtained data-driven solution is $e^r_u=9.55\times 10^{-5}$. Combined with the error density map and three time snapshots, it can be seen that the dynamic behavior of the vSG equation under the quadratic polynomial coefficients has been successfully learned.
	\begin{figure}[htpb]
    \centering
    \begin{subfigure}[b]{0.48\textwidth}
    \includegraphics[width=9cm,height=5.5cm]{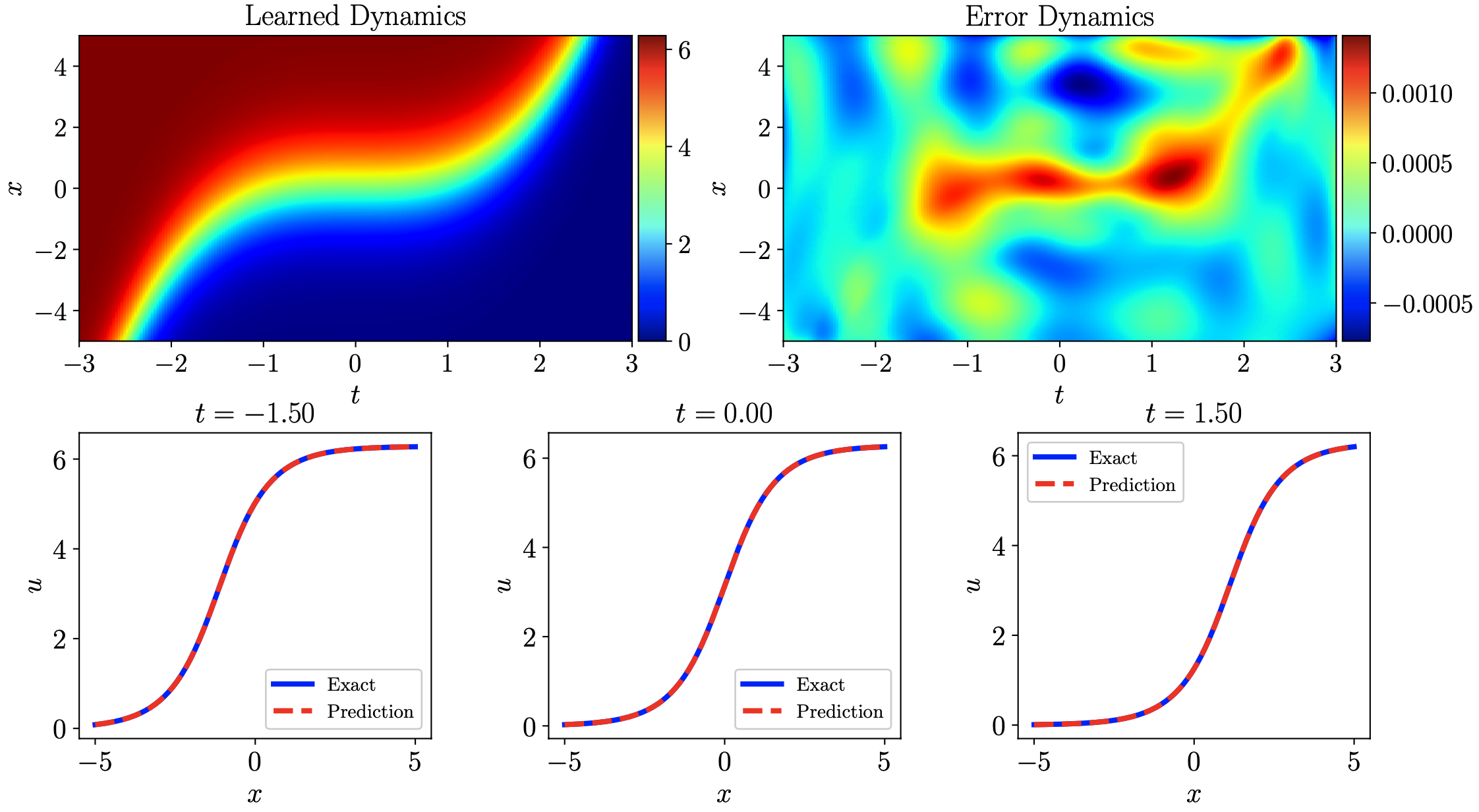}
    \caption*{(a)}
    \end{subfigure}
    \hspace{0.3cm} 
    \begin{subfigure}[b]{0.39\textwidth}
    \includegraphics[width=7cm,height=5.5cm]{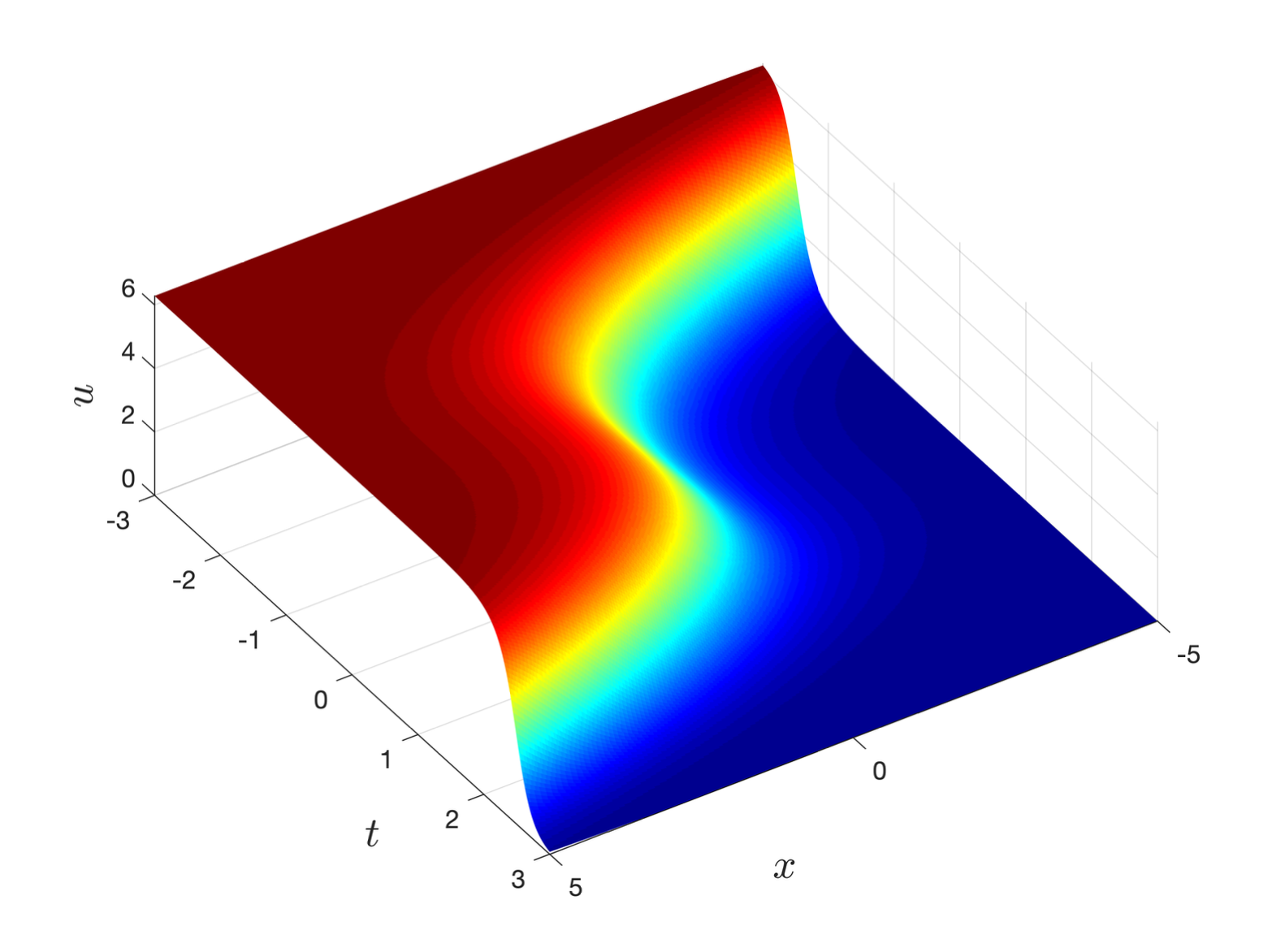}
    \caption*{(b)}
    \end{subfigure}\\
    \caption{(Color online) The kink-like solution $u_2^{(vSG)}$ of the vSG equation under quadratic coefficients ($k_1=1$). (a): The density plot of the data-driven solution and corresponding error are located in the upper part. The comparison of the three time snapshot curves of the exact solution and the data-driven solution is located in the lower part. (b): $3D$ surface plots of the data-driven solution.}
    \label{SG-p2}
    \end{figure}
	
	\item \textbf{Trigonometric function:} When the coefficient function is a cosine function with periodic properties, that is, $h(t)=3\text{cos}(2t)$, the corresponding exact solution is
	\begin{equation}
		u_3^{(vSG)}=4\text{arctan}\left(e^{-\frac{3\text{sin}\left(2t\right)}{2k_1}+k_1x}\right).
	\end{equation}
	
	The coefficient function with periodic properties determines that kink also appear periodically. The density plot of the error in Fig. \ref{SG-cos} clearly shows that where the kink occurs is accompanied by a larger error (in a relative sense), as we found in \cite{miao2022physics}, there is a strong correlation between high error and large gradient. Surprisingly, the proposed method also achieves satisfactory results under coefficients with periodic properties: the relative $L^2$ generalization error of the data-driven solution is $e^r_u=1.73\times 10^{-3}$. More detailed graphical results are displayed in Fig. \ref{SG-cos}.
	\begin{figure}[htpb]
    \centering
    \begin{subfigure}[b]{0.48\textwidth}
    \includegraphics[width=9cm,height=5.5cm]{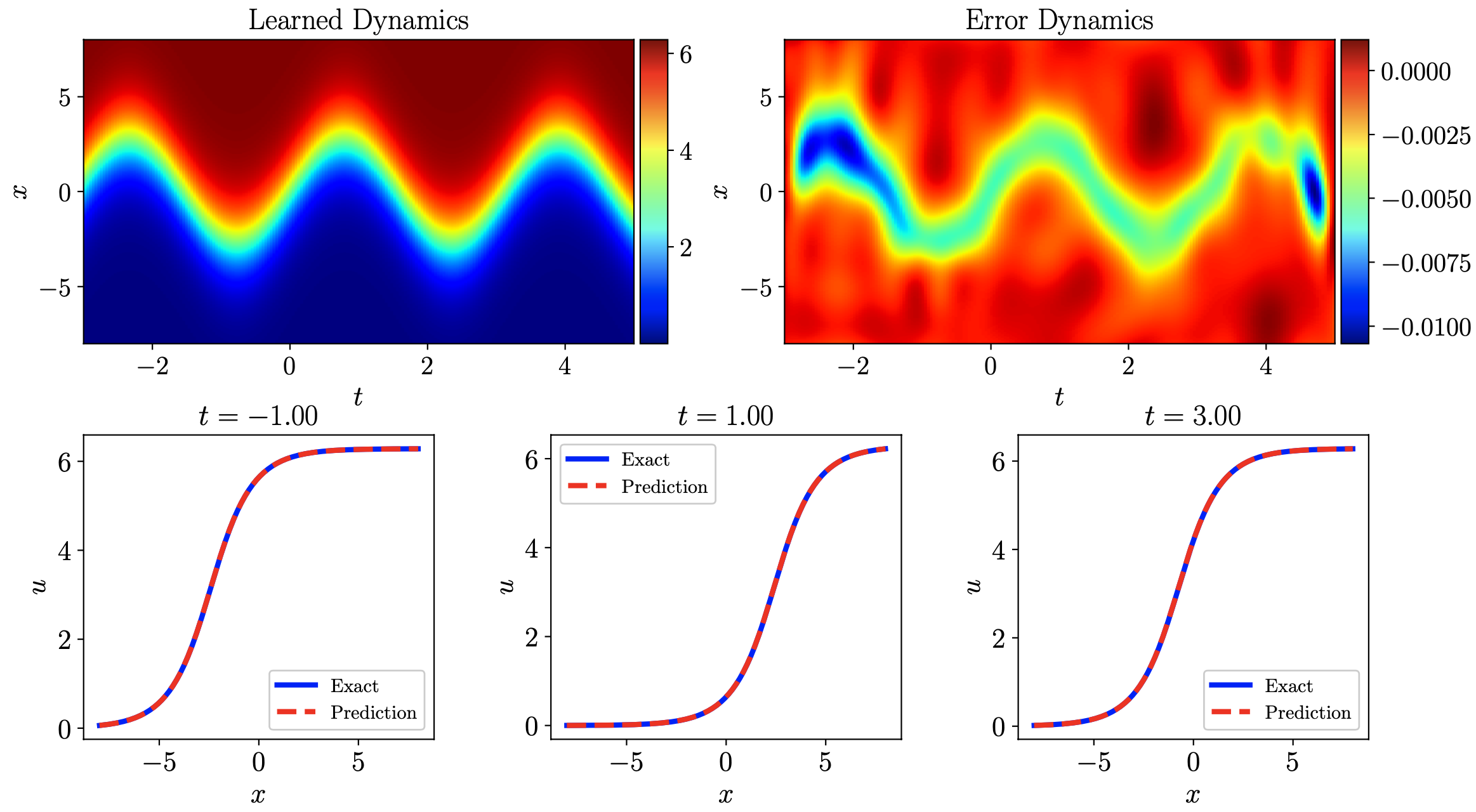}
    \caption*{(a)}
    \end{subfigure}
    \hspace{0.3cm} 
    \begin{subfigure}[b]{0.39\textwidth}
    \includegraphics[width=7cm,height=5.5cm]{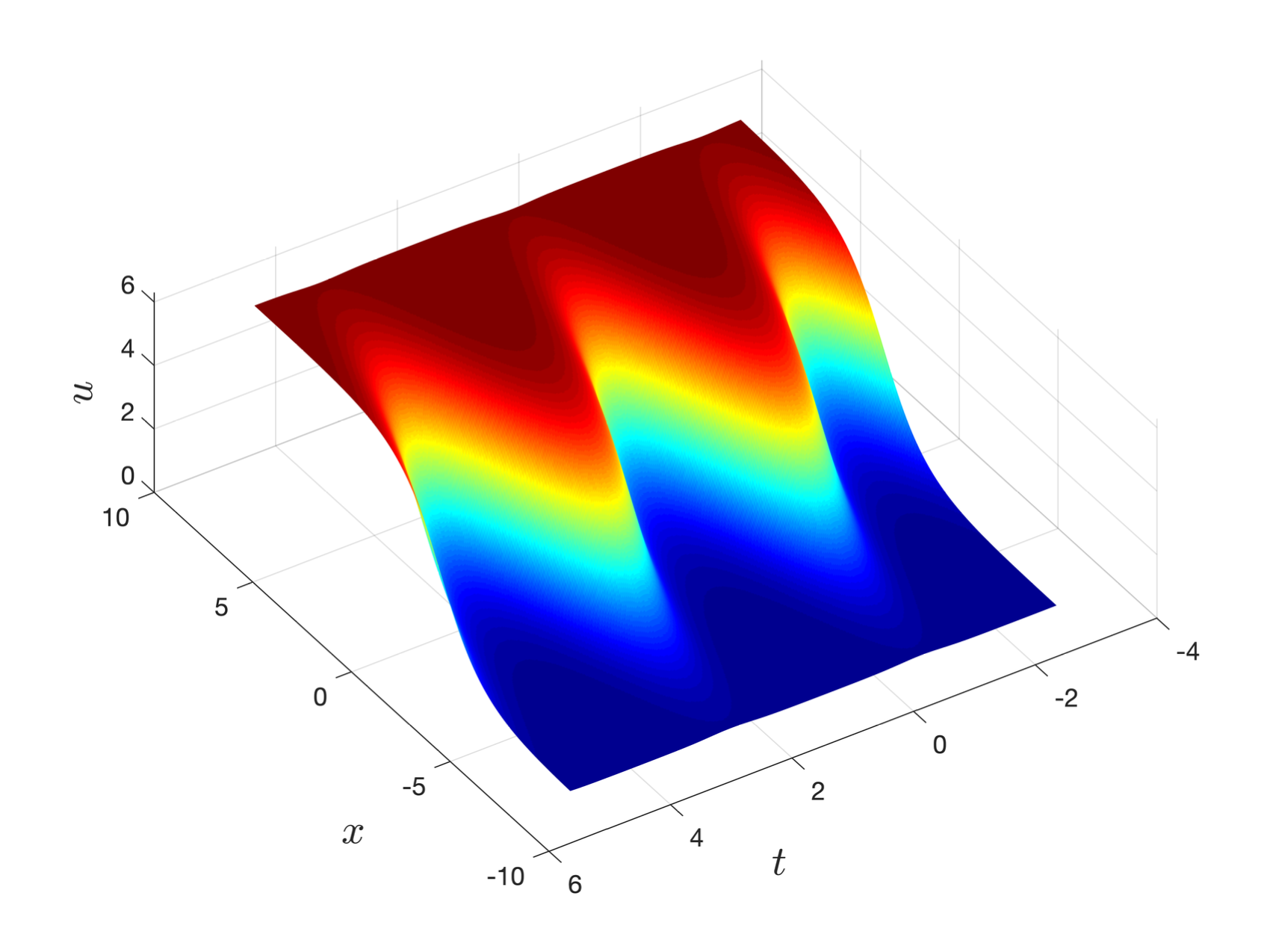}
    \caption*{(b)}
    \end{subfigure}\\
    \caption{(Color online) The kink-like solution $u_3^{(vSG)}$ of the vSG equation under the cosine coefficient ($k_1=1$). (a): The density plot of the data-driven solution and corresponding error are located in the upper part. The comparison of the three time snapshot curves of the exact solution and the data-driven solution is located in the lower part. (b): $3D$ surface plots of the data-driven solution.}
    \label{SG-cos}
    \end{figure}
\end{itemize}

In the above numerical experiments, although the spatiotemporal regions discussed in each example are different, the same $513\times 201$ equidistant discrete method is used to mesh the region to obtain the initial boundary value data preliminarily screened and $g_u$-$type$ points required for generalization error analysis. All examples use a unified Tanh activation function and a unified network structure: the trunk network is $NN_u\{10,40,4,2\}$, and the branch network is $NN_c\{6,30,2,2\}$. (exception: in the case of $k_1=1$ under the linear coefficient, the main network structure is $NN_u\{10,40\}$.) The training strategy is $L$-$BFGS$ optimization after 5000 $Adam$ iterations, and the number of various types of points involved in the loss function is set to $\{n_I+n_b, n_f, n_c\}=\{800, 20000, 60\}$.\footnote{In practice, $I$-$type$ points and $b$-$type$ points are sampled together. The same is true unless otherwise specified in the following examples.} In Appendix \ref{Appendix_vSG_forward}, more detailed preset model parameters and experimental results including random seeds, training time, number of iterations, etc. In general, the proposed method shows good performance in the forward problem of $(1+1)$ dimensional variable coefficients. In the face of various coefficient types, the $L^2$ generalization error reaches $10^{-4}$ or even $10^{-5}$ level.

\subsection{Generalized Kadomtsev–Petviashvili equation with variable coefficient.}\label{KP}
\quad

Various forms of generalized KP equations with variable coefficients have been proposed a long time ago \cite{david1987integrable, david1989solitons, gungor2002generalized}. The motivation for these models was to describe water waves propagating in straits or rivers, rather than waves propagating on unbounded surfaces like oceans. Additional terms and variable coefficients allow them to handle channels of varying width, depth, and density and even take eddies into account, providing a more realistic description of surface waves than the standard KP equations. In \cite{li2010painleve} and \cite{ye2008grammian}, the Painlevé analysis and Grammian solution of the following generalized variable coefficient KP equation are respectively given. The specific form of the equation is as follows:
\begin{equation}\label{gvKP1}
	(u_t + f(t)uu_x+g(t)u_{xxx}+l(t)u+q(t)u_x+n(t)u_y)_x+m(t)u_{yy}=0,
\end{equation}
where $f(t)\neq 0$ and $g(t)\neq 0$ represent the coefficients of nonlinearity and dispersion respectively, $l(t)$, $q(t)$ and $n(t)$ are regarded as the coefficients of perturbation effects, and $m(t)$ is the disturbed wave velocity along the y direction, and these variable coefficients are all analytical functions about t. Equation \eqref{gvKP1} can degenerate into standard KP equation \cite{ablowitz1991solitons} and cylindrical KP equation \cite{dryuma1983integration} under certain coefficients. In order to test the performance of our method in $(2+1)$-dimensional scenarios, let the variable coefficient $n(t)=q(t)=0$ in \eqref{gvKP1}, thus considering a simpler generalized KP equation with variable coefficient (gvKP):
\begin{equation}\label{gvKP2}
	(u_t + f(t)uu_x+g(t)u_{xxx}+l(t)u)_x+m(t)u_{yy}=0,
\end{equation}
where $u=u(x, y, t)$, $x$, $y$ are space variables, and $t$ is time variable. \cite{li2010painleve} gives the exact solution of equation \eqref{gvKP1} based on the auto-Bäcklund transformation. Equation \eqref{gvKP2} is a special case of equation \eqref{gvKP1}, and its exact solution can naturally be obtained. Specifically, consider the following coefficient constraints:
\begin{equation}\label{const1}
	\begin{split}
		g(t)=\gamma f(t)e^{-\int l(t)dt},\\
		m(t)=\rho f(t)e^{-\int l(t)dt},
	\end{split}
\end{equation}
where $\gamma$ and $\rho$ are arbitrary parameters. When these two parameters are fixed, it can be seen from constraint \eqref{const1} that the equation is completely determined by variable coefficients $f(t)$ and $l(t)$. Given an analytical solution of equation \eqref{gvKP2} under constraints \eqref{const1}:
\begin{equation}\label{KPsol}
u(x,y,t)=12\frac{g}{f}\frac{\partial^2}{\partial x^2}\text{ln}\ \phi
\end{equation}
with
\begin{equation}\label{phi1}
	\phi=1+e^{px+ry-\int4p^3g(t)-m(t)dt},\ r=\sqrt{\frac{3\gamma}{\rho}}p^2,
\end{equation}
where $p$ is an arbitrary constant, and $r$ is determined by $\gamma$, $\rho$ and the second formula of \eqref{phi1}. When different function combinations of $f(t)$ and $l(t)$ are selected, the solution \eqref{KPsol} presents a completely different form. Next, four coefficient combinations are discussed to test the performance of the proposed method and reveal the abundant dynamical behavior of the solution of the gvKP equation. Before that, make some settings, let the parameter $\gamma=\rho=1$, so that the variable coefficient $g(t)=m(t)$, so only three variable coefficients are involved in the discussion of the following forward problem, and they are all free. (Although the acquisition of the exact solution \eqref{KPsol} depends on constraint \eqref{const1}, this constraint is not involved in solving the forward problem with variable coefficients by VC-PINN method, and they are considered independent of each other in the neural network.) The initial boundary value data ($I$-$type$ points and $b$-$type$ points) required in the forward problem come from the discreteness of the exact solution \eqref{KPsol} at the corresponding position. Of course, the initial boundary value data at this time are distributed on an initial value surface and $4$ boundary surfaces (as we described in \cite{miao2022physics}). Then the data-driven solutions in the four cases are as follows:
\begin{itemize}
	\item \textbf{Case 1:} If $f(t)=\text{sin}(t),l(t)=\frac{1}{10}$, it is naturally derived from \eqref{const1} and $\gamma=\rho=1$:
	\begin{equation}
		g(t)=m(t)=e^{-\frac{t}{10}}\text{sin}(t),
	\end{equation}
	where the variable coefficients $g(t)$ and $m(t)$ formed by the product of the exponential function and the trigonometric function both oscillate and decay over time, while the exact solution \eqref{KPsol} becomes (let $p=1$)
\begin{equation}
	u_1^{(gvKP)}=\frac{12e^{-\frac{t}{10}+x+\sqrt{3}y+\frac{40}{101}e^{-t/10}[10\text{cos}(t)+\text{sin}(t)]}}{(1+e^{x+\sqrt{3}y+\frac{40}{101}e^{-t/10}[10\text{cos}(t)+\text{sin}(t)]})^2}.
\end{equation}

The discrete values of the coefficients required for the forward problem are directly obtained from the expression of the variable coefficients, so the data-driven solution under Case 1 is shown in Fig. \ref{KP-case1}.
\begin{figure}[htpb]
\centering
\includegraphics[width=17cm,height=5.2cm]{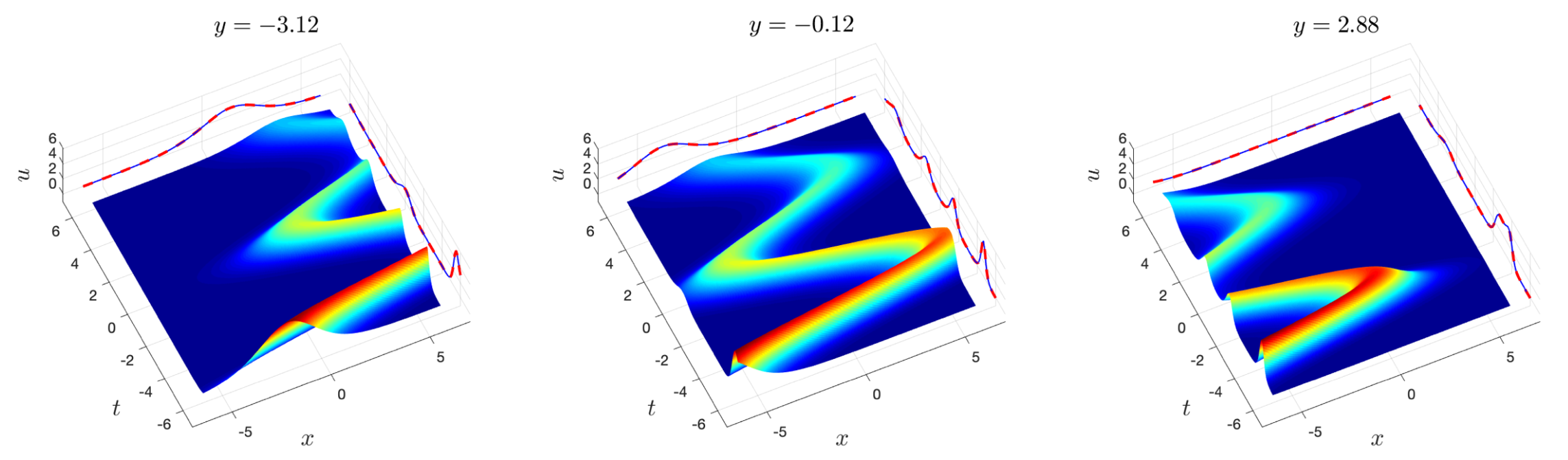}
\caption{(Color online) Data-driven solution $u_1^{(gvKP)}$ of the gvKP equation in Case 1: $3D$ plot of the data-driven solution at 3 fixed $y$-axis coordinates. The curves on both sides represent the cross-section of the data-driven solution on the central axis of the $x$ and $t$ coordinates. (The blue solid line and the red dashed line correspond to the predicted solution and the exact solution, respectively.)}
\label{KP-case1}
\end{figure}

From the expression \eqref{KPsol} of the exact solution, it can be seen that the change of the coefficient directly affects the form of the term related to the time variable $t$, but hardly affects the form of the term only related to the space variable $x$, $y$. The indefinite integral of the variable coefficient $g(t)$ and $m(t)$ is still in the form of the product of the exponential function and the trigonometric function, which is why the trend of the wave in Fig. \ref{KP-case1} displays a similar property to the variable coefficient $g(t)$ and $m(t)$. (``serpentine movement" that oscillates and decays over time). Not only the direction of the wave, when $t$ gradually increases, the amplitude of the wave also gradually decays, and the wave has the property of time localization on the positive semi-axis of $t$. In this case, the relative $L^2$ error of the obtained data-driven solution is $e^r_u=3.48\times 10^{-4}$, which preliminarily shows that the proposed method also has the expected effect in the $(2+1)$-dimension.

	\item \textbf{Case 2:} When both $f(t)$ and $l(t)$ are in the form of trigonometric functions (i.e.$f(t)=\text{sin}(t)\text{cos}(t),l(t)=\text{sin}(t)$), the other two variable coefficients are
	\begin{equation}
		g(t)=m(t)=e^{\text{cos}(t)}\text{cos}(t)\text{sin}(t),
	\end{equation}	
where it is obvious that they are both periodic functions, then the exact solution to the gvKP equation is derived from \eqref{KPsol} as follows ($p=1$):	
\begin{equation}
	u_2^{(gvKP)}=\frac{12e^{x+\sqrt{3}y+4e^{\text{cos}(t)}[-1+\text{cos}(t)]+\text{cos(t)}}}{(1+e^{x+\sqrt{3}y+4e^{\text{cos}(t)}[-1+\text{cos}(t)]})^2}.
\end{equation}

The indefinite integral of the variable coefficients $g(t)$ and $m(t)$ still maintains the periodic nature, which is consistent with the phenomenon that the wave appears periodically along the t direction seen in Fig. \ref{KP-case2}. And the solution presents a ``swallowtail" waveform in each time period, which is quite different from the soliton or breather in the constant coefficient equation. The reason for the formation of this waveform is closely related to the form of the indefinite integral of $g(t)$, which is completely symmetrical in each time period, but $g(t)$ is not. The predicted and exact curves on both sides of the $3D$ graph fit perfectly, which is very clear in Fig. \ref{KP-case2}, and the numerical results show that the relative $L^2$ error of the data-driven solution is $e^r_u=4.83\times 10^{-4}$. The above evidence fully demonstrates that our method predicts the dynamic behavior of $u^{(gvKP)}_2$ with high accuracy.
\begin{figure}[htpb]
\centering
\includegraphics[width=17cm,height=5.2cm]{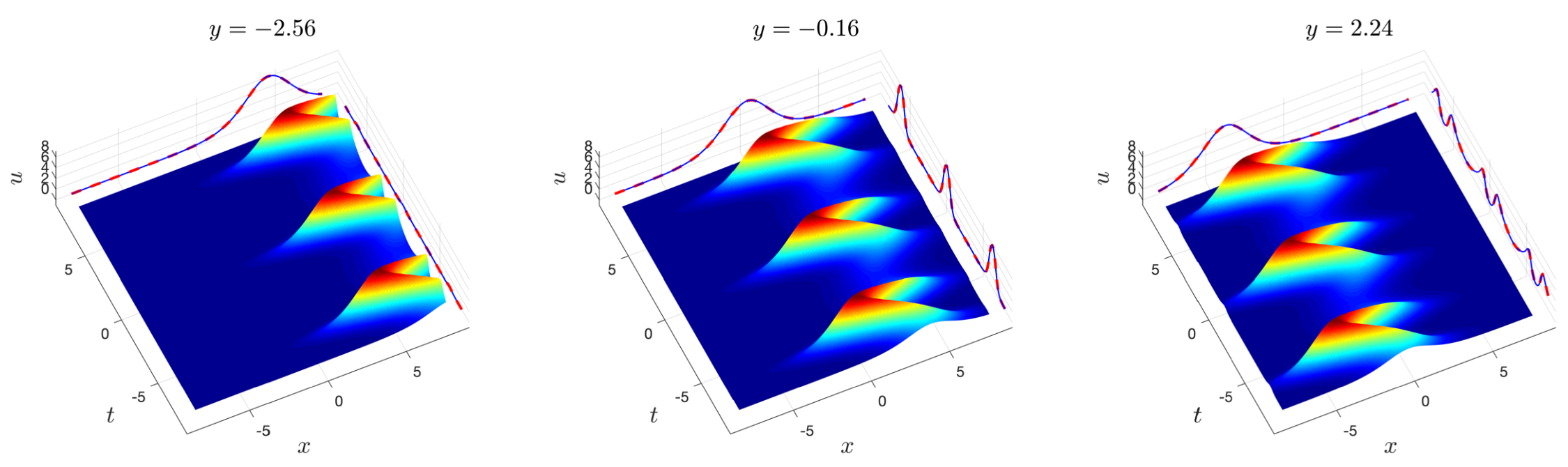}
\caption{(Color online) Data-driven solution $u_2^{(gvKP)}$ of the gvKP equation in Case 2: $3D$ plot of the data-driven solution at 3 fixed $y$-axis coordinates. The curves on both sides represent the cross-section of the data-driven solution on the central axis of the $x$ and $t$ coordinates. (The blue solid line and the red dashed line correspond to the predicted solution and the exact solution, respectively.)}
\label{KP-case2}
\end{figure}
	\item \textbf{Case 3:} When $f(t)$ and $l(t)$ are both linear functions ($f(t)=l(t)=t$), $g(t)$ and $m(t)$ are the product of the exponential function and the polynomial, which is
	\begin{equation}
		g(t)=m(t)=e^{-\frac{t^2}{2}}t.
	\end{equation}
Then substitute them into the exact solution \eqref{KPsol} to obtain ($p=1$)
\begin{equation}
	u_3^{(gvKP)}=\frac{6e^{-\frac{t^2}{2}}}{1+\text{cosh}(4e^{-\frac{t^2}{2}+x+\sqrt{3}y})}.
\end{equation}

Fig.\ref{KP-case3} depicts the data-driven solution under Case 3, and the waveform at this time seems to be very similar to the waveform restricted to a single time period in Case 2. It is an interesting finding that the solutions under linear coefficients and triangular periodic coefficients have such similar waveforms.
\begin{figure}[htpb]
\centering
\includegraphics[width=17cm,height=5.2cm]{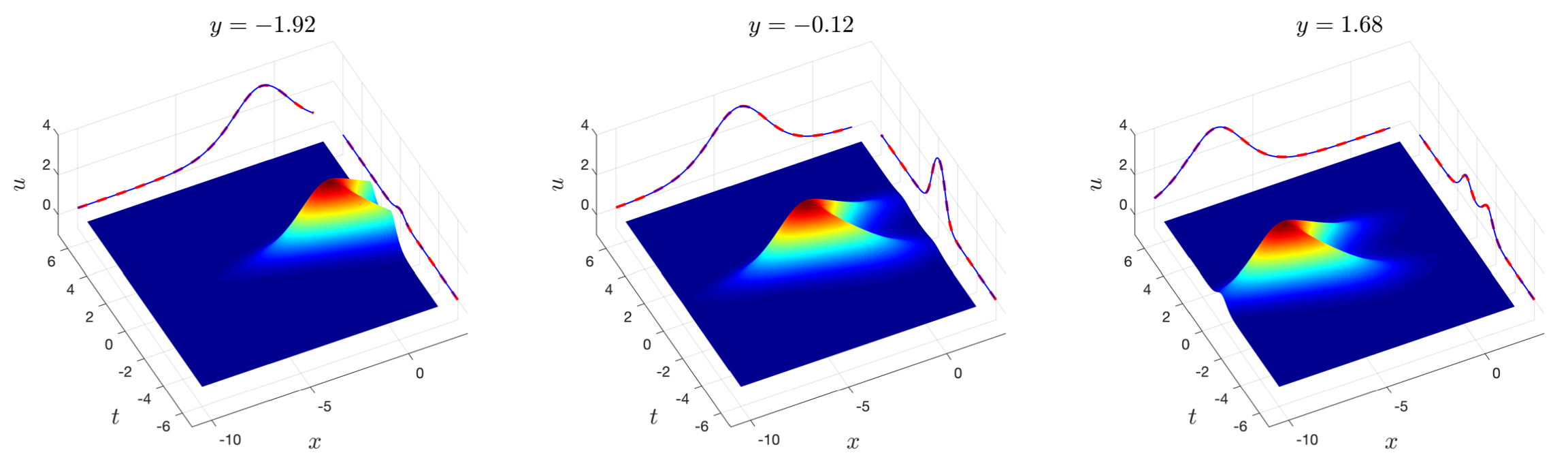}
\caption{(Color online) Data-driven solution $u_3^{(gvKP)}$ of the gvKP equation in Case 3: $3D$ plot of the data-driven solution at 3 fixed $y$-axis coordinates. The curves on both sides represent the cross-section of the data-driven solution on the central axis of the $x$ and $t$ coordinates. (The blue solid line and the red dashed line correspond to the predicted solution and the exact solution, respectively.)}
\label{KP-case3}
\end{figure}

Going back to the result we are most concerned about, under the proposed method, the $L^2$ relative generalization error of the data-driven solution is $e^r_u=2.23\times 10^{-3}$, and the dynamic behavior of the solution of the gvKP equation is successfully restored again.
	\item \textbf{Case 4:} When we reselect the variable coefficient $f(t)$ in Case 3 as a quadratic function (i.e.$f(t)=t^2$, $l(t)=t$), variable coefficients $g(t)$ and $m(t)$ become
	\begin{equation}\label{KP4g}
		g(t)=m(t)=e^{-\frac{t^2}{2}}t^2.
	\end{equation}
The corresponding exact solution also becomes
\begin{equation}\label{KPsol4}
	u_4^{(gvKP)}=\frac{6 e^{-\frac{t^2}{2}}}{1+\text{cosh}\left[4e^{-\frac{t^2}{2}}t+x+\sqrt{3}y-2\sqrt{2 \pi}\text{erf}(\frac{t}{\sqrt{2}})\right]}
\end{equation}
where $\text{erf}(\cdot)$ represents the Gaussian error function, which is defined as
\begin{equation}
	\text{erf}(t)=\frac{2}{\sqrt{\pi}}\int_0^{t}e^{-\eta^2}d\eta.
\end{equation}
The reason for the appearance of this non-elementary function in solution \eqref{KPsol4} is the indefinite integral with variable coefficients in \eqref{KP4g}. Among all the examples of the gvKP equation, only the waveform in this example is the closest to the shape of the soliton under the constant coefficient equation. But when we focus on the $3D$ diagram in Fig. \ref{KP-case4}, we find that the wave is different from the soliton, and actually presents the shape of a cubic function, which is inseparable from the fact that the variable coefficient $f(t)$ is a quadratic function. Combined with the example of quadratic coefficients in the vSG equation (Fig. \ref{SG-p2}), variable coefficients of the same form can find connections even in completely different equations. Then give the relative $L^2$ generalization error in this case as $e^r_u=1.30\times 10^{-3}$.
\begin{figure}[htpb]
\centering
\includegraphics[width=17cm,height=5.2cm]{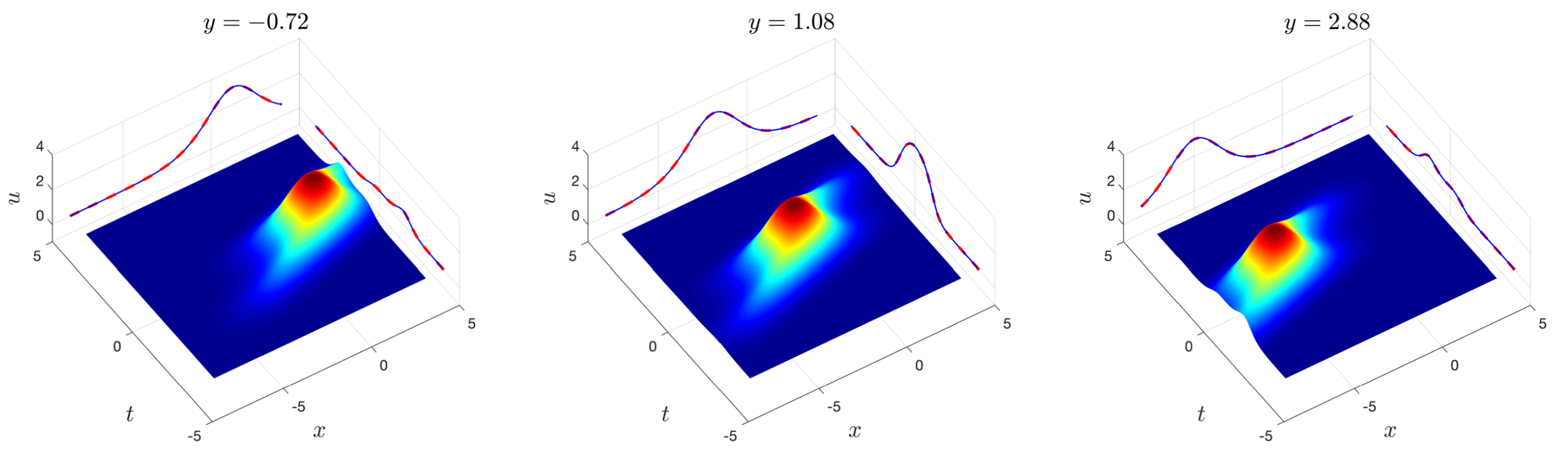}
\caption{(Color online) Data-driven solution $u_4^{(gvKP)}$ of the gvKP equation in Case 4: $3D$ plot of the data-driven solution at 3 fixed $y$-axis coordinates. The curves on both sides represent the cross-section of the data-driven solution on the central axis of the $x$ and $t$ coordinates. (The blue solid line and the red dashed line correspond to the predicted solution and the exact solution, respectively.)}
\label{KP-case4}
\end{figure}
\end{itemize}

In all the above examples, we display the 3D map at different position coordinates of $y$ rather than at different time $t$. This is because the variable coefficients of the discussed equation \eqref{gvKP2} are functions only related to $t$. If we fix the time $t$, what we can see is a traveling wave in space, and its traveling direction and speed change with time. But these changes are difficult for us to feel through the $3D$ map. In these numerical examples, we fully feel the magical power of variable coefficients, which makes the waveform ever-changing to meet the requirements of natural phenomena for mathematical models.

At the end of this section, we declare the parameter settings in the method. An equidistant discretization of $101\times 101 \times 101$ is used for all examples. The other settings are also uniform in all examples: the trunk network and the branch network are $NN_u\{10,40,4,2\}$ and $NN_c\{8,30,2,2\}$ respectively, the activation function is Tanh, 5000 $Adam$ iterations are performed before using $L$-$BFGS$, and other parameters are $\{n_I+n_b, n_f, n_c\}=\{6000, 50000, 100\}$. More detailed numerical results and parameter settings are in Appendix \ref{Appendix_gvKP_forward}. In general, the results of 4 numerical examples prove that in $(2+1)$-dimensional variable coefficient equations, our proposed method is not inferior and performs satisfactorily. We have reason to believe that it can still perform well in higher-dimensional equations, and this may also be our future work.

\section{Numerical experiments on inverse problems}\label{Inverse_ex}
This section presents numerical examples of the VC-PINN method in variable coefficient inverse problems. In addition to the most common $(1+1)$-dimensional equations, we also tried inverse problems in high-dimensional situations, and inverse problems with multiple variable coefficients simultaneously. This section involves the previously discussed gvKP equation as well as two new equations: the variable-coefficient Korteweg-de Vries equation and the variable-coefficient Sawada-Kotera equation.

\subsection{Korteweg-de Vries equation with variable coefficient.}\label{KdV}
\subsubsection{Single variable coefficient.}\label{KdV-1v}
\quad

The Korteweg-de Vries equation is one of the most important equations in the field of integrable systems. It was first used to describe waves on shallow water surfaces, and it is a completely solvable model. (solved by inverse scattering transformation \cite{gardner1967method}.) The equation we discuss in this section is its variable coefficient version, that is, the variable coefficient Korteweg-de Vries equation (vKdV), which was first proposed by Grimshaw \cite{grimshaw1979slowly}. The specific form is
\begin{equation}\label{vKdV}
	u_t+f(t)uu_x+g(t)u_{xxx}=0,
\end{equation}
where $f(t)$ and $g(t)$ are arbitrary analytic functions. Then in the case of polynomial coefficients, the auto-Bäcklund transformation, Painlevé property, and similarity reduction of this equation are obtained by techniques such as the WTC method and classical Lie group method \cite{nirmala1986auto, nirmala1986variable}. In addition, Fan also gives a Lax pair, a symmetry, two conservation laws, and an analytical solution to the vKdV equation by means of homogeneous balance \cite{fan2002auto}. This analytical solution is of concern in the numerical practice of this section, and it is an important sample of the inverse problem of VC-PINN. Assume that the variable coefficients in equation \eqref{vKdV} satisfy the constraints
\begin{equation}\label{vKdVc}
	g(t)=cf(t),
\end{equation}
where $c$ is an arbitrary constant. Then under this constraint, the exact solution given in \cite{fan2002auto} has the following form:
\begin{equation}\label{KdVsol}
	u(x,t)=3c\alpha^2\text{sech}^2\left[\frac{1}{2}\alpha (x-c\alpha^ 2 \int f(t) dt)\right],
\end{equation}
where $\alpha$ is a free parameter. An obvious fact is that once the variable coefficient $f(t)$, parameters $c$ and $\alpha$ are determined, the analytical solution \eqref{KdVsol} is fully determined. Let $c=\alpha=1$, under this parameter setting, we discuss $3$ different forms of f to test the performance of the proposed method on the inverse problem. As the first attempt of VC-PINN on the inverse problem, our example is also the simplest and most general. (neither high-dimension nor coexistence of multiple variable coefficients.) The internal data ($s$-$type$ points) required in the inverse problem are completely derived from the discretization of the exact solution \eqref{KdVsol}, and the coefficient information provided in this example only contains two endpoints (boundary), but no other derivative information (that is, does not include \eqref{hdc}). The following shows the discovery of function parameters in three variable coefficient forms:
\begin{itemize}
	\item \textbf{First-degree polynomial:} When the coefficient $f(t)$ is linear (i.e. $f(t)=t$), the exact solution \eqref{KdVsol} becomes
\begin{equation}
	u_1^{(vKdV)}=3\text{sech}^2\left[\frac{1}{4}(t^2-2x)\right].
\end{equation}

Fig.\ref{KdV-p1}(b) shows a parabolic soliton with linear coefficients, which has a completely different shape from a line soliton with constant coefficients. The reason for the formation of the parabolic shape is consistent with the example of the vSG equation: the indefinite integral of a linear function is of quadratic polynomial type. But the difference is that the convex hull or concave hull in the vSG equation evolves from kink, but here it evolves from line solitons, so it is localized. The comparison of the exact value and the predicted value of the coefficient $f(t)$ in Fig. \ref{KdV-p1}(a) tells us that the proposed method is also successful in the inverse problem, and the $L^2$ relative error of the coefficient $f(t)$ is $e^r_c=2.82\times 10^{-4}$.
\begin{figure}[htpb]
\centering
\begin{subfigure}[b]{0.46\textwidth}
\includegraphics[width=8.5cm,height=5cm]{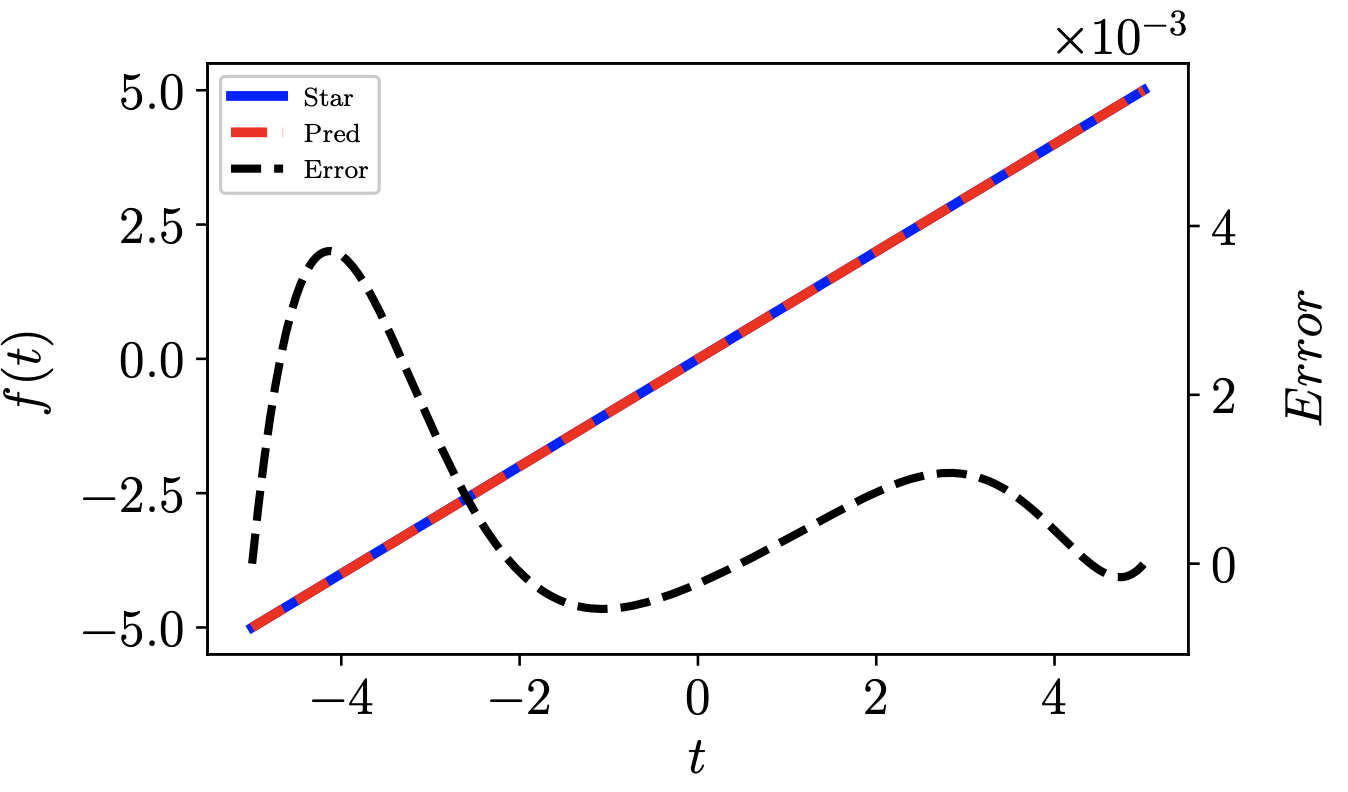}
\caption*{(a)}
\end{subfigure}
\hspace{0.3cm} 
\begin{subfigure}[b]{0.46\textwidth}
\includegraphics[width=7.5cm,height=5cm]{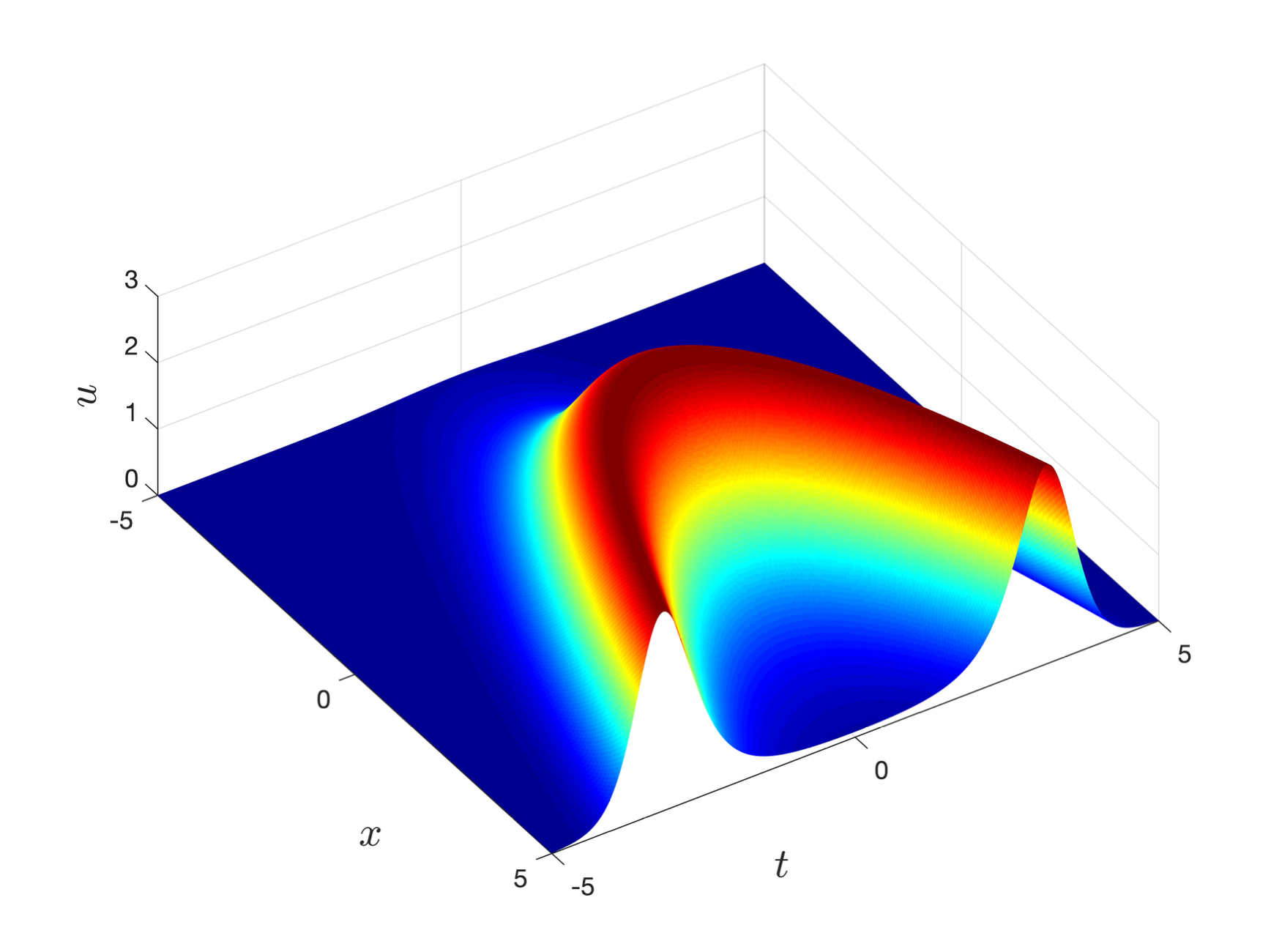}
\caption*{(b)}
\end{subfigure}
\caption{(Color online) Function parameter discovery for the vKdV equation under linear coefficients: (a) The real solution (red dotted line), predicted solution (blue solid line) and error curve (black dotted line, real minus predicted) of the function parameter $f(t)$, the former two follow the left coordinates, the latter follow right coordinates. (b) Data-driven dynamics of solution $u_1^{(vKdV)}$.}
\label{KdV-p1}
\end{figure}
	\item \textbf{Cubic polynomial:} Suppose the variable coefficient $f(t)$ is a cubic polynomial, that is, $f(t)=t^3$, and the exact solution of equation \eqref{vKdV} is
\begin{equation}\label{vKdVsol2}
	u_2^{(vKdV)}=3\text{sech}^2\left[\frac{1}{8}(t^4-4x)\right].
\end{equation}

The quartic term of $t$ in solution \eqref{vKdVsol2} is obtained by the indefinite integral of the coefficient of the cubic term, which directly leads to the approximation of the wave direction of the soliton in Fig. \ref{KdV-p3}(b) to a quartic curve. Compared with the parabolic soliton under the linear coefficient, the trajectory of the soliton in this case is more convex (note that the $t$-axis coordinate ranges of Fig. \ref{KdV-p1} and Fig. \ref{KdV-p3} are different). Another notable point is that the error increases from the order of $10^{-3}$ to the order of $10^{-2}$ when we change the coefficients from linear to cubic polynomial. And the error curve is more fluctuating than the case of linear coefficients (other network settings have achieved control variables), which shows that the inverse problem in this case is more difficult. Finally, the $L^2$ relative error of the coefficient $f(t)$ is given as $e^r_c=2.56\times 10^{-4}$.

\begin{figure}[htpb]
\centering
\begin{subfigure}[b]{0.46\textwidth}
\includegraphics[width=8.5cm,height=5cm]{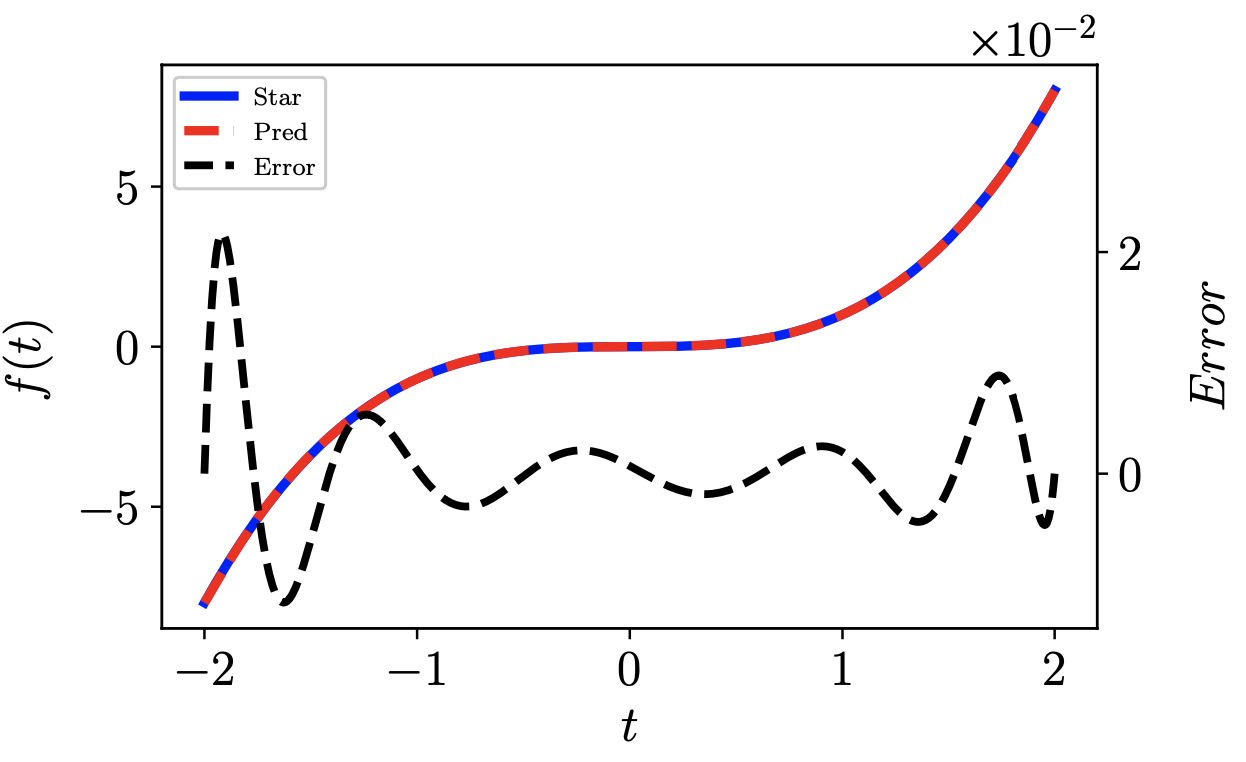}
\caption*{(a)}
\end{subfigure}
\hspace{0.3cm} 
\begin{subfigure}[b]{0.46\textwidth}
\includegraphics[width=7.5cm,height=5cm]{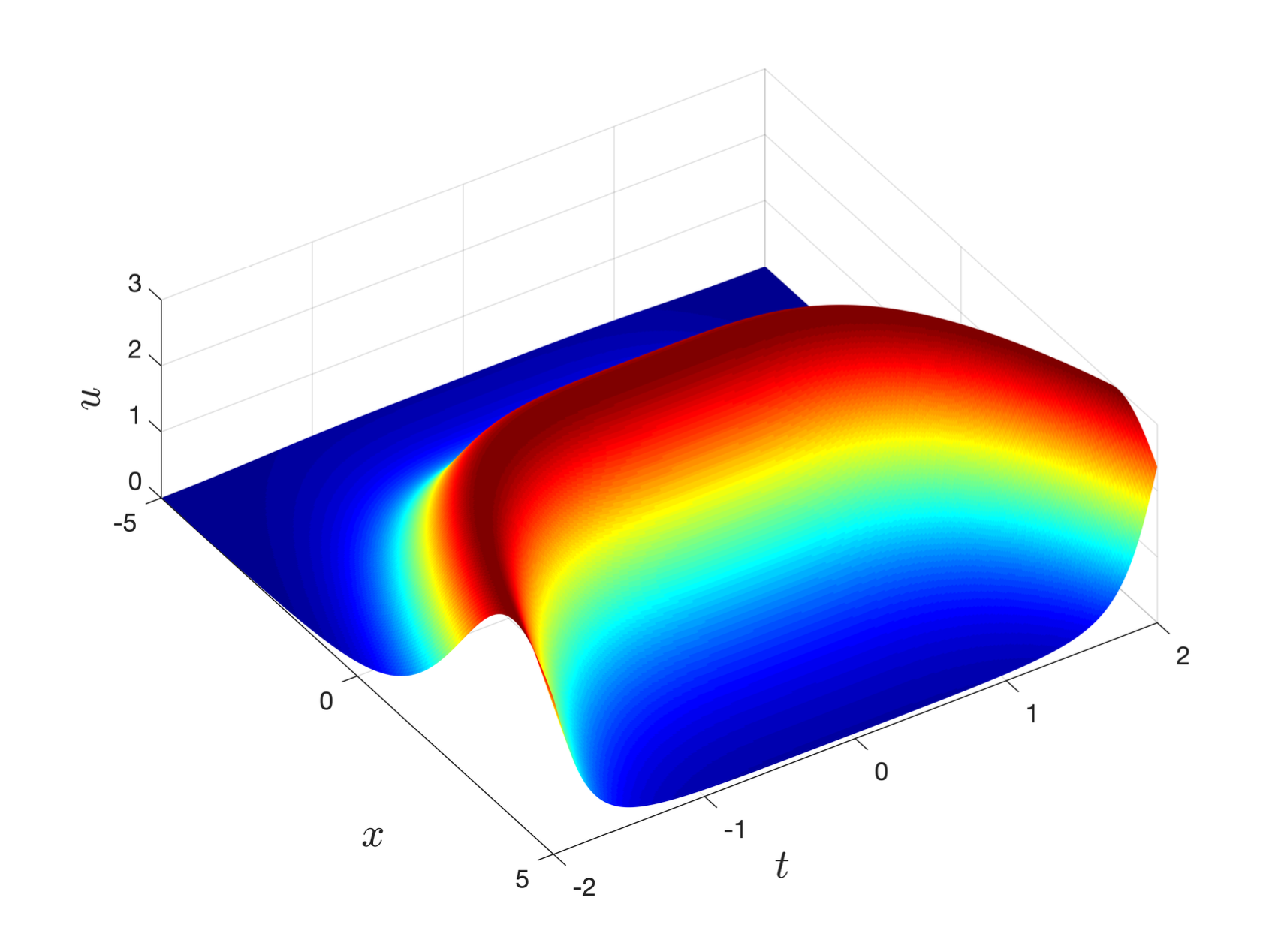}
\caption*{(b)}
\end{subfigure}
\caption{(Color online) Function parameter discovery for the vKdV equation under Cubic polynomial: (a) The real solution (red dotted line), predicted solution (blue solid line) and error curve (black dotted line, real minus predicted) of the function parameter $f(t)$, the former two follow the left coordinates, the latter follow right coordinates. (b) Data-driven dynamics of solution $u_2^{(vKdV)}$.}
\label{KdV-p3}
\end{figure}

	\item \textbf{Trigonometric functions:} When the coefficient function is a cosine function, that is, $f(t)=\text{cos}(t)$, the exact solution of the corresponding vKdV equation is
\begin{equation}
	u_3^{(vKdV)}=3\text{sech}^2\left[\frac{1}{2}(x-\text{sin}(t))\right].
\end{equation}

The periodic coefficient function determines that the direction of the soliton we see in Fig. \ref{KdV-cos} is also periodic, and the period length or amplitude of the coefficient directly affects the evolution behavior of the soliton. The error curve in Fig. \ref{KdV-cos}(a) is the most fluctuating among the above examples, which is inseparable from the periodicity of the variable coefficient $f(t)$. The error curve remains on the order of $10^{-3}$, and the evidence that the relative error of the coefficient function $L^2$ is $e^r_c=2.26\times 10^{-4}$. Suggests that the proposed method successfully inverts the variation of the coefficient.

\begin{figure}[htpb]
\centering
\begin{subfigure}[b]{0.46\textwidth}
\includegraphics[width=8.5cm,height=5cm]{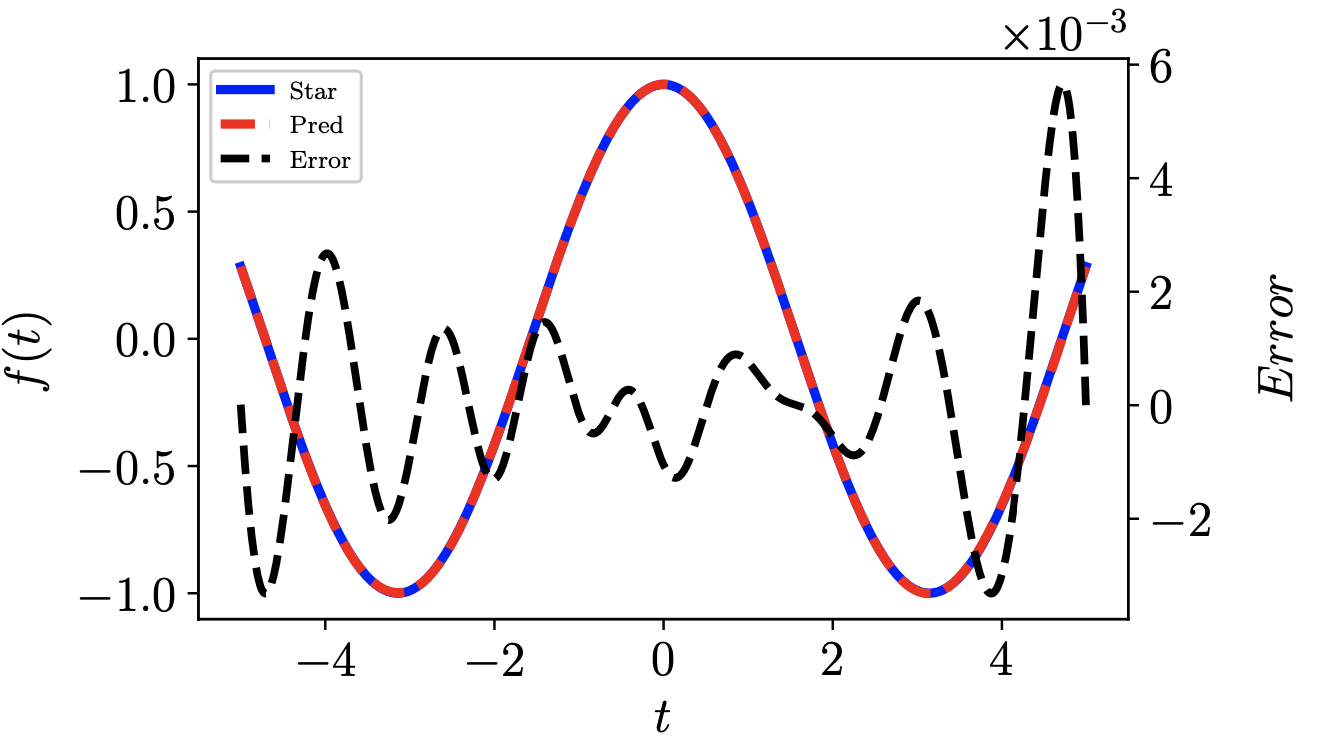}
\caption*{(a)}
\end{subfigure}
\hspace{0.3cm} 
\begin{subfigure}[b]{0.46\textwidth}
\includegraphics[width=7.5cm,height=5cm]{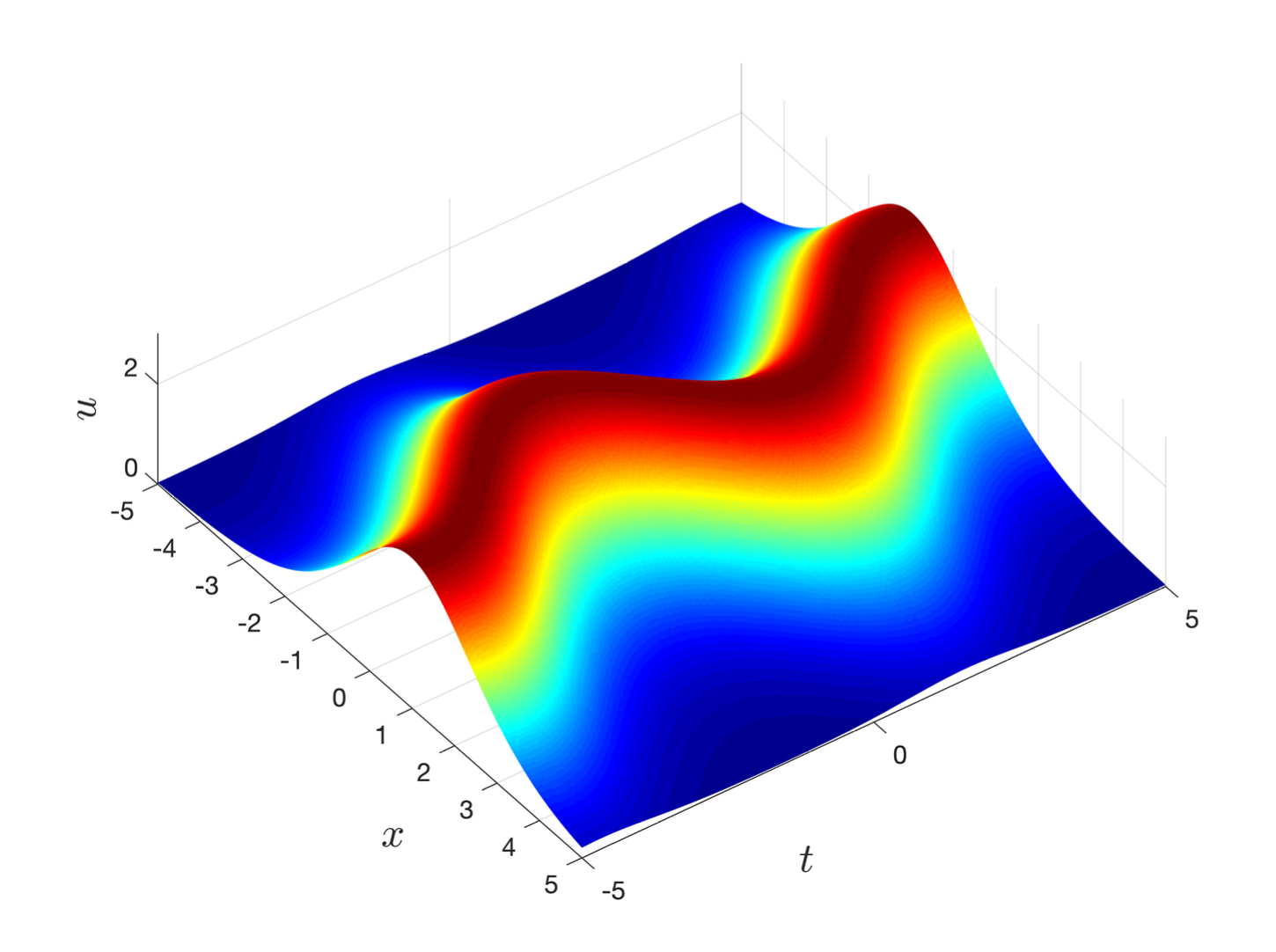}
\caption*{(b)}
\end{subfigure}
\caption{(Color online) Function parameter discovery for the vKdV equation under cosine coefficients: (a) The real solution (red dotted line), predicted solution (blue solid line) and error curve (black dotted line, real minus predicted) of the function parameter $f(t)$, the former two follow the left coordinates, the latter follow right coordinates. (b) Data-driven dynamics of solution $u_3^{(vKdV)}$.}
\label{KdV-cos}
\end{figure}
\end{itemize}

The following same settings are applied in all examples: variable coefficients are equidistantly divided into 500 equal parts, the activation function is Tanh, trunk network and branch network are $NN_u\{8,40,3,2\}$ and $NN_c\{6,30,2,2\}$ respectively, before using $L$-$BFGS$ optimization $5000$ $Adam$ iterations are performed, and the other parameters are $\{n_s, n_f\}=\{2000, 20000\}$. More detailed numerical results and parameter settings are in Appendix \ref{Appendix_vKdV_Inverse_1}. Our method shines in the first attempt of the variable coefficient inverse problem. Under different forms of coefficients, it can successfully invert the change of coefficients with time. In the following chapters, we look forward to its performance in multiple variable coefficients and high-dimensional situations.

\subsubsection{Multiple variable coefficients.}\label{KdV_2v_ex}
\quad

In the previous section, we discussed the inverse problem of the vKdV equation in the case of a single variable coefficient. Although there are two variable coefficients in equation \eqref{vKdV}, what is discussed is their solution under constraint \eqref{vKdVc}, and the constraint \eqref{vKdVc} is substituted into the network, which is why it is only a single variable coefficient problem for the network. At the beginning of this section, we plan not to impose constraint \eqref{vKdVc} into the network, and rerun the experiments on the $3$ examples from the previous section to test the performance of the proposed method under two variable coefficients.

\begin{figure}[htpb]
\centering
\begin{subfigure}[b]{0.32\textwidth}
\includegraphics[width=5.9cm,height=4.5cm]{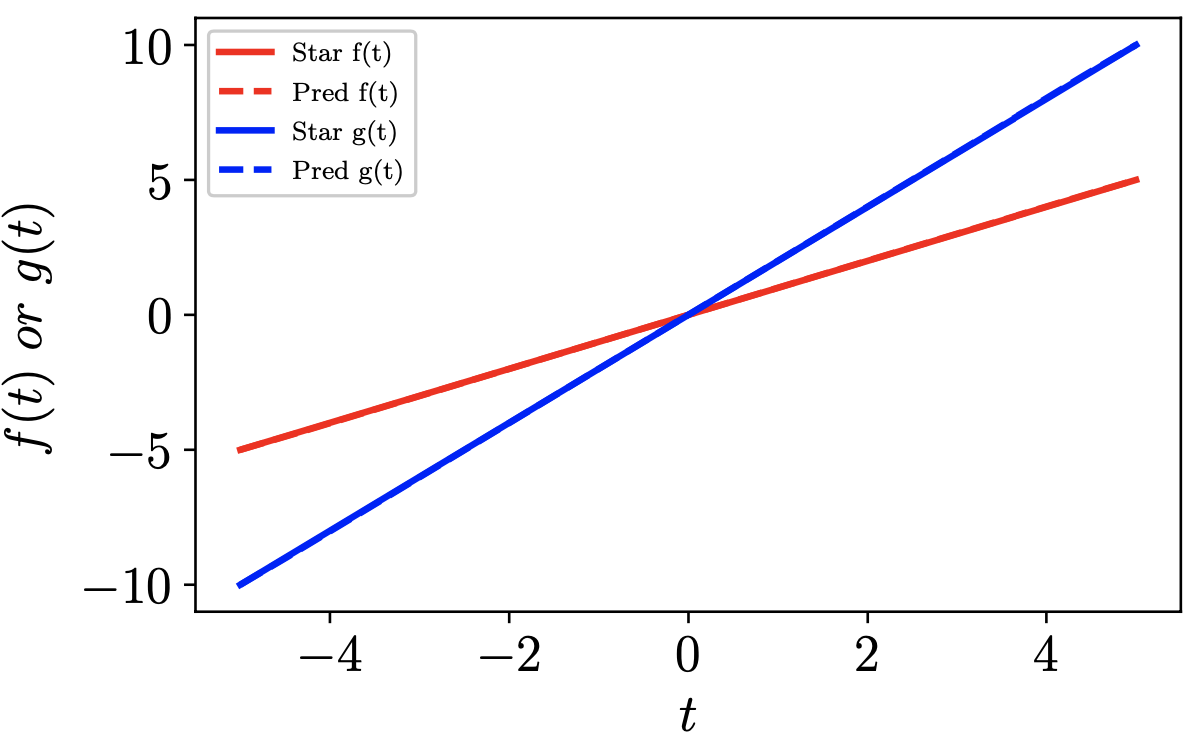}
\end{subfigure}
\begin{subfigure}[b]{0.32\textwidth}
\includegraphics[width=5.9cm,height=4.5cm]{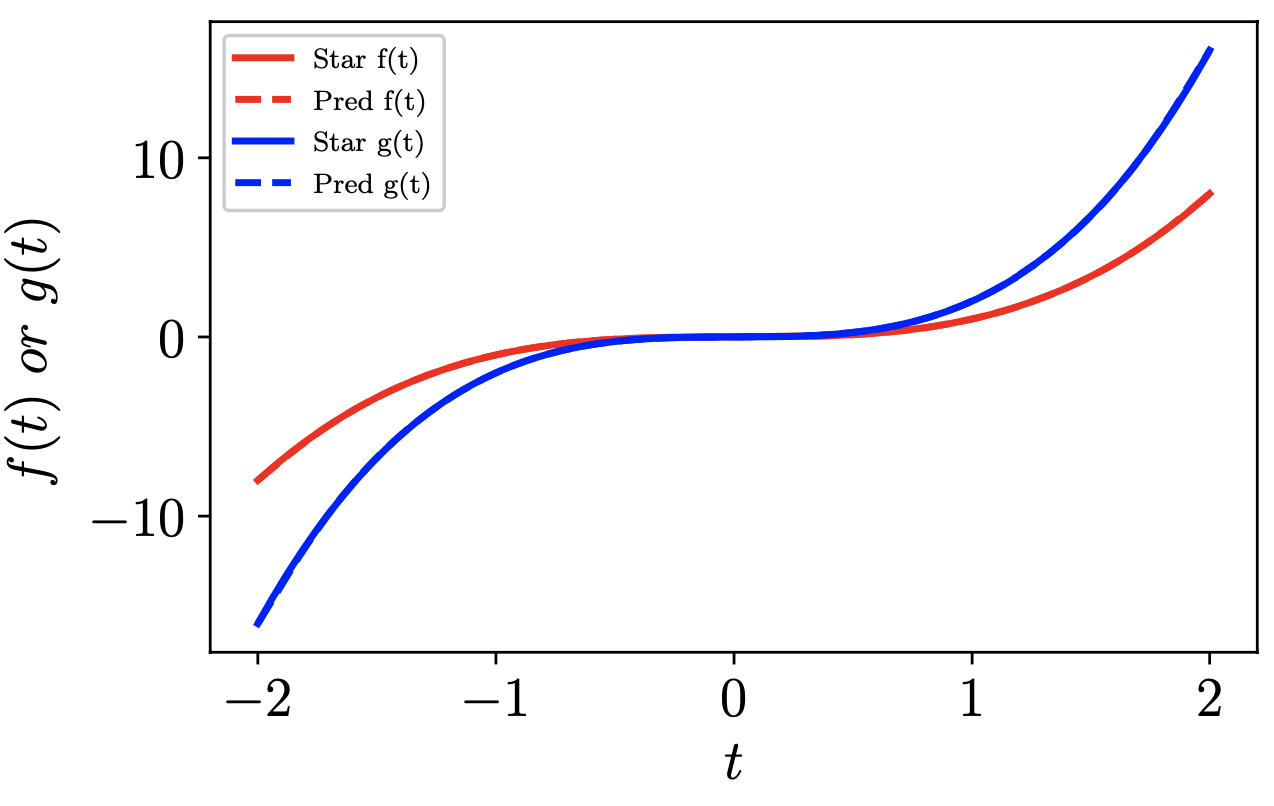}
\end{subfigure}
\hspace{0.3cm} 
\begin{subfigure}[b]{0.32\textwidth}
\includegraphics[width=5.9cm,height=4.5cm]{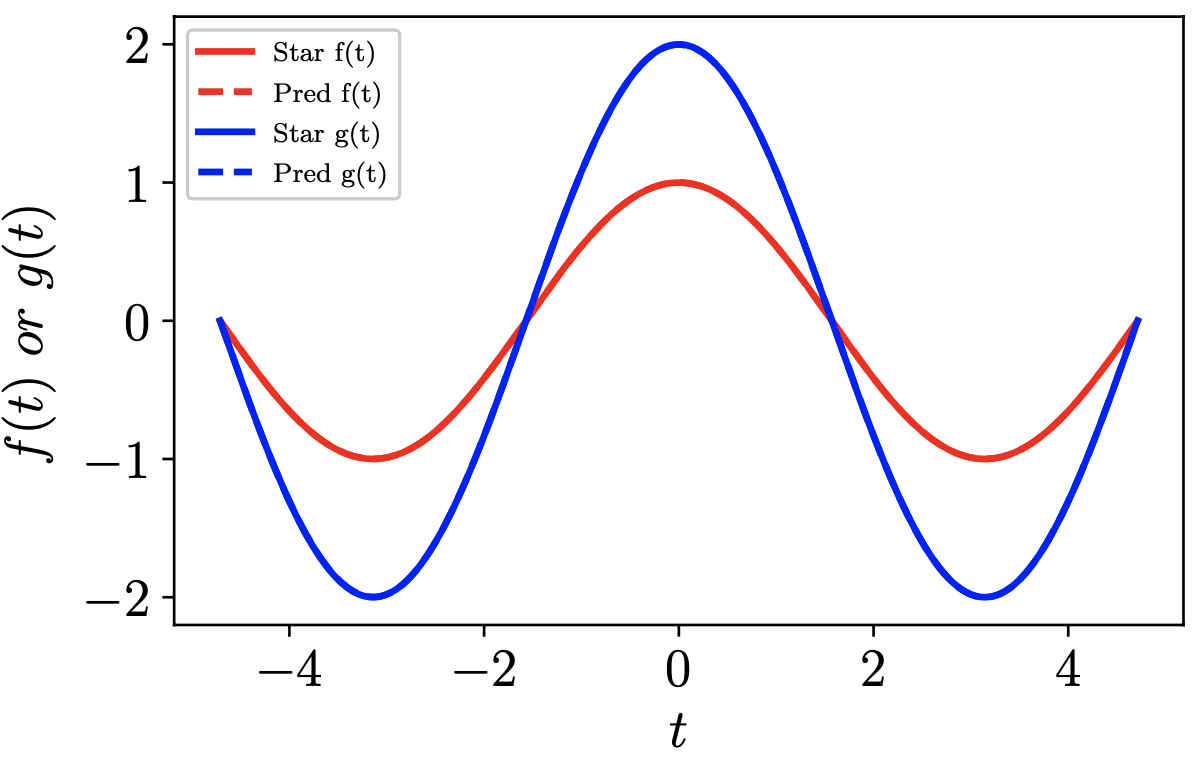}
\end{subfigure}\\
\begin{subfigure}[b]{0.32\textwidth}
\includegraphics[width=5.9cm,height=4.5cm]{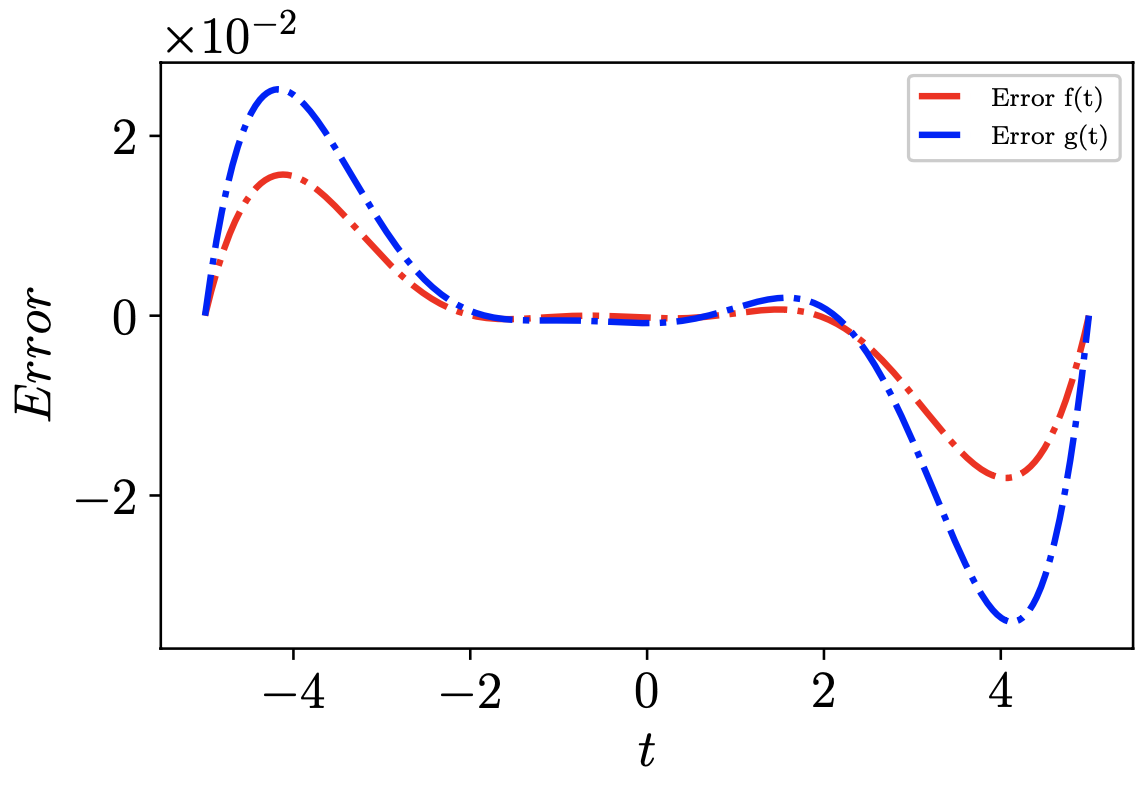}
\caption*{(a)}
\end{subfigure}
\begin{subfigure}[b]{0.32\textwidth}
\includegraphics[width=5.9cm,height=4.5cm]{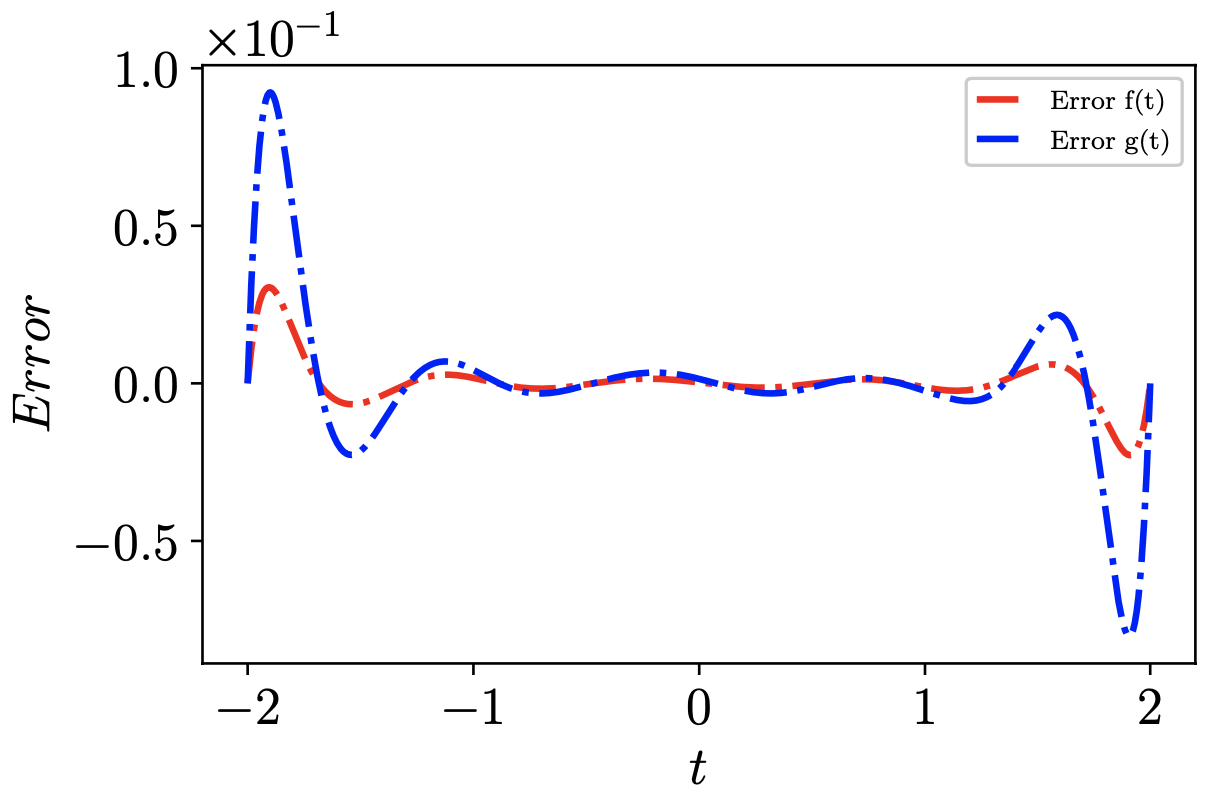}
\caption*{(b)}
\end{subfigure}
\hspace{0.3cm} 
\begin{subfigure}[b]{0.32\textwidth}
\includegraphics[width=5.9cm,height=4.5cm]{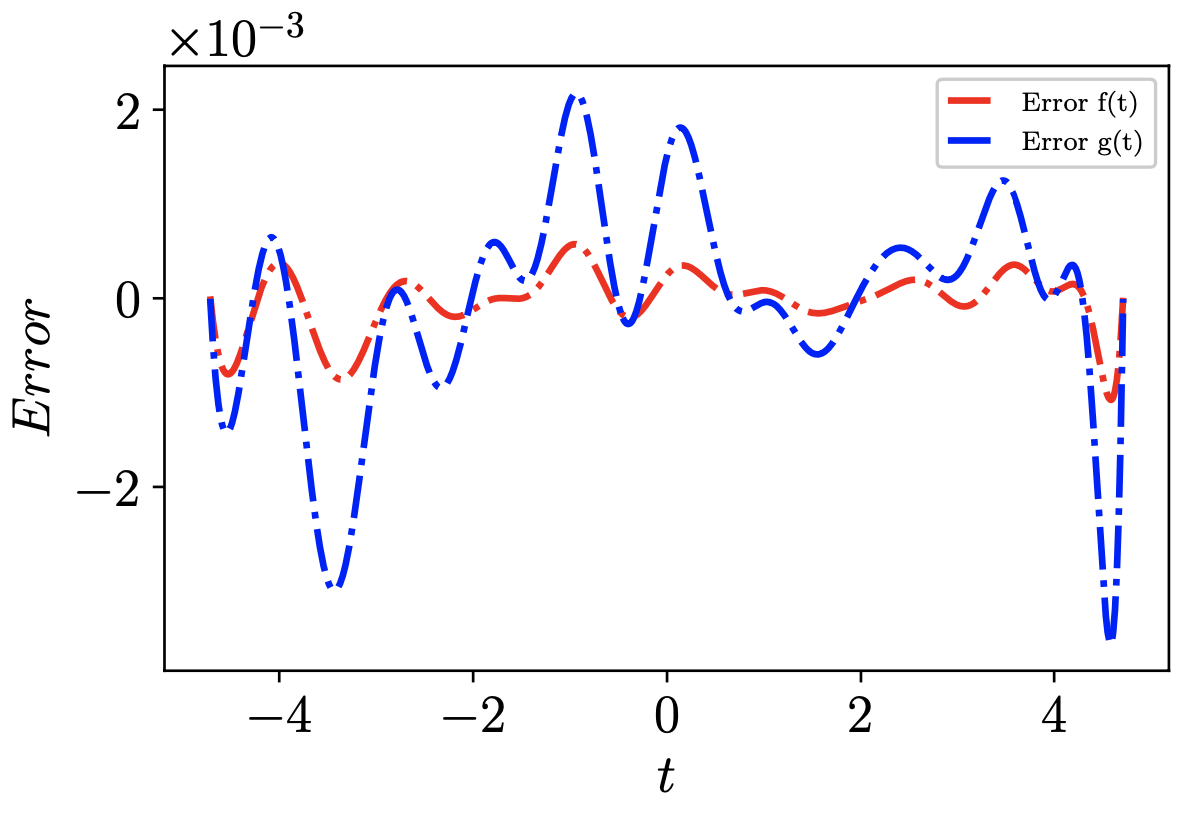}
\caption*{(c)}
\end{subfigure}\\
\caption{(Color online) The function parameters of the vKdV equation under two variable coefficients are found (the first line represents the true curve and the predicted curve of $f(t)$ and $g(t)$, and the second line represents the error curve): (a) linear coefficients. (b) cubic polynomial coefficients. (c) cosine coefficients.}
\label{KdV-2v}
\end{figure}

All settings including the exact solution and network parameters are kept the same as in Section \ref{KdV-1v} (except $c$ changed from $1$ to $2$), and Fig. \ref{KdV-2v} shows the numerical results of the discovery of the function parameters of the vKdV equation under two variable coefficients.

The results in Fig. \ref{KdV-2v} display that when increasing the number of function parameters to be discovered for the examples in Section \ref{KdV-1v} to $2$, the proposed method is still able to invert the changes of all function parameters over time. From the error graph, it can be found that for linear and cubic polynomials, the main part of the error is distributed near the left and right boundary areas, while for the cosine coefficient, the error still maintains a relatively high-frequency oscillation. Table \ref{KdVt}  presents more detailed $L^2$ relative error results.

\begin{table}[H]
\caption{$L^2$ relative error of variable coefficient $f(t)$ and $g(t)$}
\label{KdVt}
\begin{center}
\begin{tabular}{c||c|c|c}
\hline
\hline
 & Linear Coefficients &  Cubic Polynomial & Cosine Coefficients \\
\hline
\hline
Error of $f(t)$ & 2.95$\times 10^{-3}$ & 2.42$\times 10^{-3}$ & 4.53$\times 10^{-4}$\\
Error of $g(t)$ & 2.56$\times 10^{-3}$ & 3.90$\times 10^{-3}$ & 7.88$\times 10^{-4}$\\
\hline
\hline
\end{tabular}
\end{center}
\end{table}

In addition to re-experimenting the examples in Section \ref{KdV-1v}, we also discuss the problem of function parameter discovery in two other coefficient forms. The specific situation is as follows (set $c=2, \alpha=1$):
\begin{itemize}
	\item \textbf{Case 1:} Assuming that the variable coefficients $f(t)$ and $g(t)$ are both fractional, that is, $g(t)=2f(t)=\frac{1}{t}$, the exact solution is
\begin{equation}
	u_4^{(vKdV)}=6\text{sech}^2\left[\frac{1}{2}(x-4\text{ln}(t))\right],
\end{equation}
which is not an analytical solution, since it is undefined at $t=0$, so we only discuss it in the positive half of the $t$-axis in the experiment of the inverse problem.

\begin{figure}[htpb]
\centering
\begin{subfigure}[b]{0.45\textwidth}
\includegraphics[width=8cm,height=5cm]{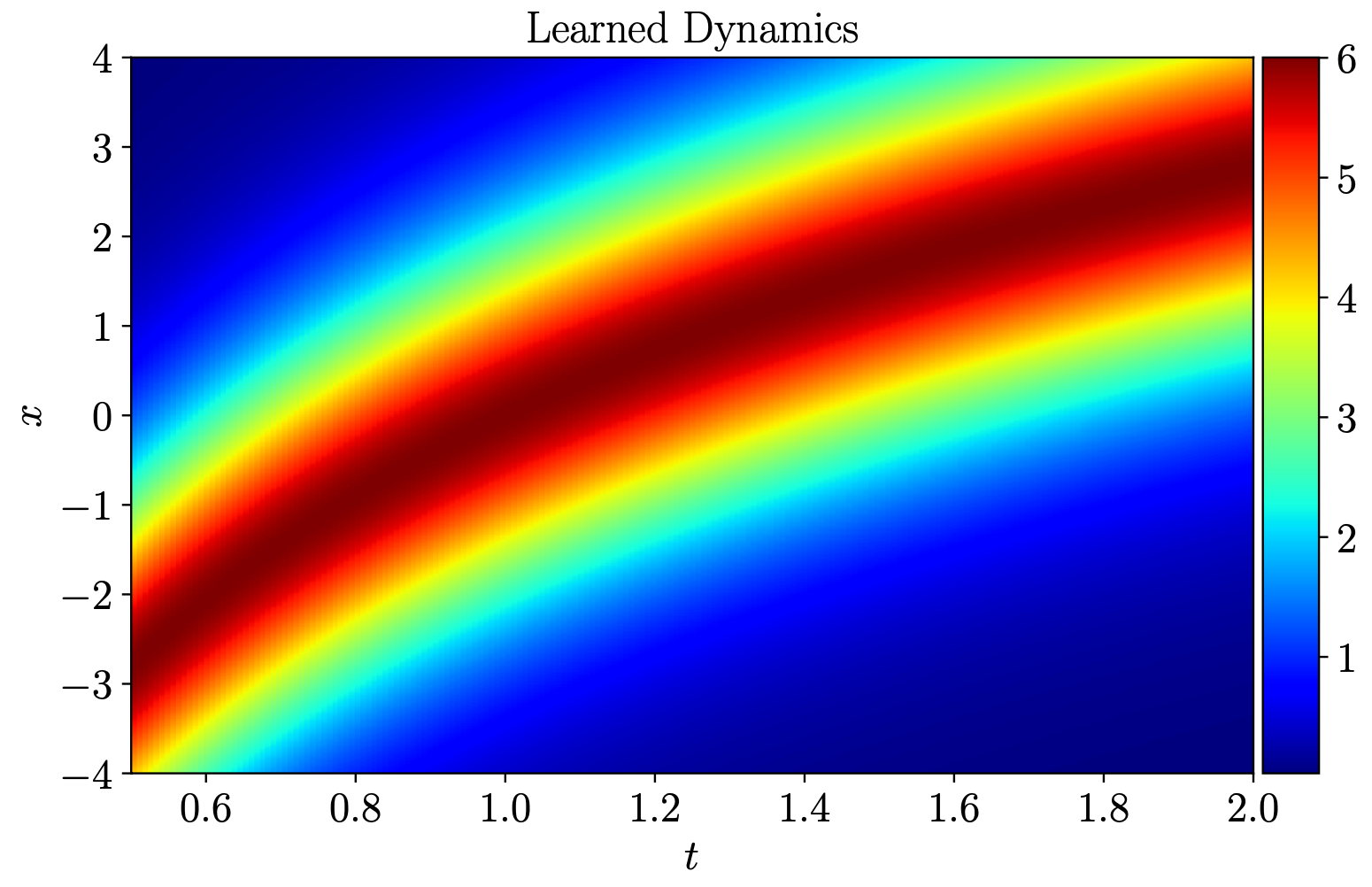}
\caption*{(a)}
\end{subfigure}
\begin{subfigure}[b]{0.45\textwidth}
\includegraphics[width=8cm,height=5cm]{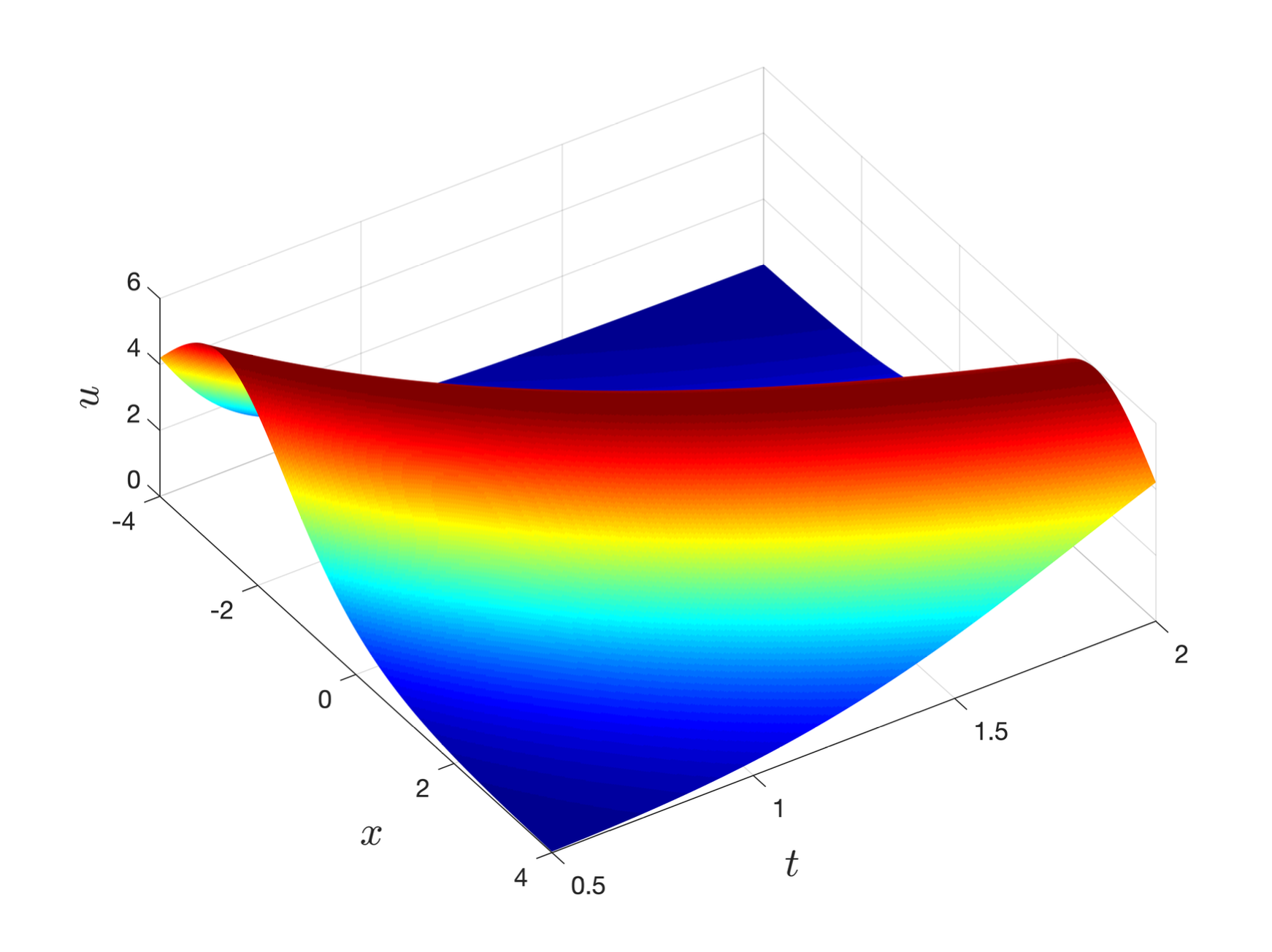}
\caption*{(b)}
\end{subfigure}\\
\begin{subfigure}[b]{0.45\textwidth}
\includegraphics[width=8cm,height=5.5cm]{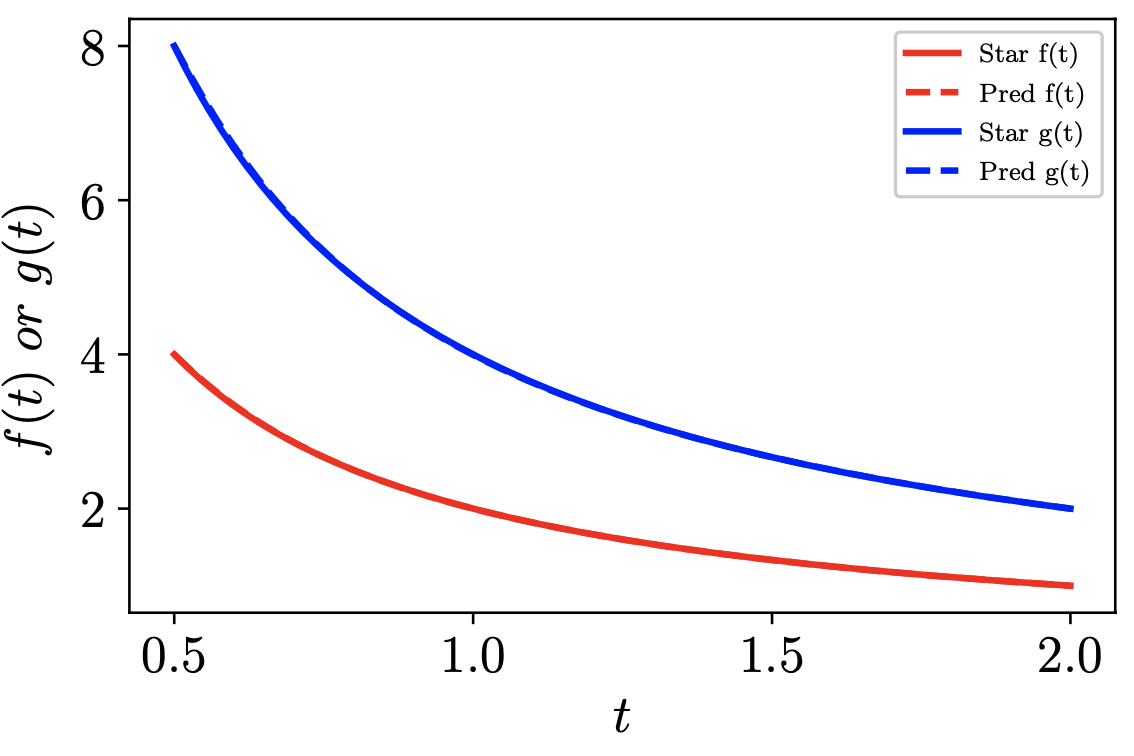}
\caption*{(c)}
\end{subfigure}
\begin{subfigure}[b]{0.45\textwidth}
\includegraphics[width=8cm,height=5.5cm]{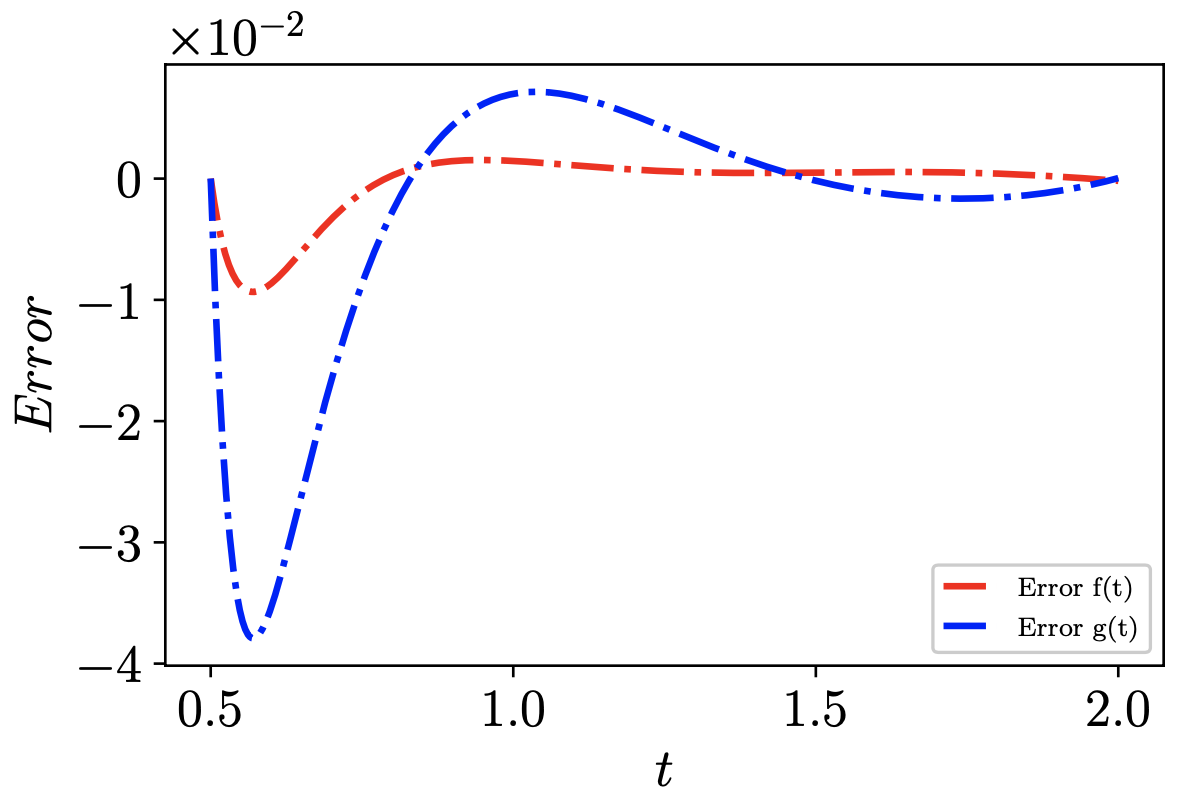}
\caption*{(d)}
\end{subfigure}\\
\caption{(Color online) Functional coefficients discovery (inverse problem) of the vKdV equation in case $1$: (a) Density plot of the data-driven solution of $u_4^{(vKdV)}$. (b) $3D$ plot of the data-driven solution of $u_4^{(vKdV)}$. (c) Comparison of predicted and true curves for two variable coefficients. (d) Error curves for two coefficients.}
\label{KdV-2v-f1}
\end{figure}
Limiting the time interval to $[0.5,2]$ avoids the singularity problem. It should be noted that the motion curve of the soliton shown in Fig. \ref{KdV-2v-f1} is consistent with the logarithmic function, not the reciprocal function. (although the two function curves are very similar.) The error curve tells us that for the case of fractional coefficients, the error mainly comes from the area near the time starting point ($t=0.5$). Combined with the error curve graph in Fig. \ref{KdV-2v}, it can be concluded that the errors in the discovery of function coefficients are distributed in the region of large coefficients and the region of rapid coefficient changes. (This is very similar to the distribution of errors for solutions in the forward problem.) Finally, the $L^2$ relative errors of $f(t)$ and $g(t)$ are $1.35\times 10^{-3}$ and $2.86\times 10^{-3}$, respectively.

	\item \textbf{Case 2:} Let the variable coefficients $f(t)$ and $g(t)$ be in the form of the product of an exponential function and a cosine function (i.e. $g(t)=2f(t)=e^{-\frac{t}{2}}\text{cos}(t)$), the exact solution \eqref{KdVsol} becomes
\begin{equation}
	u_5^{(vKdV)}=6\text{sech}^2\left[\frac{1}{2}(x+\frac{4}{5}e^{-\frac{t}{2}}(\text{cos}(t)-2\text{sin}(t))\right].
\end{equation}

\begin{figure}[htpb]
\centering
\begin{subfigure}[b]{0.45\textwidth}
\includegraphics[width=8cm,height=5cm]{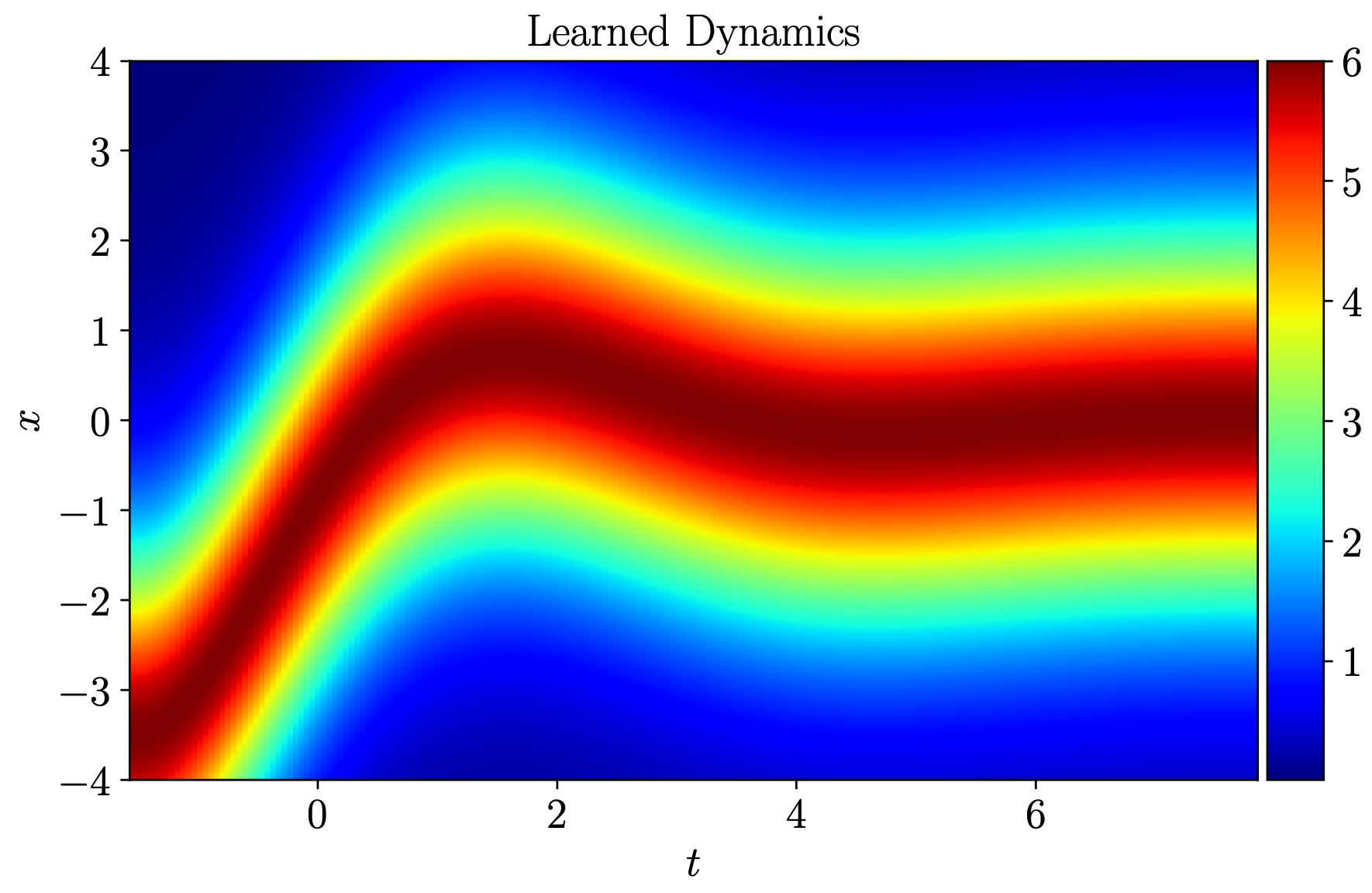}
\caption*{(a)}
\end{subfigure}
\begin{subfigure}[b]{0.45\textwidth}
\includegraphics[width=8cm,height=5cm]{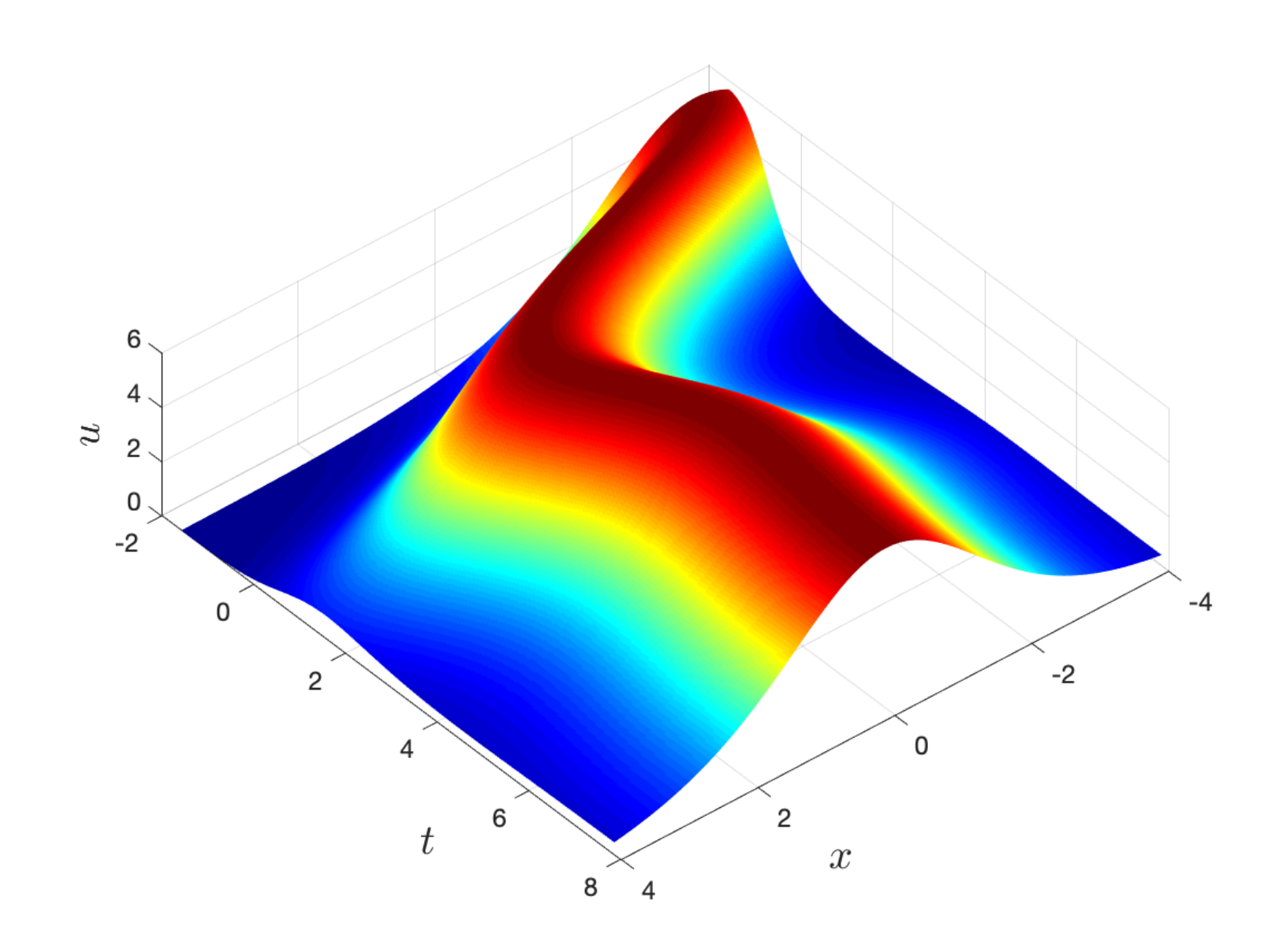}
\caption*{(b)}
\end{subfigure}\\
\begin{subfigure}[b]{0.45\textwidth}
\includegraphics[width=8cm,height=5.5cm]{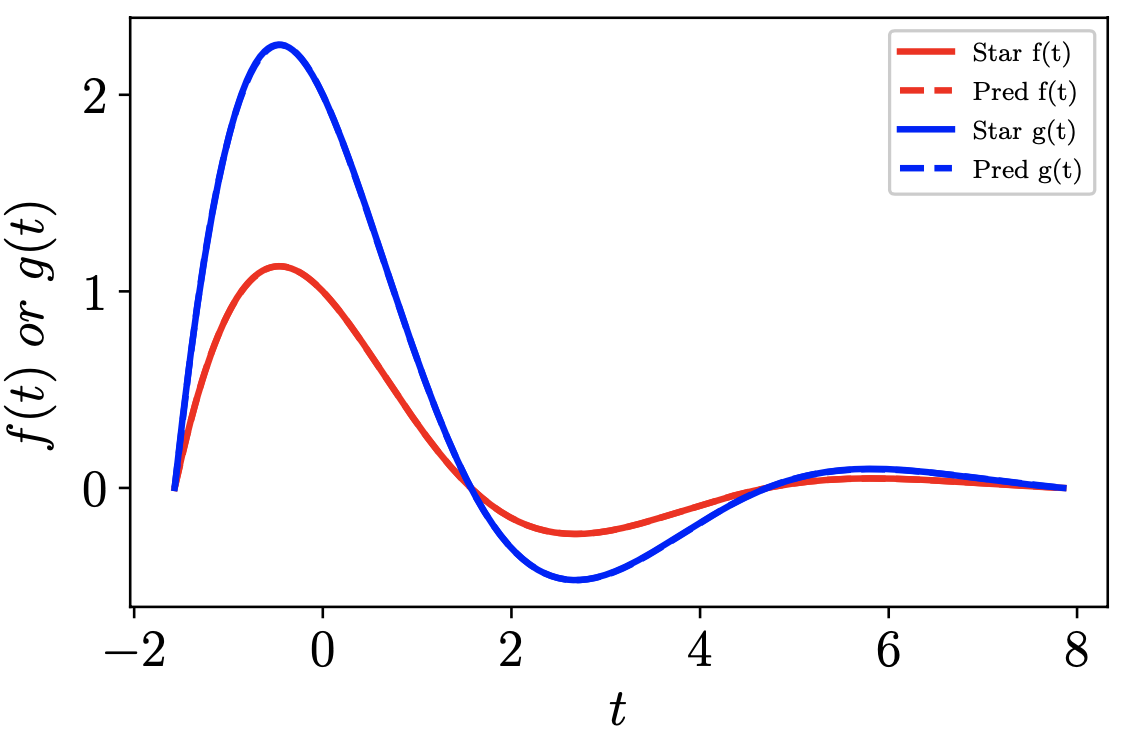}
\caption*{(c)}
\end{subfigure}
\begin{subfigure}[b]{0.45\textwidth}
\includegraphics[width=8cm,height=5.5cm]{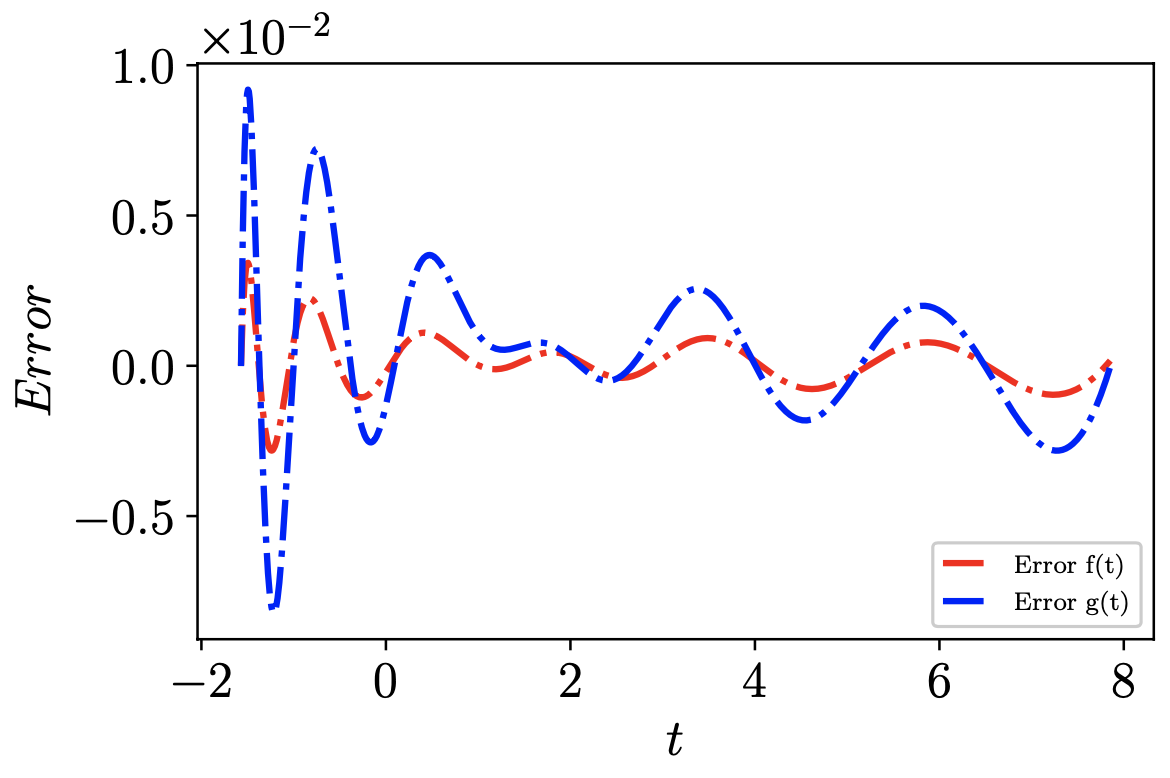}
\caption*{(d)}
\end{subfigure}\\
\caption{(Color online) Functional coefficients discovery (inverse problem) of the vKdV equation in case $2$: (a) Density plot of the data-driven solution of $u_5^{(vKdV)}$. (b) $3D$ plot of the data-driven solution of $u_5^{(vKdV)}$. (c) Comparison of predicted and true curves for two variable coefficients. (d) Error curves for two coefficients.}
\label{KdV-2v-f1}
\end{figure}
Because the indefinite integral of $f(t)$ and $g(t)$ is still in the form of the product of the exponential function and the trigonometric function, the trajectory of the soliton also presents the shape of oscillation decay consistent with the variable coefficient. In the previous examples we have found that the oscillations of the coefficients are related to the high-frequency fluctuations of the errors. Although the coefficient in this case is not strictly a periodic function, the cosine function brings the high-frequency oscillation properties similar to the periodic function to the entire coefficient $f(t)$(or $g(t)$), which can explain why the error curve fluctuates so much. In addition, the conclusions already observed in Case 1 are backed up here again, the error of the coefficient $g(t)$ with larger (larger absolute value) value and faster rate of change is significantly larger than that of $f(t)$. Then, the $L^2$ relative errors of the variable coefficients $f(t)$ and $g(t)$ are $1.85\times 10^{-3}$ and $2.77\times 10^{-3}$.
\end{itemize}

In these two new examples, the proposed method again demonstrates outstanding capabilities in the inverse problem with two variable coefficients. In our experiments, this method can handle multiple variable coefficients of various types with ease, and the relative error is on the order of $10^{-3}$ to $10^{-4}$. At the end of this section, a unified setting is given: the grid size of the coefficient is $500$, the trunk network and the branch network are $NN_u\{11,40,3,3\}$ and $NN_c\{8,30,3,2\}$, the activation function is Tanh, and 5000 $Adam$ iterations are performed before using $L$-$BFGS$. Other parameters are $\{n_s, n_f\}=\{2000, 20000\}$. More detailed numerical results and parameter settings are in Appendix \ref{Appendix_vKdV_Inverse_2}.

\subsection{Sawada-Kotera equation with variable coefficient.}\label{vSK}
\quad

In this section, consider continuing to increase the number of variable coefficients in the network to examine the capability limit of our proposed method. First, the importance of the KdV equation is self-evident, and the application of VC-PINN to the inverse problem of the vKdV equation is also discussed in Section \ref{KdV}. Another important equation, the Sawada-Kotera (SK) equation, was obtained by extending the KdV equation to the fifth order by Sawada and Kotera \cite{sawada1974method}. Their work also gives the $N$-soliton solution of the SK equation by inverse scattering transformation. The SK equation has important applications in the fields of shallow water waves and nonlinear lattices, and will not be repeated here.

What we care about in this section is the variable coefficient version of the SK equation, that is, the generalized SK equation with variable coefficient (gvSK), and its specific form is given as follows:
\begin{equation}\label{gvSK}
	u_t+\alpha(t)uu_{xxx}+\beta(t)u_xu_{xx}+\gamma(t)u^2u_x+\rho(t)u_{xxxxx}=0,
\end{equation}
where $\alpha(t),\beta(t),\gamma(t)$ and $\rho(t)$ are arbitrary analytic functions about $t$. Model \eqref{gvSK} is often referred to when describing the interaction between a water wave and a floating ice cover or gravity-capillary waves in fluid dynamics. The gvSK equation passes the Painlevé test \cite{xu2013painleve} and is proven to have the following integrability: Lax pair, $N$-soliton solution \cite{yu2010n} and conservation laws \cite{osman2019one}. In addition, it also includes many important equations, such as Lax equation, Kaup-Kupershmidt equation, and Ito equation. The increase in the order of the derivative increases the demand for computing power exponentially, but what we are really interested in is the problem of the coexistence of multiple variable coefficients rather than the problem of high-order derivatives. Based on these facts, to reduce the computational cost as much as possible, we decided to ignore the 5th order term ($\rho(t)=0$), the simplified equation is
\begin{equation}\label{gvSK2}
	u_t+\alpha(t)uu_{xxx}+\beta(t)u_xu_{xx}+\gamma(t)u^2u_x=0.
\end{equation}
where $u=u(x,t)$. Unless otherwise specified, gvSK refers to equation \eqref{gvSK2} instead of equation \eqref{gvSK} in the following descriptions of this article. \cite{el2023lie} adopts the symmetry method to obtain many new periodic wave solutions and solitary wave solutions of equation \eqref{gvSK}. To test the proposed method, we focus on some of these solutions, first let
\begin{equation}
	\xi =x-\frac{c_4}{c_2}\int\alpha(t)dt,
\end{equation}
where $c_2$ and $c_4$ are arbitrary constants, and $\xi$ is a new variable related to the original variables $x$ and $t$. Given the integrability condition as follows:
\begin{equation}\label{gvSKc}
	\beta(t)=\frac{\beta_0}{c_2}\alpha(t),\ \gamma(t)=\frac{\gamma_0}{c_2}\alpha(t),
\end{equation}
where $\beta_0$ and $\gamma_0$ are arbitrary constants. (They are constants of integration.) Recalling an exact solution in \cite{el2023lie} given under the integrability condition \eqref{gvSKc}, the expression is:
\begin{equation}\label{gvSKsol}
	u(x,t)=\frac{1}{\gamma_0}(B+4c_2-6c_2\text{tanh}^2(\xi)), 
\end{equation}
and satisfies
\begin{equation}\label{gvSKc2}
	B=\sqrt{4c_2^2+c_4\gamma_0},\ \beta_0=-c_2.
\end{equation}
In fact, conditions \eqref{gvSKc} and \eqref{gvSKc2} restrict some degrees of freedom of solution \eqref{gvSKsol}. More specifically, the gvSK equation and solution \eqref{gvSKsol} are completely determined by parameters $c_2,c_4,\gamma_0$ and variable coefficient $\alpha(t)$. Let $c_2=2,c_4=1,\gamma_0=4$, and then we change the form of the variable coefficient $\alpha(t)$ to test whether the proposed method can have the expected performance in the case of the coexistence of three variable coefficients. What needs special attention is that the integrable condition \eqref{gvSKc} restricts the variable coefficients $\alpha(t),\beta(t)$ and $\gamma(t)$ to be linearly related, which can be regarded as only one variable coefficient from a mathematical point of view. However, we did not impose integrability conditions on the network. For the network, these three variable coefficients are completely independent, so it can be regarded as a situation where multiple variable coefficients coexist. In addition, in view of the fact that the increase in the number of variable coefficients may bring greater difficulty, we decided to give the variable coefficients more boundary information (including the boundary value and the first-order derivative information of the boundary), which is derived from \eqref{ip} and \eqref{hdc}:
\begin{equation}
	\begin{split}
		&c_1(T_0)=\mathcal{C}_0,\ c_1(T_1)=\mathcal{C}_1,\\
		&\partial _t c_1\big|_{t=T_0}=\mathcal{C}_0^{(1)},	\partial _t c_1\big|_{t=T_1}=\mathcal{C}_1^{(1)},
	\end{split}
\end{equation}
where $c_1(t)$ is referred to as three variable coefficients $\alpha(t),\beta(t)$ and $\gamma(t)$. Then, the discretization of the exact solution \eqref{gvSKsol} provides internal data points ($s$-$type$ points) for the inverse problem, so that the results of the inverse problem under the polynomial coefficients of three different powers are shown as follows:
\begin{itemize}
	\item \textbf{Linear coefficient:} Assuming variable coefficient $\alpha(t)=t$, then:
\begin{equation}
	\beta(t)=-\alpha(t)=-t,\ \gamma(t)=2\alpha(t)=2t.
\end{equation}
The exact solution \eqref{gvSKsol} at this time becomes
\begin{equation}
	u_1^{(gvSK)}=\frac{1}{4}\left[8+2\sqrt{5}-12\text{tanh}^2(\frac{t^2}{4}-x)\right].
\end{equation}

Fig. \ref{SK-3v-p1} displays that for the case of linear coefficients, although the proposed method can invert the overall approximate changes of the three coefficients, the predicted curve and the accurate curve do not match well in some intervals. In addition, the error curves also exhibit a strange phenomenon, the three error curves seem to intersect at the same point at $t=0$. Although this seems to be somehow related to the intersection of the three variable coefficients at $(0,0)$, we did not find a plausible explanation. The error level shown in Fig. \ref{SK-3v-p1}(b) has also reached an unprecedented $10^{-1}$, but it is gratifying that the relative error can still maintain a level of about $10^{-2}$. Specifically, the $L^2$ relative errors of the variable coefficients $\alpha(t),\beta(t)$ and $\gamma(t)$ are $3.05\times 10^{-2}$, $4.18\times 10^{-2}$ and $2.94\times 10^{-2}$, respectively.

\begin{figure}[htpb]
\centering
\begin{subfigure}[b]{0.45\textwidth}
\includegraphics[width=8cm,height=5.5cm]{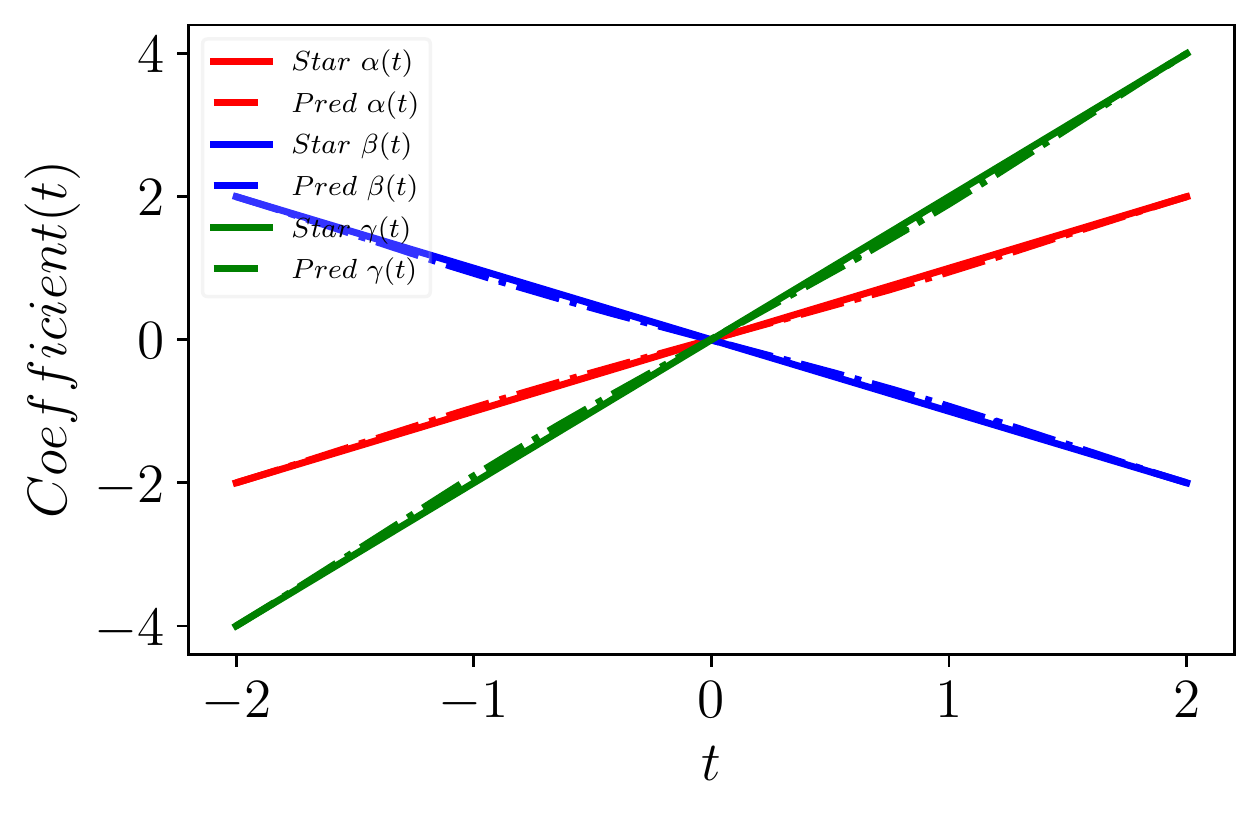}
\caption*{(a)}
\end{subfigure}
\begin{subfigure}[b]{0.45\textwidth}
\includegraphics[width=8cm,height=5.5cm]{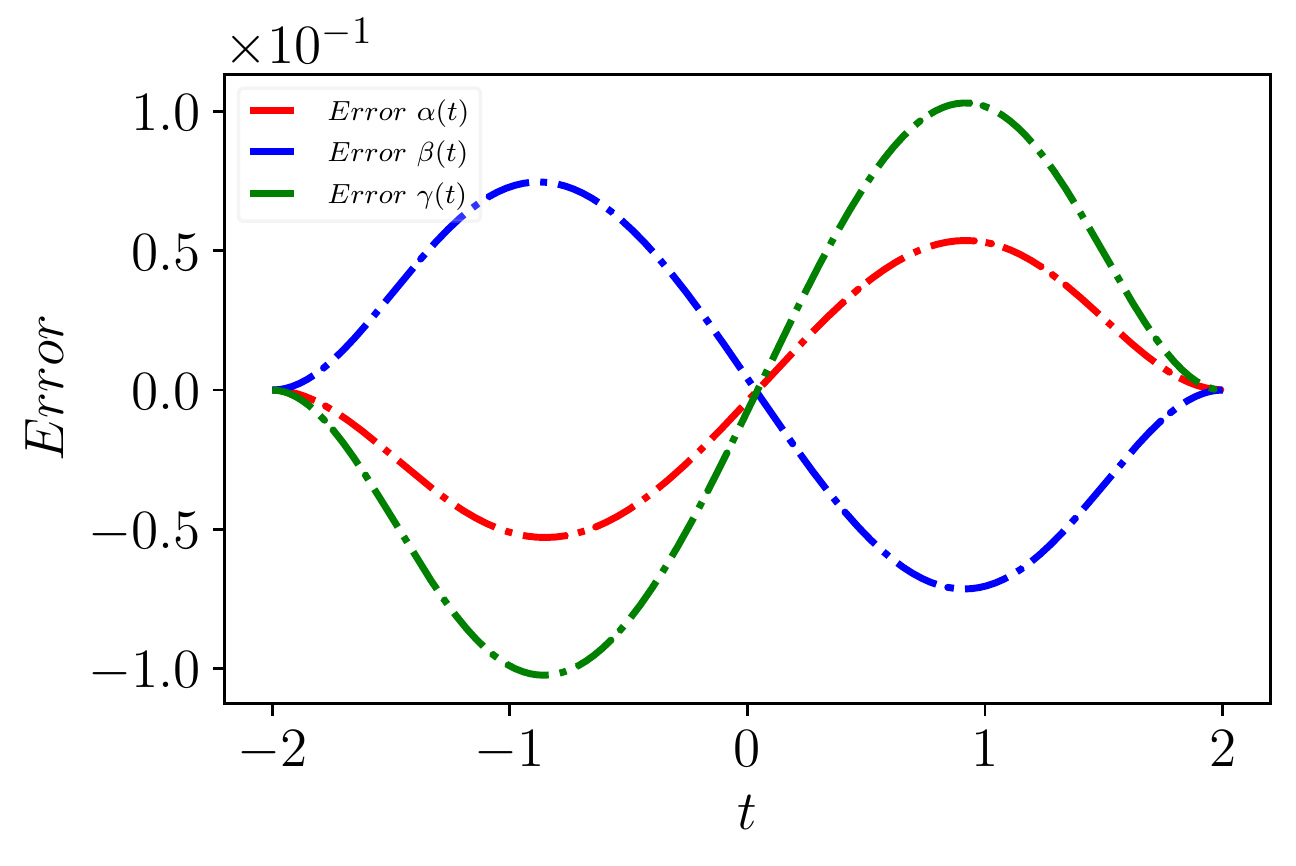}
\caption*{(b)}
\end{subfigure}\\
\caption{(Color online) Functional coefficient discovery (inverse problem) of the gVSK equation under linear coefficients: (a) Comparison of predicted and true values of three variable coefficients. (b) Error curves for three variable coefficients.}
\label{SK-3v-p1}
\end{figure}

	\item \textbf{Quadratic polynomial coefficient:} Assuming variable coefficient $\alpha(t)=\frac{t^2}{4}$, then the other two variable coefficients are
\begin{equation}
	\beta(t)=-\alpha(t)=-\frac{t^2}{4},\ \gamma(t)=2\alpha(t)=\frac{t^2}{2}.
\end{equation}
Under the above variable coefficient setting, the corresponding exact solution is
\begin{equation}
	u_2^{(gvSK)}=\frac{1}{4}\left[8+2\sqrt{5}-12\text{tanh}^2{\frac{t^3}{24}-x}\right].
\end{equation}

The results presented in Fig. \ref{SK-3v-p2} show that the results of the inverse problem under the quadratic polynomial coefficients are unexpectedly better than the linear coefficients. But inverting the coefficients of quadratic polynomials will be more difficult, which is a natural idea. After excluding the cause of chance, we guess that the reason for the counterintuitive phenomenon here is that even though ResNet can make multi-layer nonlinear layers more effectively approximate linear mapping, its ability seems to have an upper limit, at least in this example is such that. In this example, the $L^2$ relative errors of variable coefficients $\alpha(t),\beta(t)$ and $\gamma(t)$ are $1.68\times 10^{-2}$, $1.49\times 10^{-2}$ and $1.63\times 10^{-2}$, respectively.

\begin{figure}[htpb]
\centering
\begin{subfigure}[b]{0.45\textwidth}
\includegraphics[width=8cm,height=5.5cm]{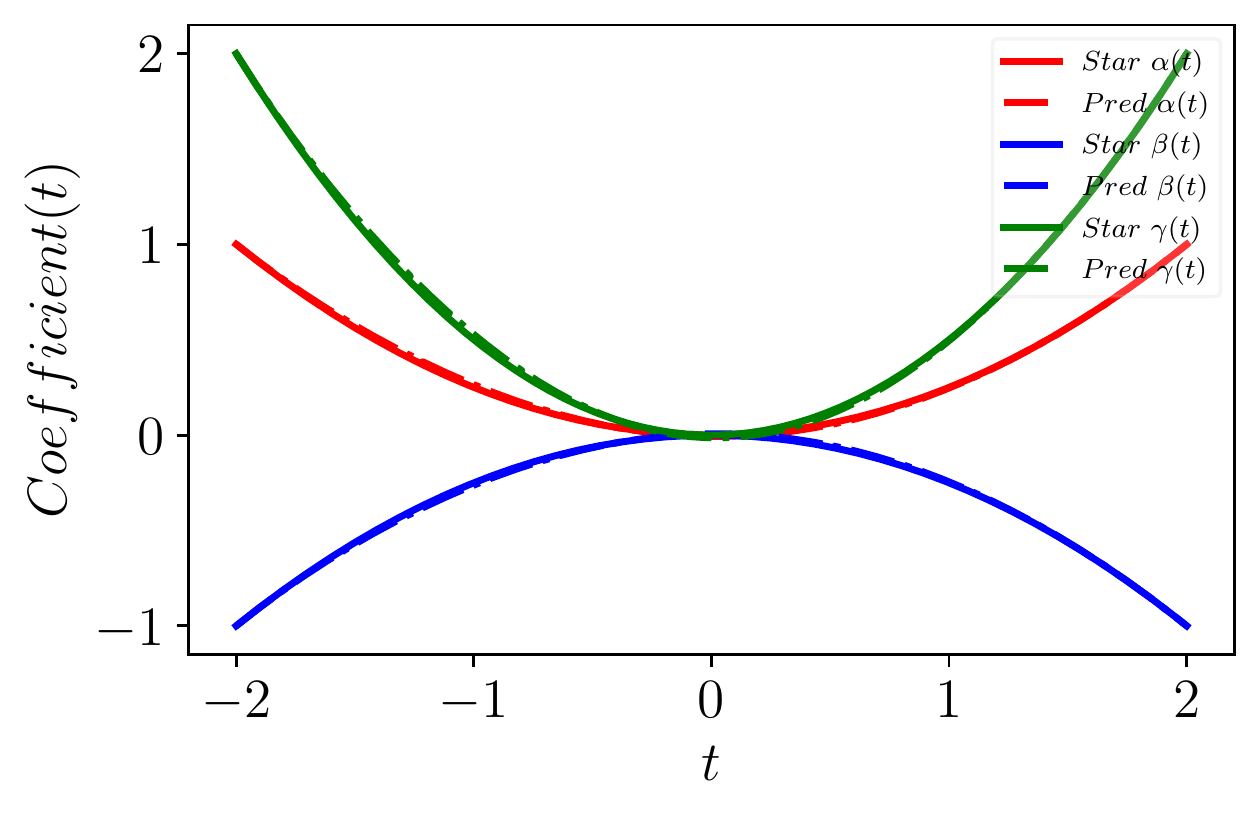}
\caption*{(a)}
\end{subfigure}
\begin{subfigure}[b]{0.45\textwidth}
\includegraphics[width=8cm,height=5.5cm]{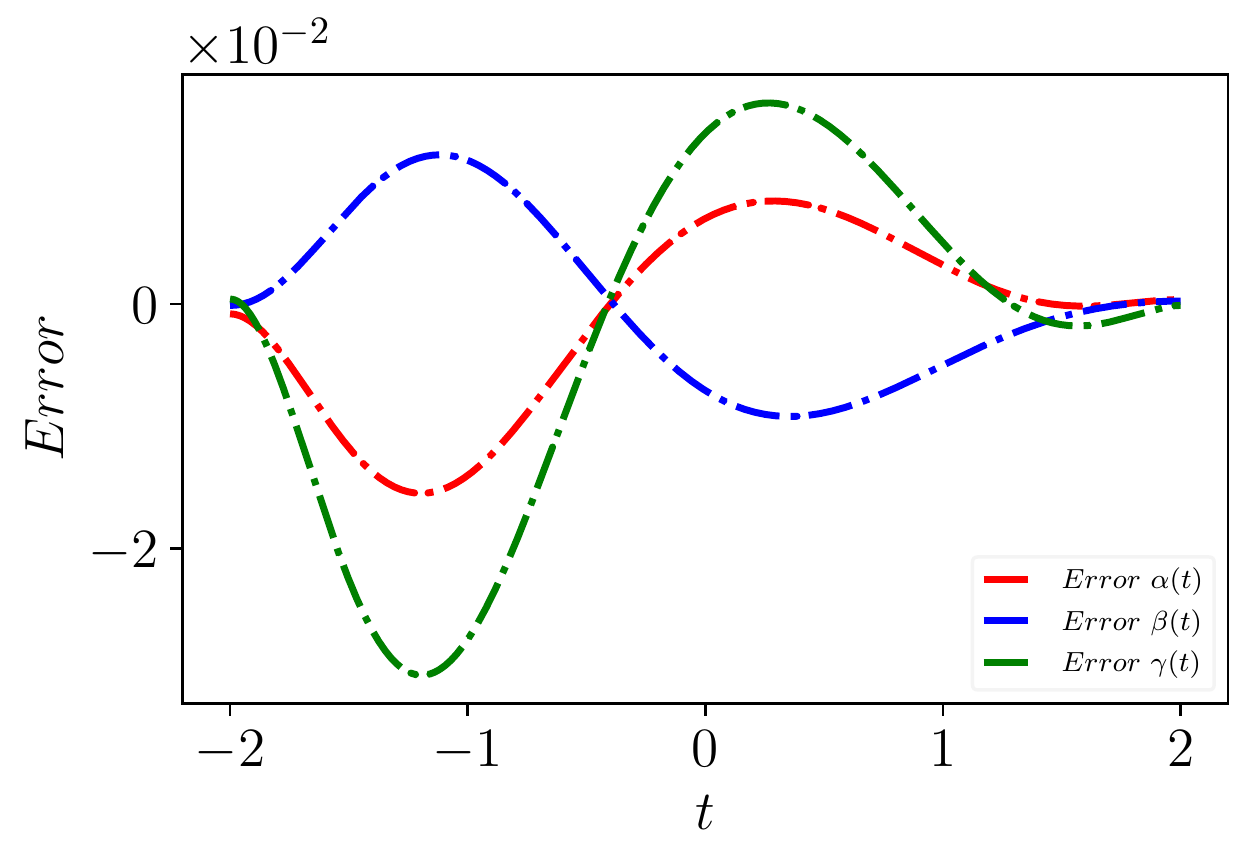}
\caption*{(b)}
\end{subfigure}\\
\caption{(Color online) Functional coefficient discovery (inverse problem) of the gVSK equation under quadratic polynomial coefficients: (a) Comparison of predicted and true values of three variable coefficients. (b) Error curves for three variable coefficients.}
\label{SK-3v-p2}
\end{figure}

	\item \textbf{Cubic polynomial coefficient:} Assuming that the variable coefficient $\alpha(t)$ is in the form of a cubic polynomial, that is, $\alpha(t)=\frac{t^3}{4}$, then the variable coefficients $\beta(t)$ and $\gamma(t)$ are
\begin{equation}
	\beta(t)=-\alpha(t)=-\frac{t^3}{4},\ \gamma(t)=2\alpha(t)=\frac{t^3}{2},
\end{equation}
thus the exact solution \eqref{gvSKsol} becomes
\begin{equation}
	u_3^{(gvSK)}=\frac{1}{4}\left[8+2\sqrt{5}-12\text{tanh}^2(\frac{t^4}{32}-x)\right].
\end{equation}

All three examples, including this one, show the unexplainable phenomenon we mentioned earlier, that is, the error curves of the three coefficients have common intersection point(s). The results in this example tell us that there can even be multiple intersection points, and they even seem to have nothing to do with the intersection points of the original coefficient curve. This also reflects the ``black box" problem of the neural network to a certain extent. I believe this will be one of the works we explore in the future. Another thing worth noting is that the error of $\gamma(t)$ (green line) with a larger change rate and coefficient value is also the largest among the three, which provides a factual basis for our previous empirical conclusions. Of course, the first two examples also clearly have such a phenomenon. Then give more quantified numerical results, the $L^2$ relative errors of the variable coefficients $\alpha(t),\beta(t)$ and $\gamma(t)$ are $1.54\times 10^{-2}$, $1.93\times 10^{-2}$ and $1.45\times 10^{-2}$, respectively.

\begin{figure}[htpb]
\centering
\begin{subfigure}[b]{0.45\textwidth}
\includegraphics[width=8cm,height=5.5cm]{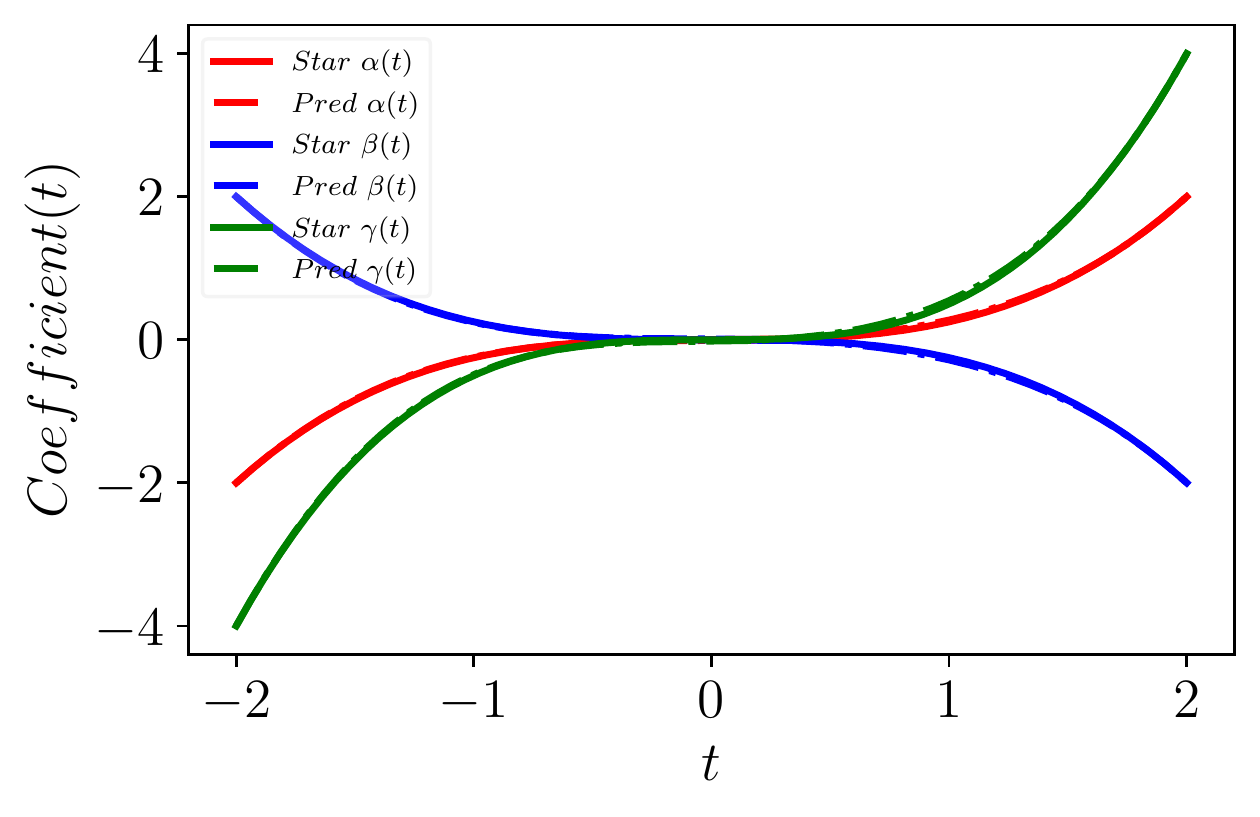}
\caption*{(a)}
\end{subfigure}
\begin{subfigure}[b]{0.45\textwidth}
\includegraphics[width=8cm,height=5.5cm]{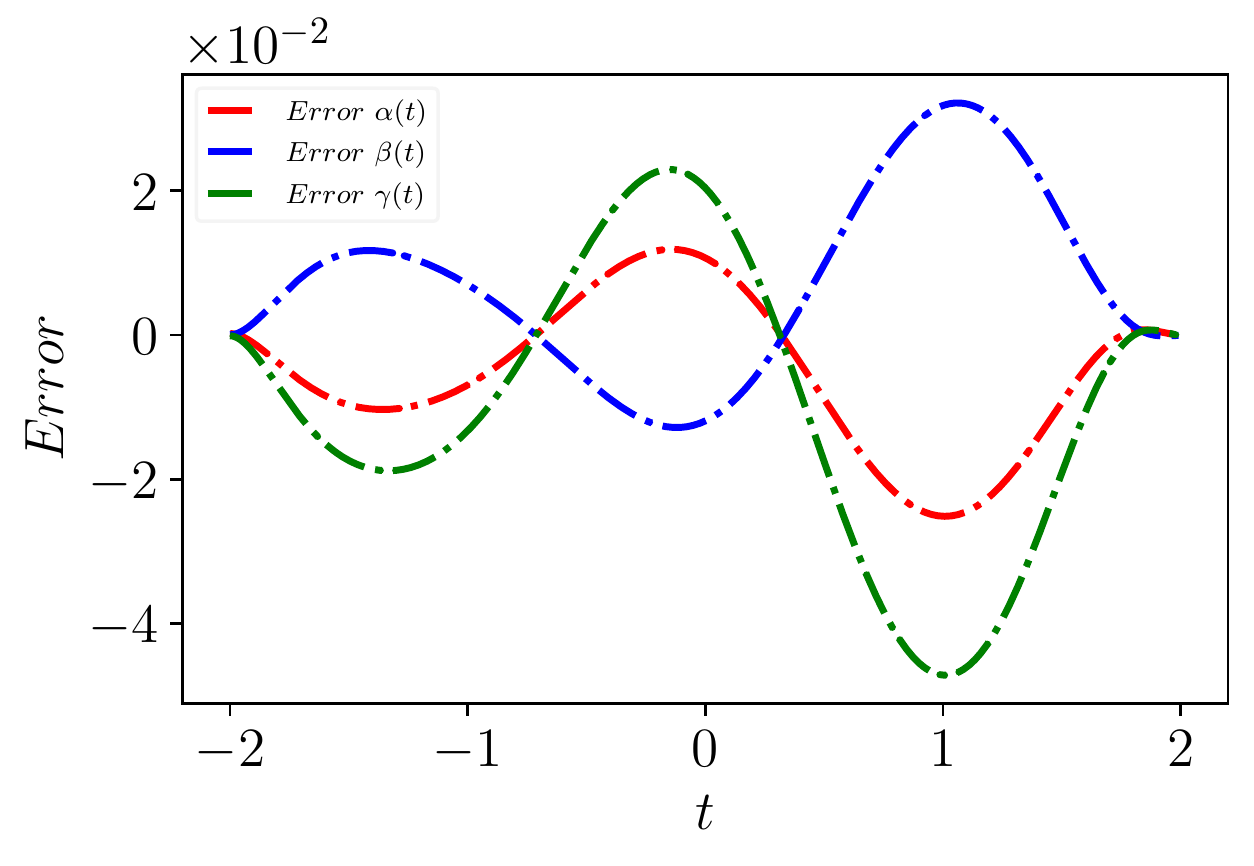}
\caption*{(b)}
\end{subfigure}\\
\caption{(Color online) Functional coefficient discovery (inverse problem) of the gVSK equation under cubic polynomial coefficient: (a) Comparison of predicted and true values of three variable coefficients. (b) Error curves for three variable coefficients.}
\label{SK-3v-p3}
\end{figure}
\end{itemize}

In order to more directly compare the errors of variable coefficients under different types of coefficients, we unify the above results into Table \ref{SKt}.

\begin{table}[H]
\caption{$L^2$ relative error of variable coefficient $\alpha(t)$, $\beta(t)$ and $\gamma(t)$}
\label{SKt}
\begin{center}
\begin{tabular}{c||c|c|c}
\hline
\hline
 & Linear Coefficients &  Quadratic polynomial & Cubic polynomial \\
\hline
\hline
Error of $\alpha(t)$ & 3.05$\times 10^{-2}$ & 1.68$\times 10^{-2}$ & 1.54$\times 10^{-2}$\\
Error of $\beta(t)$ & 4.18$\times 10^{-2}$ & 1.49$\times 10^{-2}$ & 1.93$\times 10^{-2}$\\
Error of $\gamma(t)$ & 2.94$\times 10^{-2}$ & 1.63$\times 10^{-2}$ & 1.45$\times 10^{-2}$\\
\hline
\hline
\end{tabular}
\end{center}
\end{table}
The results of these numerical examples show that, compared with the single variable coefficient or the coexistence of two coefficients of the KdV equation in Section \ref{KdV}, the error of the coexistence of three coefficients in the gvSK equation in this section is significantly increased. However, the relative $L^2$ error can still be maintained above the $10^{-2}$ level, which is acceptable to us. Combined with the numerical results of the gvKP equation in the next section, it will be found that the form of the equation may also be an important factor affecting the accuracy. Generally speaking, the proposed method has also withstood the test when the three variable coefficients coexist, but how to further increase the capability limit of the method is the direction we need to think about. Finally, some uniform settings in the experiment are given: the grid size of the coefficients is $500$, the trunk network and the branch network are $NN_u\{11,40,3,3\}$ and $NN_c\{8,30,3,2\}$, the activation function is Tanh, and 5000 $Adam$ iterations are performed before using $L$-$BFGS$, and other the parameter is $\{n_s,n_f\}=\{2000, 20000\}$. More detailed numerical results and parameter settings are in Appendix \ref{Appendix_gvSK_Inverse}.

\subsection{Generalized Kadomtsev–Petviashvili equation with variable coefficient.}\label{gKP_inverse_1}
\quad

\begin{figure}[htpb]
\centering
\begin{subfigure}[b]{0.45\textwidth}
\includegraphics[width=8cm,height=5.5cm]{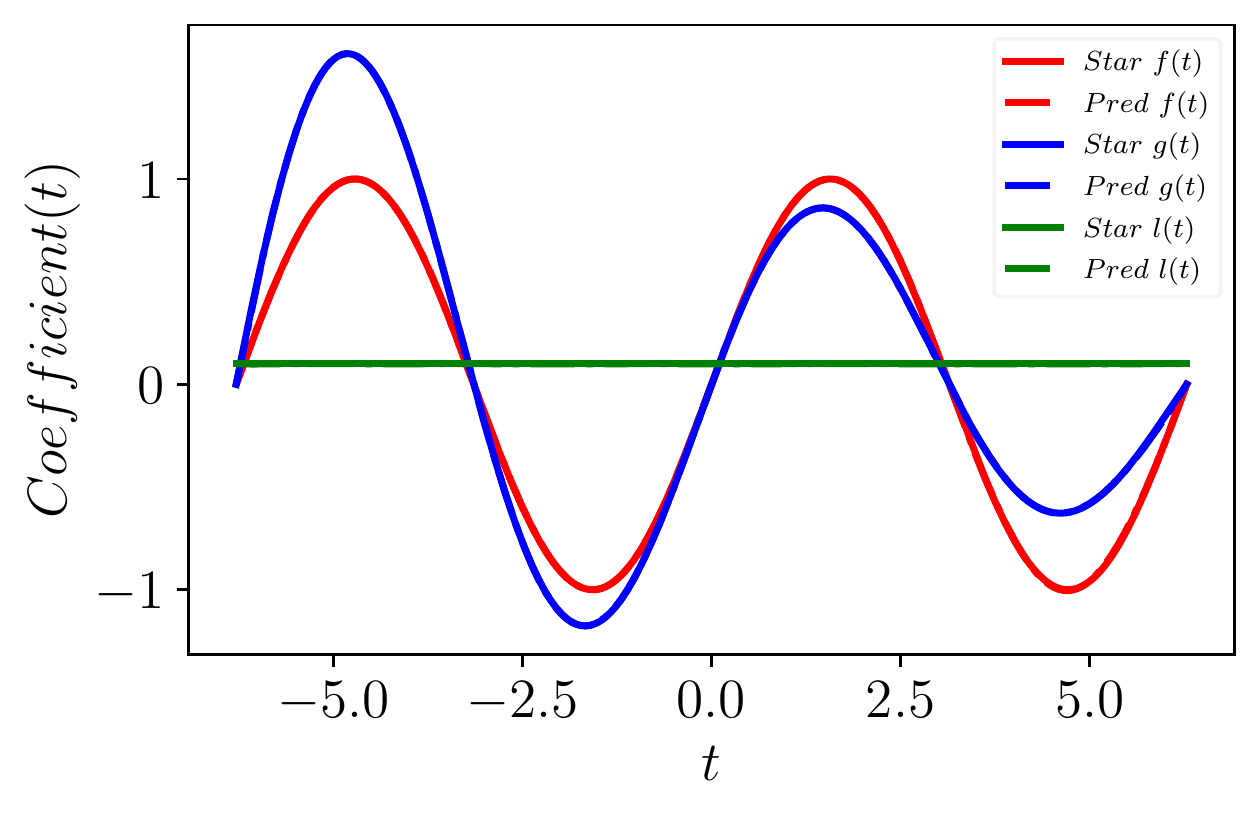}
\end{subfigure}
\begin{subfigure}[b]{0.45\textwidth}
\includegraphics[width=8cm,height=5.5cm]{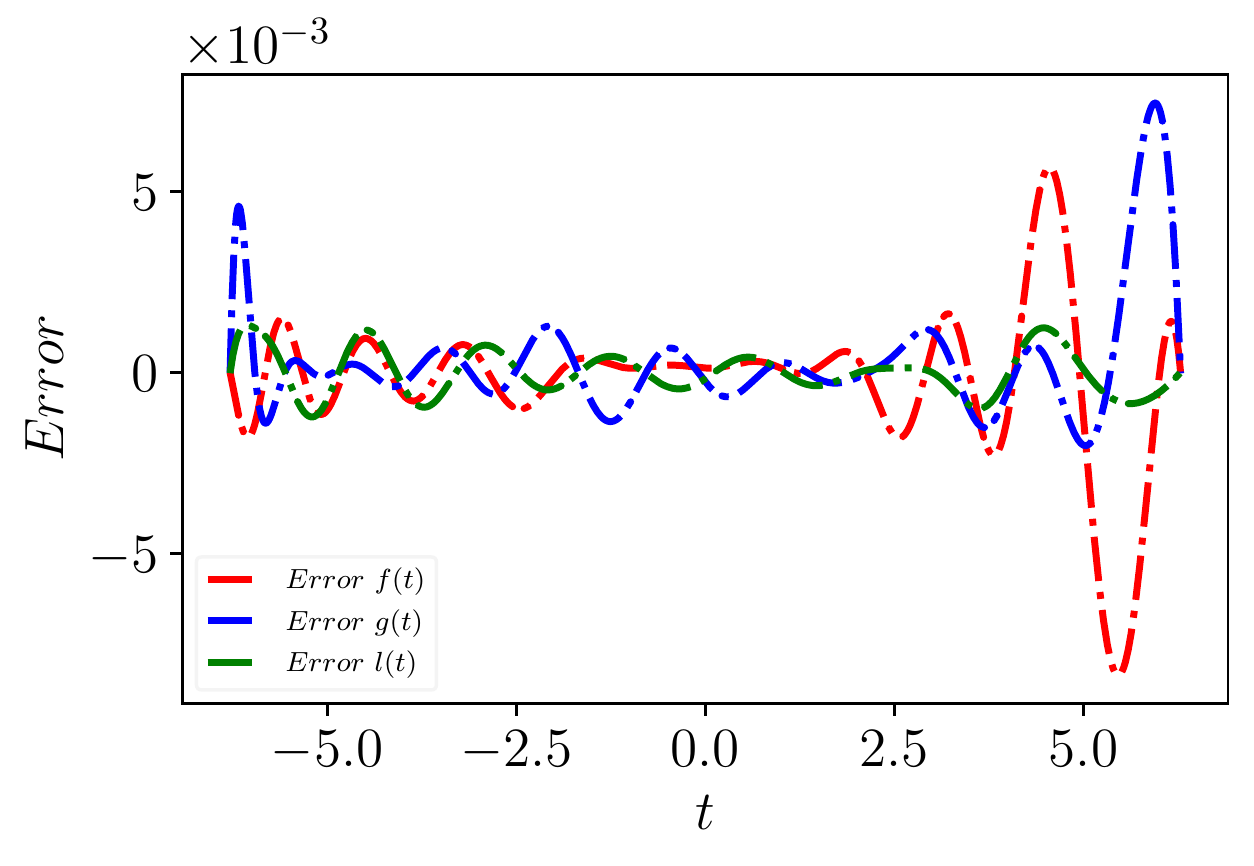}
\end{subfigure}\\
\begin{subfigure}[b]{0.45\textwidth}
\includegraphics[width=8cm,height=5.5cm]{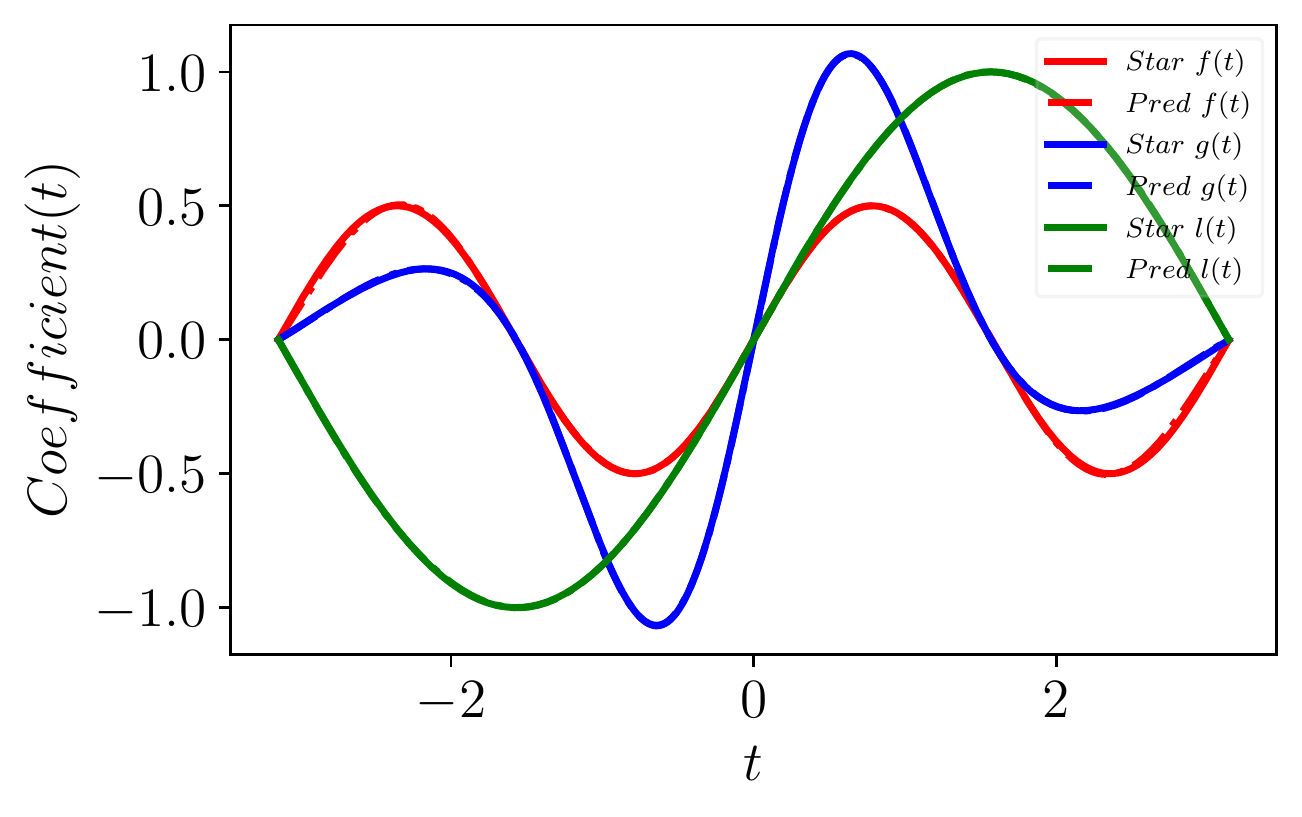}
\end{subfigure}
\begin{subfigure}[b]{0.45\textwidth}
\includegraphics[width=8cm,height=5.5cm]{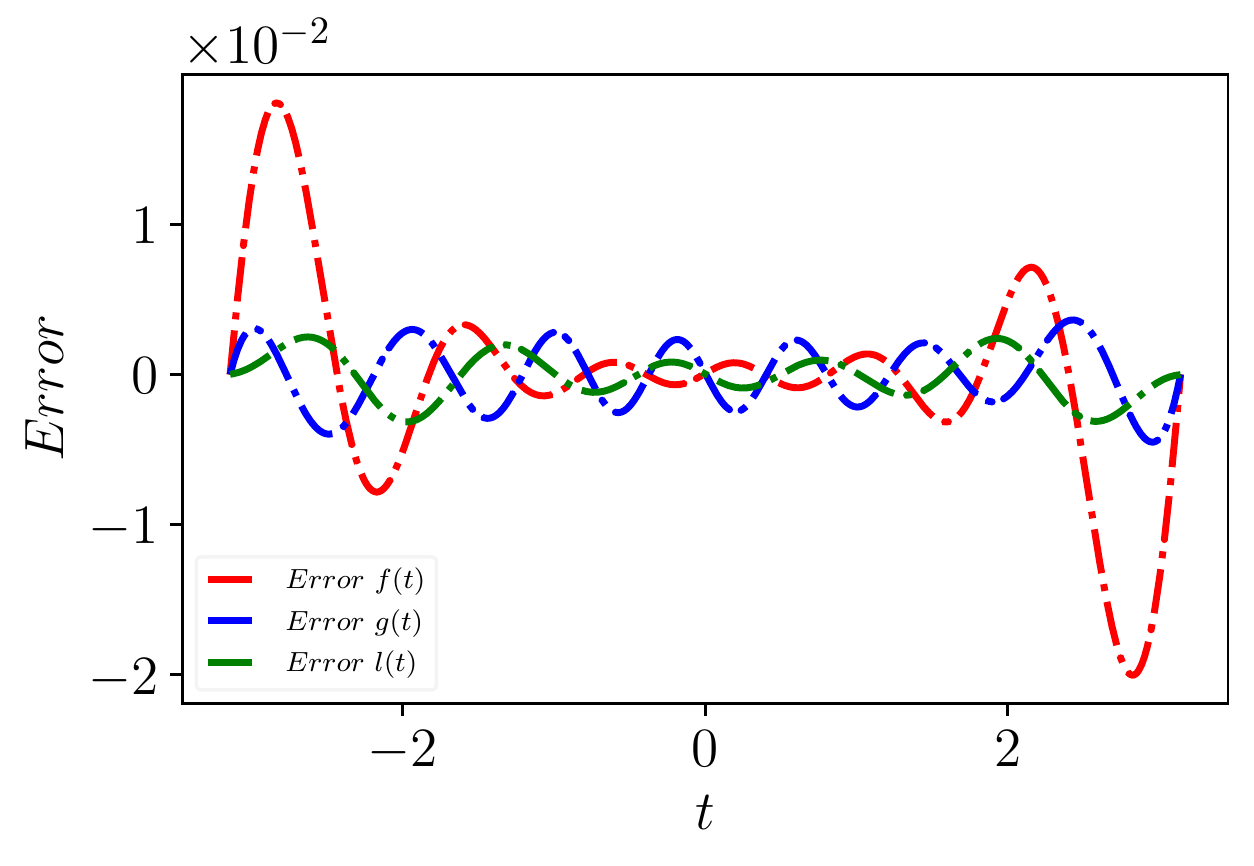}
\end{subfigure}\\
\begin{subfigure}[b]{0.45\textwidth}
\includegraphics[width=8cm,height=5.5cm]{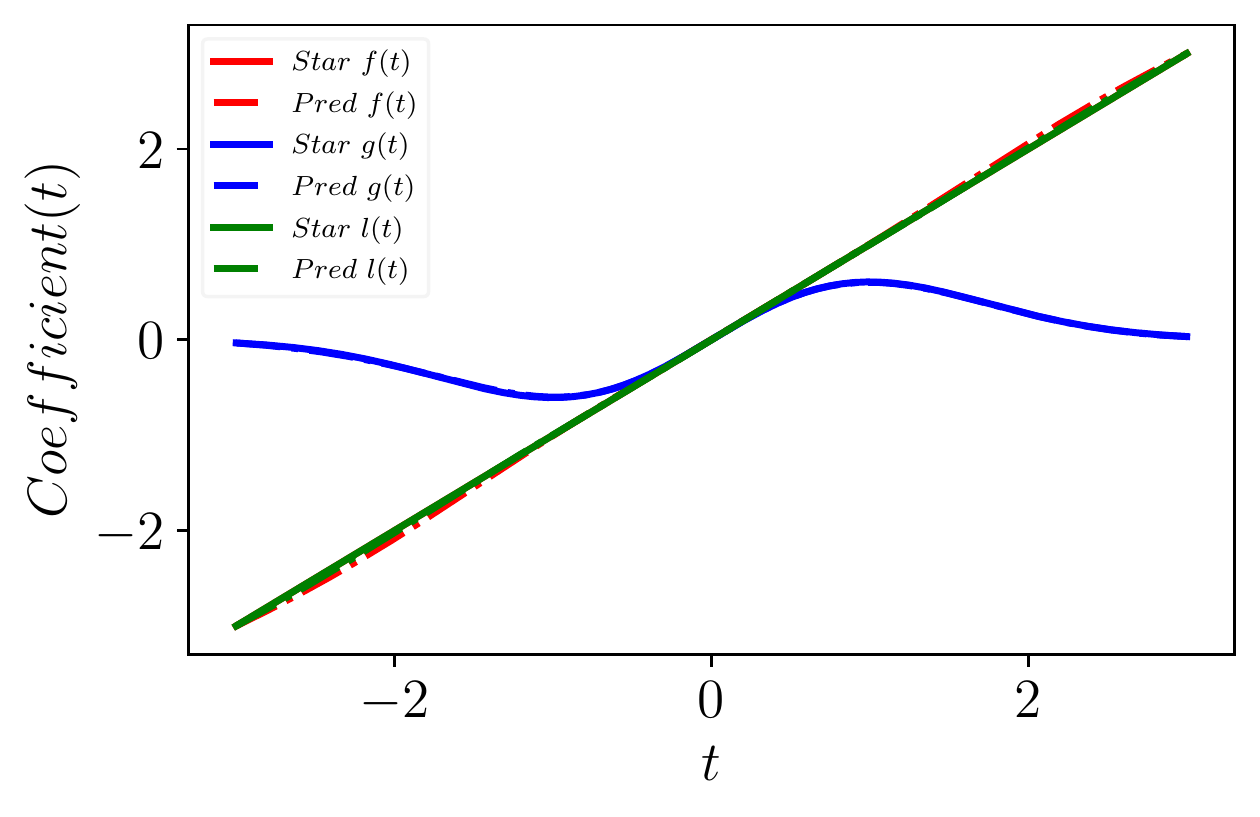}
\end{subfigure}
\begin{subfigure}[b]{0.45\textwidth}
\includegraphics[width=8cm,height=5.5cm]{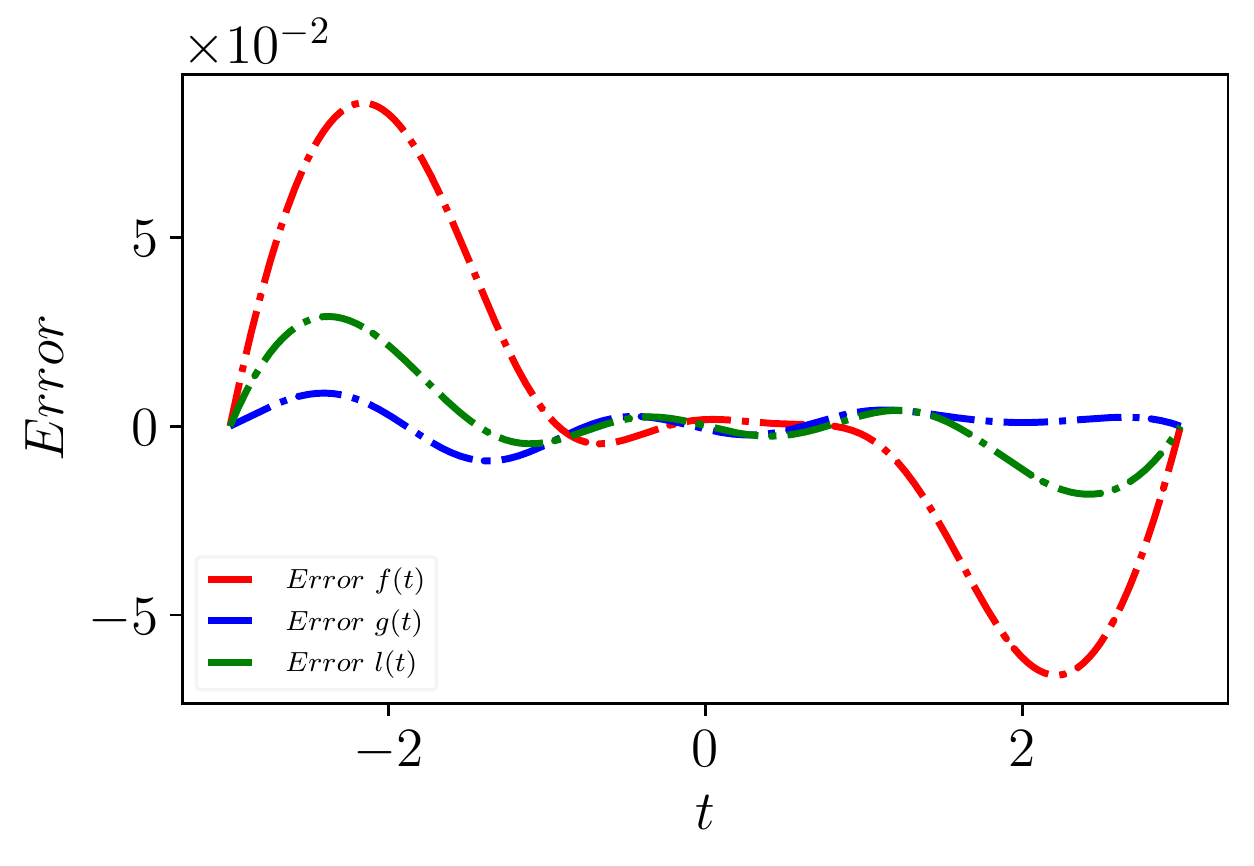}
\end{subfigure}\\
\begin{subfigure}[b]{0.45\textwidth}
\includegraphics[width=8cm,height=5.5cm]{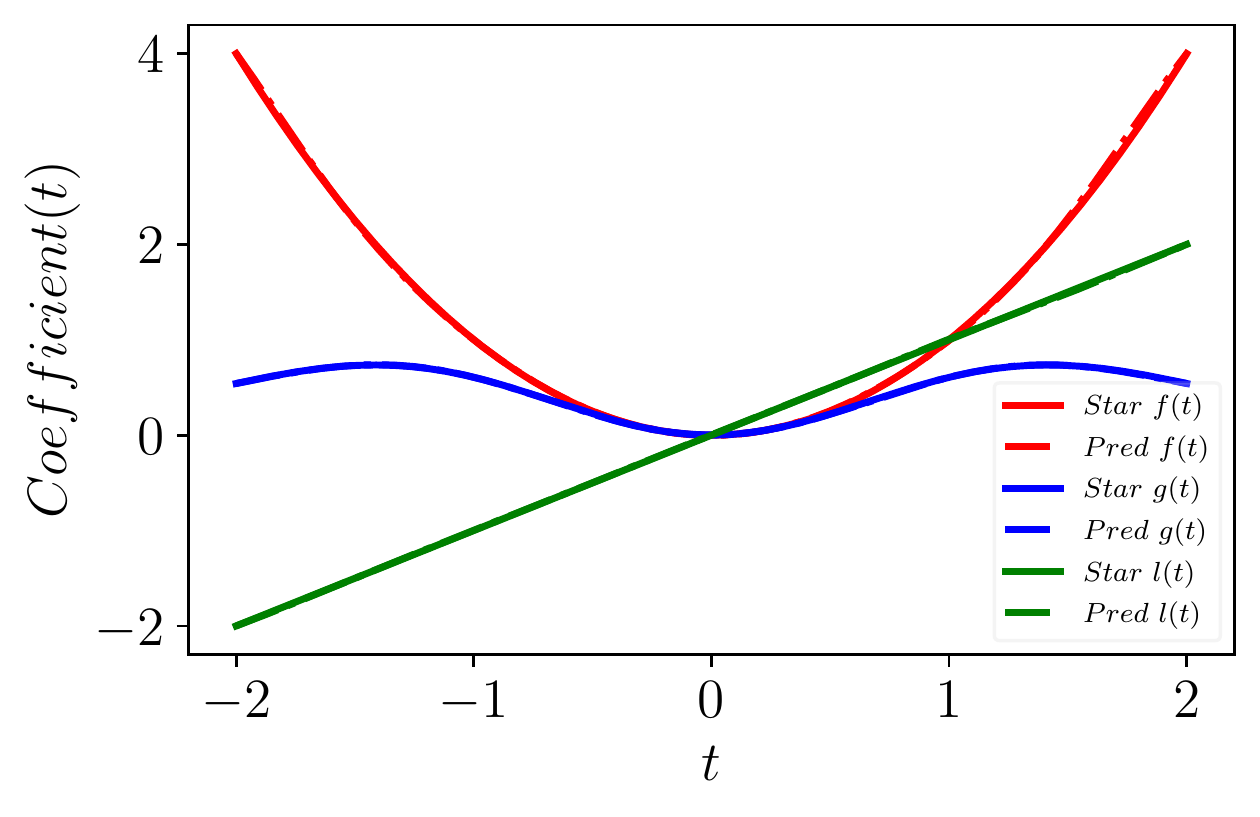}
\caption*{(a)}
\end{subfigure}
\begin{subfigure}[b]{0.45\textwidth}
\includegraphics[width=8cm,height=5.5cm]{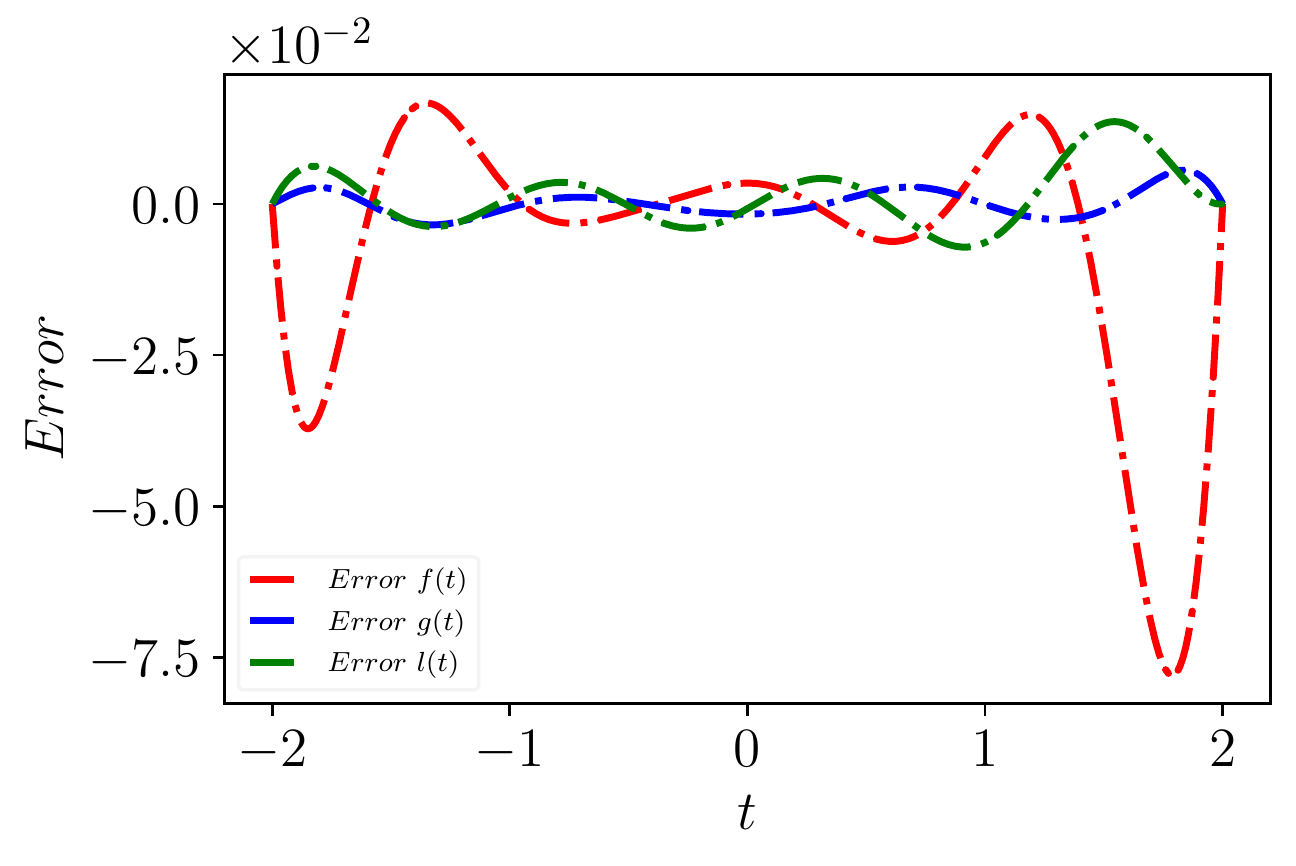}
\caption*{(b)}
\end{subfigure}\\
\caption{(Color online) The discovery of the function coefficients of the gvKP equation under four cases (inverse problem), each row corresponds to the results of a case (from top to bottom are Case 1 to Case 4): (a) The predicted value of the three variable coefficients compared with the real value. (b) Error curves for the three coefficients.}
\label{KP-3v}
\end{figure}

In this section, we wish to further complicate the problem from another angle, specifically, we consider increasing the dimensionality of the equation while keeping the number of co-existing variable coefficients at three. The $(2+1)$-dimensional KP equation discussed in Section \ref{KP} contains three variable coefficients, which fully meets our requirements. Therefore, the exact solutions $u_i^{(gvKP)},i=1,2,3,4$ under the four cases in Section \ref{KP} will be samples of the inverse problem in this section. In the following discussion, the free parameters and variable coefficients in these exact solutions are the same as in Section \ref{KP}, the only difference is the space-time region discussed. (In order to better present the changes in the coefficients.) Fig. \ref{KP-3v} shows the results of the inverse problem in four cases.

\begin{table}[H]
\caption{$L^2$ relative error of variable coefficient $f(t)$, $g(t)$ and $l(t)$}
\label{KPt}
\begin{center}
\begin{tabular}{c||c|c|c|c}
\hline
\hline
 & Case 1 &  Case 2 & Case 3 & Case 4 \\
\hline
\hline
Error of $f(t)$ & 2.98$\times 10^{-3}$ & 1.95$\times 10^{-2}$ & 2.45$\times 10^{-2}$ & 1.29$\times 10^{-2}$\\
Error of $g(t)$ & 2.06$\times 10^{-3}$ & 4.01$\times 10^{-3}$ & 1.10$\times 10^{-2}$ & 3.98$\times 10^{-3}$\\
Error of $l(t)$ & 5.94$\times 10^{-3}$ & 2.10$\times 10^{-3}$ & 6.98$\times 10^{-3}$ & 4.42$\times 10^{-3}$\\
\hline
\hline
\end{tabular}
\end{center}
\end{table}

Quantified numerical results are more helpful for analysis and comparison. Table  \ref{KPt} shows the more detailed error results of the inverse problem. The order of magnitude of the $l^2$ relative error is basically at the level of $10^{-2}$ to $10^{-3}$. This result is better than the result of the vSK equation in Section \ref{vSK}, which is very surprising. Although such comparisons were not performed under strict control of variables, some attempts after tuning hyperparameters tell us that this result seems to be general. Therefore, it is reasonable to guess that some properties of the vSK equation itself affect the generalization of the network. This further admits that although the ultra-universal PINN framework and its variants can be applied to most equations almost equally, it is also necessary to design a more targeted network for equations with specific structures. In addition, observation of the error curves tells us that there are some exceptions to our proposed empirical conclusions in complex scenarios. For example, the largest error in Case 2 is the variable coefficient $f(t)$ (red). Overall, in the inverse problem where multiple variable coefficients coexist in $(2+1)$-dimensions, the proposed method can still invert the variation of coefficients with acceptable accuracy.

Compared with the forward problem, the network parameters in the inverse problem of the gvKP equation have been slightly modified, and a unified setting is given: the grid size of the coefficients is $500$, the trunk network and the branch network are $NN_u\{10,40,4,2\}$ and $NN_c\{8,30,3,2\}$, the activation function is Tanh, and $5000$ $Adam$ iterations are performed before using $L$-$BFGS$, and other the parameter is $\{n_s,n_f\}=\{20000, 50000\}$. More detailed numerical results and parameter settings are in Appendix \ref{Appendix_gvKP_Inverse}.

\section{Analysis and Discussion}\label{analy_Dis}
In the numerical experiments of the forward and inverse problems in Section \ref{Forward_ex} and Section \ref{Inverse_ex}, the performance of VC-PINN is obvious to all. However, this section will make a further in-depth analysis of the proposed method from the perspectives of principle and results. It mainly includes the following four aspects: 1. The necessity of ResNet; 2. The relationship between the convexity of variable coefficients and learning; 3. Anti-noise analysis; 4. The unity of forward and inverse problems/relationship with standard PINN.
\subsection{The necessity of ResNet}\label{nece_ResNet}
\quad

The proposed method adopts the structure of ResNet, and this design is mainly based on two considerations. On the one hand, ResNet itself can alleviate the ``vanishing gradient", on the other hand, in the variable coefficient problem, it unifies linearity and nonlinearity. The following will further explain why ResNet is a suitable choice in our network from these two aspects.
\subsubsection{Mitigating the problem of vanishing gradients}
\quad

``Vanishing gradient" is an important issue in the training process of deep learning, which was first formally proposed by Hochreite (1991) in his graduation thesis\cite{hochreiter1991untersuchungen}. The reason for this phenomenon is that the small gradient value gradually accumulates during the backpropagation process, and finally decays exponentially as the number of network layers increases. In order to explain how ResNet alleviates the disappearance of gradients from a theoretical level, we will analyze the propagation of gradients in the network (only the gradient of a single sample is given). In the subsequent derivation, in order to distinguish the trunk network and the branch network, we put subscripts on $X^{[i]}_{[\cdot]}$, $R^{[i]}_{[\cdot]}$, $W^{[i]}_{[\cdot]}$ and put superscripts on $D^{[\cdot]}$, $N_B^{[\cdot]}$, $N_h^{[\cdot]}$, $\mathcal{K}^{[\cdot]}$ where the labels $u$ and $c$ represent the corresponding quantities in the trunk network and branch network, respectively. First, similar to the result in \cite{he2016identity}, the calculation formula of the gradient of the loss item $Loss_s(\theta)$ to $X^{[j]}_u$ is given as ($iN_h+1\le j< [(i+1)N_h+1]$)
\begin{align}\label{grad_loss_s}
	\frac{\partial Loss_s}{\partial X^{[j]}_u}&=\frac{\partial Loss_s}{\partial R^{[N_B^u]}_u}\cdot \prod_{k=j}^{(i+1)N_h^u}\frac{\partial X^{[k+1]}_u}{\partial X^{[k]}_u}\cdot \prod_{k=i+1}^{N_B^u-1}\left[\mathcal{K}^u+\frac{1}{\partial R^{[k]}_u}\left(\mathcal{L}_k(R^{[k]}_u)\right)\right],\end{align}
other definitions involved here are consistent with \eqref{eq2} and \eqref{eq3}. In fact, the gradient of $Loss_s$ to $W^{[j-1]}_u$ only needs to be multiplied by $\frac{\partial X^{[j]}_u}{\partial W^{[j-1]}_u}$ on \eqref{grad_loss_s}. The gradient calculation formulas of the loss items $Loss_I(\theta)$, $Loss_b(\theta)$ and $Loss_c(\theta)$ involved in the forward and inverse problems are similar to the above, while the loss item $Loss_f(\theta)$ is different. The gradient of the loss item $Loss_f(\theta)$ to the weights $W^{[j]}_u$ and $W^{[j]}_c$ of the trunk network and the branch network is given as:
\begin{align}
	\frac{\partial [(\tilde{u})_t]}{\partial W^{[j]}_u}&=W^{[D^u-1]}_u\cdot\frac{\partial R^{[0]}_u}{\partial t}\cdot \prod_{\substack{0\le k \le N^u_B-1\\k\neq i}} \left[\mathcal{K}^u+\frac{1}{\partial R^{[k]}_u}\left(\mathcal{L}_k(R^{[k]}_u)\right)\right]\cdot \frac{1}{\partial W^{[j]}_u}\left[\mathcal{K}^u+\frac{1}{\partial R^{[i]}_u}\left(\mathcal{L}_i(R^{[i]}_u)\right)\right],\label{grad_u_w}\\
	\frac{\partial Loss_f}{\partial W^{[j]}_c}&=\frac{\partial Loss_f}{\partial f}\cdot\frac{\partial X^{[j+1]}_c}{\partial W^{[j]}_c}\cdot\mathcal{N}[\tilde{u}]\cdot W^{[D^c-1]}_c\cdot \prod_{k=j+1}^{(i+1)N_h^c}\frac{\partial X^{[k+1]}_c}{\partial X^{[k]}_c}\cdot \prod_{k=i+1}^{N_B^c-1}\left[\mathcal{K}^c+\frac{1}{\partial R^{[k]}_c}\left(\mathcal{L}_k(R^{[k]}_c)\right)\right]\label{grad_loss_f}.
\end{align}
where the conditions for \eqref{grad_u_w} and \eqref{grad_loss_f} to hold are $iN_h^u+1\le j<[(i+1)N_h^u+1]$ and $iN_h^u\le j<[(i+1)N_h^u]$, respectively. It is easy to find that \eqref{grad_u_w} is not the complete gradient of $Loss_f(\theta)$ to $W^{[j]}_u$, but is only calculated with the $\tilde{u}_t$ term in $f$. But it is completely similar for other terms in $f$, the final gradient is the sum of the gradients of all terms, so it doesn't hurt to focus on a specific term.

The loss items $Loss_s(\theta)$, $Loss_I(\theta)$, $Loss_b(\theta)$ and $Loss_c(\theta)$ in the forward and inverse problem only contribute to the gradient of the weight in the trunk network, while the loss item $Loss_f(\theta)$ contributes to the gradient of the weight in both the trunk network and the branch network. \eqref{grad_loss_s}, \eqref{grad_u_w} and \eqref{grad_loss_f} are the gradient formula (or the main part of the gradient formula) of the loss function to the weights. The part involving multiplication ($\prod$) among them is what we care about, because this is the item most likely to cause the gradient to tend to zero. It can be inferred from the above formula that if we adopt the ResNet structure for both the trunk network and the branch network (i.e. $\mathcal{K}^u=\mathcal{K}^c=1$), and ensure that the number of network layers ($N_h^u$ and $N_h^c$) contained in each residual block is not too large, then the multiplication part in formulas \eqref{grad_loss_s}, \eqref{grad_u_w} and \eqref{grad_loss_f} is unlikely to be a small amount, so the ``vanishing gradients" is alleviated to a certain extent. It is worth mentioning that if batch normalization (BN) technology is used on the basis of the ResNet structure, it may produce more surprising effects, but this is not the focus of this article. More specific details on the derivation are presented in Appendix \ref{A_grad}.
\subsubsection{Unity of linear and nonlinear}\label{sec_linear_nonlinear}
\quad

It is well known that nonlinear phenomena are an important part of natural science. But for a physical model (equation) with variable coefficients, the nonlinearity of the solution and the nonlinearity of variable coefficients are not the same thing. Even linear coefficients can generate nonlinear waves to explain natural phenomena, as we saw in the numerical experiments in Section \ref{Forward_ex} and Section \ref{Inverse_ex} of parabolic solitons and kinks evolving along parabolic curves. In order to illustrate the importance of linear coefficients, a few specific examples are listed below:
\begin{itemize}
	\item The heat equation describes the diffusion behavior of heat in a region, and the following is its variable coefficient version in $3D$ space \cite{costin2012borel}
\begin{equation}\label{heat_eq}
	u_t=\alpha(x,y,z)\Delta u,
\end{equation}
where $\Delta$ represents the Laplace operator ($\Delta u = u_{xx}+u_{yy}+u_{zz}$), and the variable coefficient $\alpha(x,y,z)$ represents the thermal conductivity, which is related to the temperature and the nature of the medium. Therefore, it is entirely possible that the variable coefficient is a linear function of a certain spatial component (such as $x$) in a heterogeneous medium. Of course, due to the constant positive property of thermal conductivity, it can only maintain linearity in a certain interval. In addition, the form of the diffusion equation is very similar to \eqref{heat_eq}, and the diffusion coefficient may also show a linear change in some media with inhomogeneous concentrations.
\item The wave equation is used to describe the propagation of waves in classical physics, including mechanical waves, electromagnetic waves, and so on. When there is an external force term, the wave equation with variable coefficients is given as \cite{radu2010decay}
\begin{equation}
	u_{tt}=\text{div}(p({\bm x}, t)\nabla u)+f({\bm x},t),
\end{equation}
where $\text{div}$ and $\nabla$ represent divergence and gradient, respectively, while $p({\bm x},t)$ and $f(x,t)$ represent medium parameters and external force terms, respectively. In different physical backgrounds, $p({\bm x},t)$ can represent both medium parameters and wave velocity. The inhomogeneity and dynamics of the medium may cause the wave velocity to vary at different locations and times. For example, when a sound wave propagates in a gas, factors such as the temperature and density of the gas may affect its propagation speed; in the propagation of seismic waves, physical properties such as the density and elastic modulus of the medium may also affect the propagation speed of seismic waves. Therefore, whether $p({\bm x},t)$ is a medium parameter or a wave velocity, it is entirely possible to show a linear function with time or a certain space variable. Also, for freer external force terms, a linear variation is more likely.
\end{itemize}

\begin{figure}[htpb]
\centering
\begin{subfigure}[b]{0.45\textwidth}
\includegraphics[width=8cm,height=5.5cm]{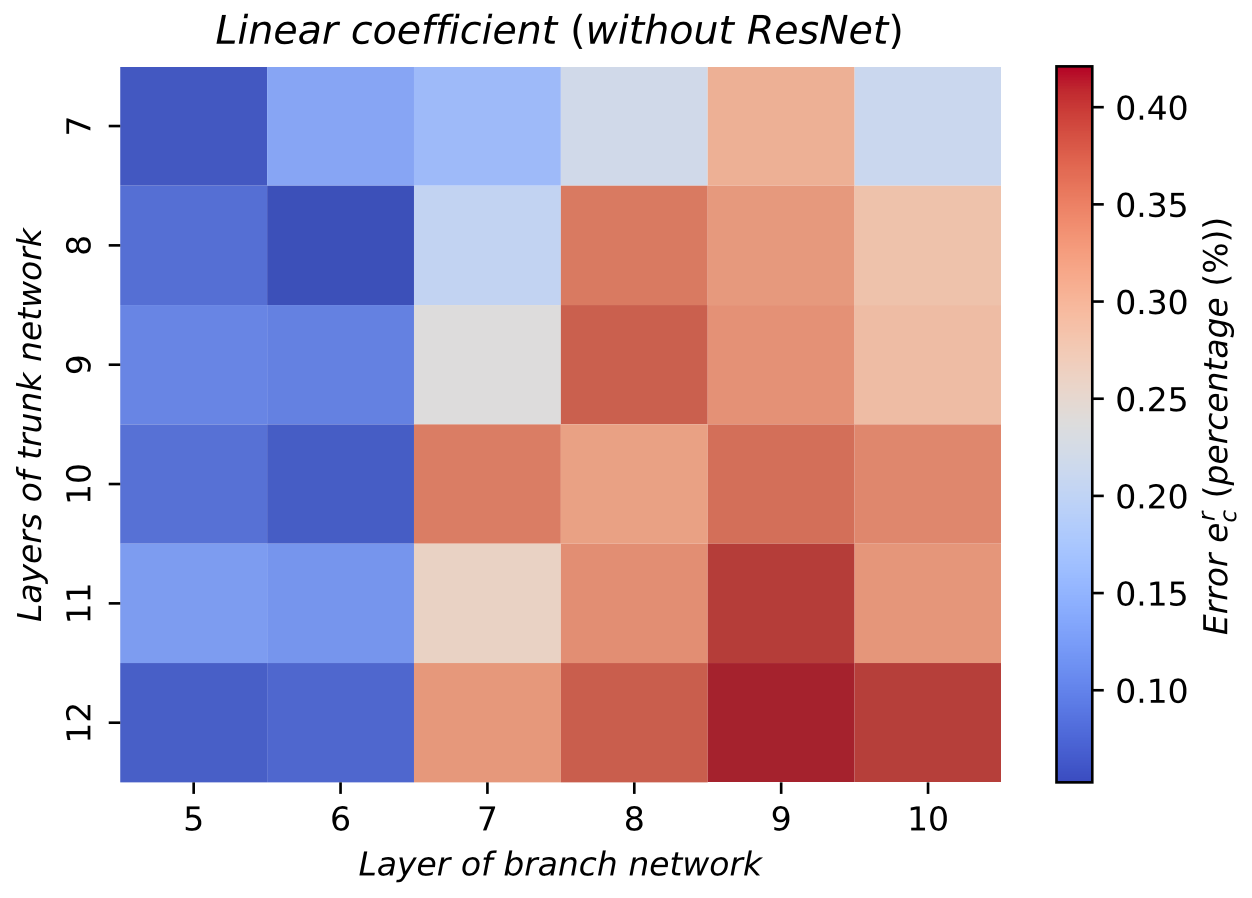}
\caption*{(a)}
\end{subfigure}
\begin{subfigure}[b]{0.45\textwidth}
\includegraphics[width=8cm,height=5.5cm]{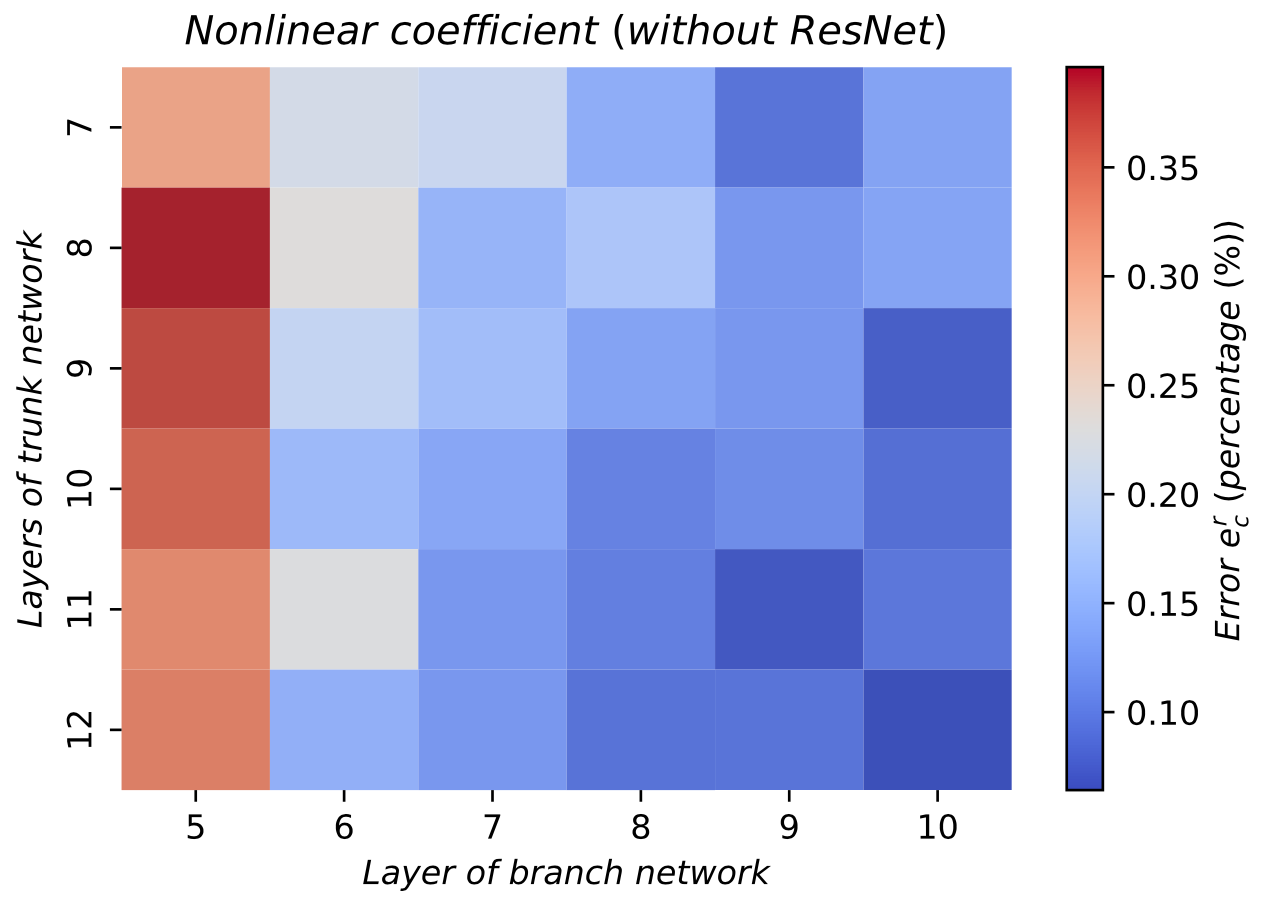}
\caption*{(b)}
\end{subfigure}\\
\begin{subfigure}[b]{0.45\textwidth}
\includegraphics[width=8cm,height=5.5cm]{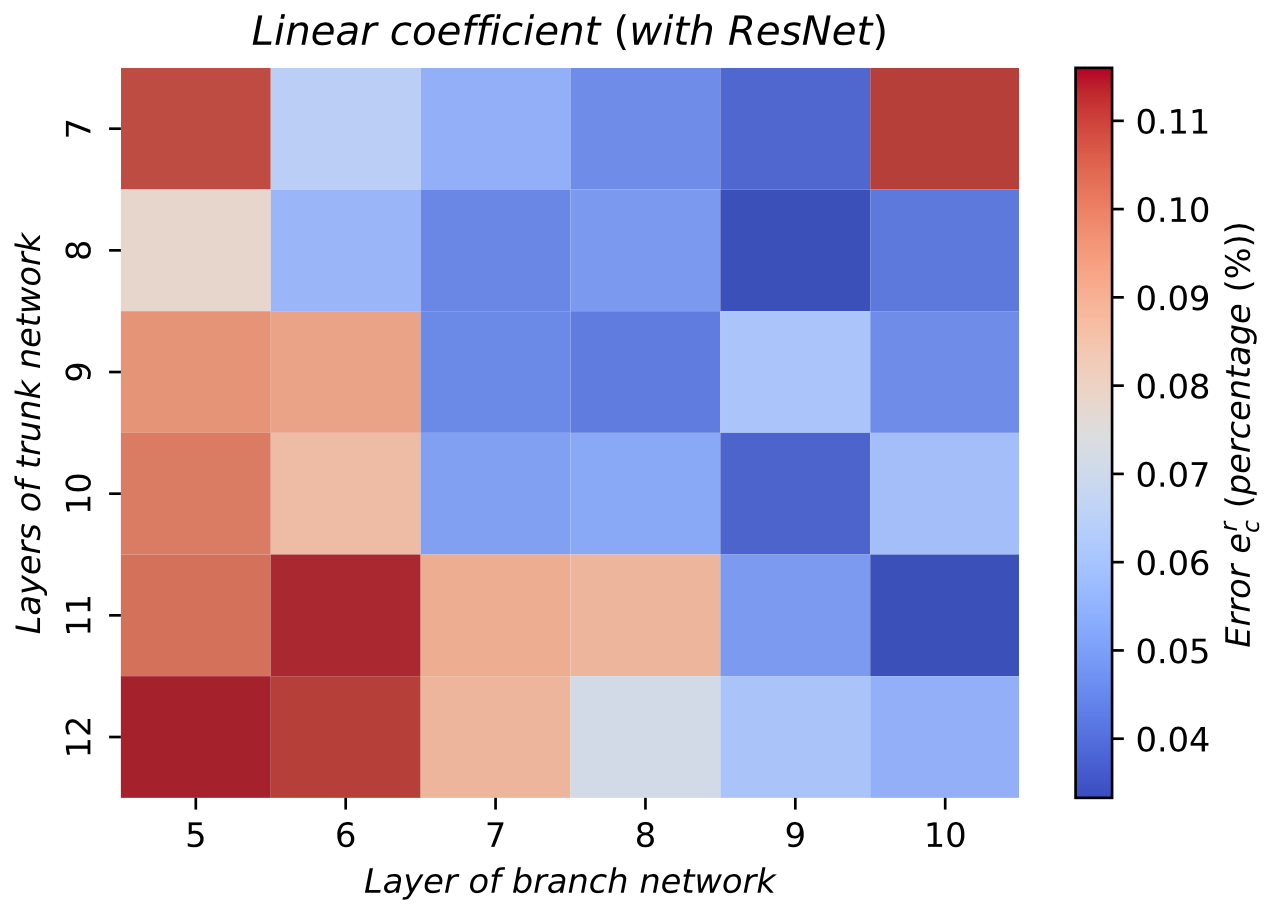}
\caption*{(c)}
\end{subfigure}
\begin{subfigure}[b]{0.45\textwidth}
\includegraphics[width=8cm,height=5.5cm]{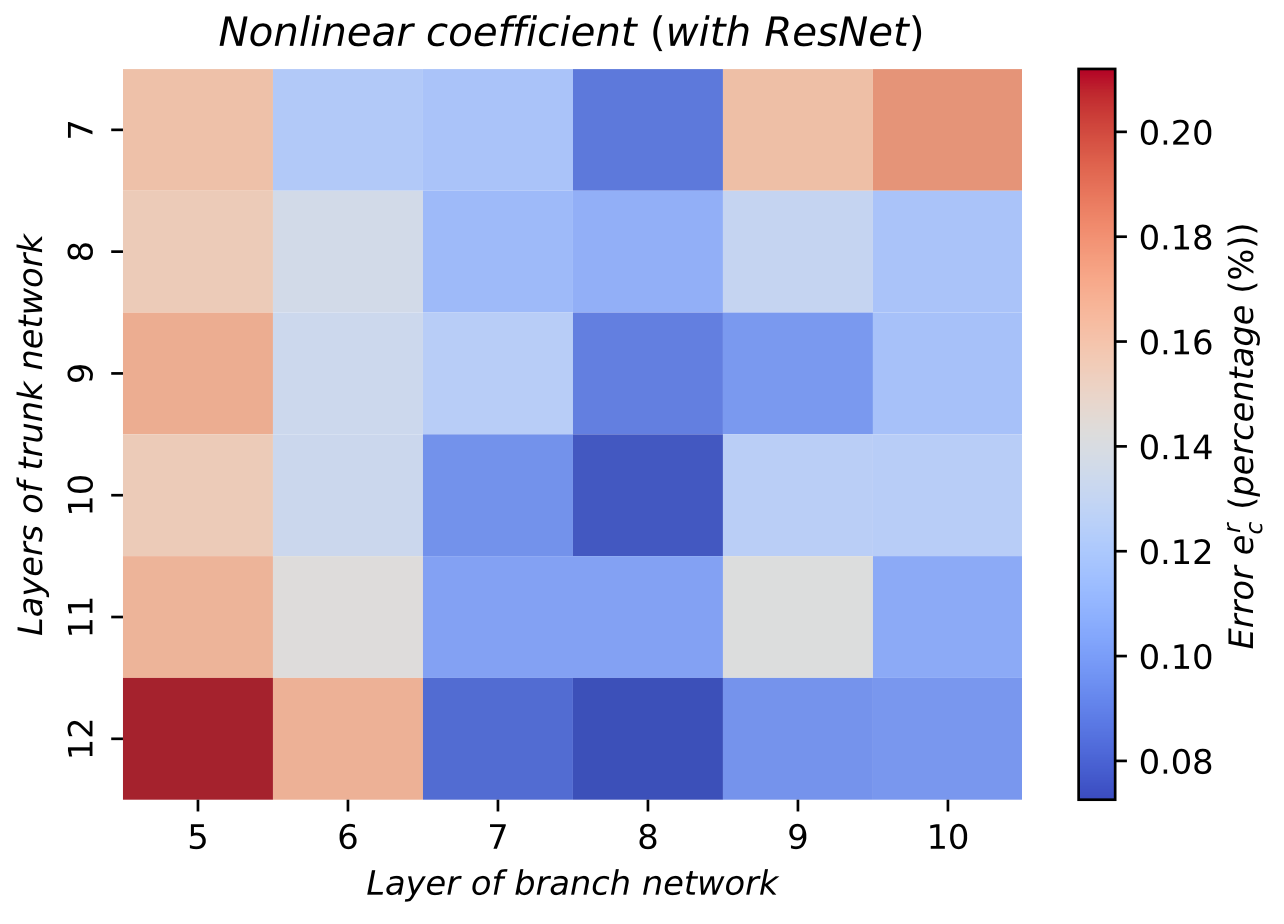}
\caption*{(d)}
\end{subfigure}\\
\caption{(Color online) The heat map of the coefficient $L^2$ error under different trunk network and branch network layers. This figure is based on the experiments of solution $u^{(vKdV)}_1$ (linear coefficient) and solution $u^{(vKdV)}_2$ (nonlinear coefficient) of the vKdV equation in Section \ref{KdV}. (a) and (b) do not use the ResNet structure. (c) and (d) use a dual ResNet structure. The data in the figure comes from repeated experiments under $5$ sets of random seeds, and more detailed settings and results are shown in Appendix \ref{A_linear_nonlinear}}.
\label{heat_map}
\end{figure}

Although in the above examples, some variable coefficients are linear functions of a certain spatial variable, from the perspective of our method, they are essentially the same as linear functions of a time variable. Overall, these examples tell us that linear variable coefficients are also important and cannot be underestimated. But in the numerical experiments of the inverse problem, we found that the increase in the number of network layers in the standard PINN does not seem to improve the accuracy of the linear variable coefficients, but is counterproductive. As shown in Fig. \ref{heat_map}.

Fig. \ref{heat_map} (a) and (b) very clearly display that as the number of layers representing the branch network of variable coefficients increases, the accuracy of linear variable coefficients gradually decreases. On the contrary, the accuracy of the non-linear variable coefficient gradually becomes higher (red represents low precision, blue represents high precision). The completely opposite behavior exhibited by the linear and nonlinear coefficients is very annoying. Because the linearity of the coefficients cannot be predicted in advance before the network is trained, it means that the choice of deep or shallow network is difficult.

\begin{figure}[htpb]
\centering
\begin{subfigure}[b]{0.45\textwidth}
\includegraphics[width=8cm,height=5.5cm]{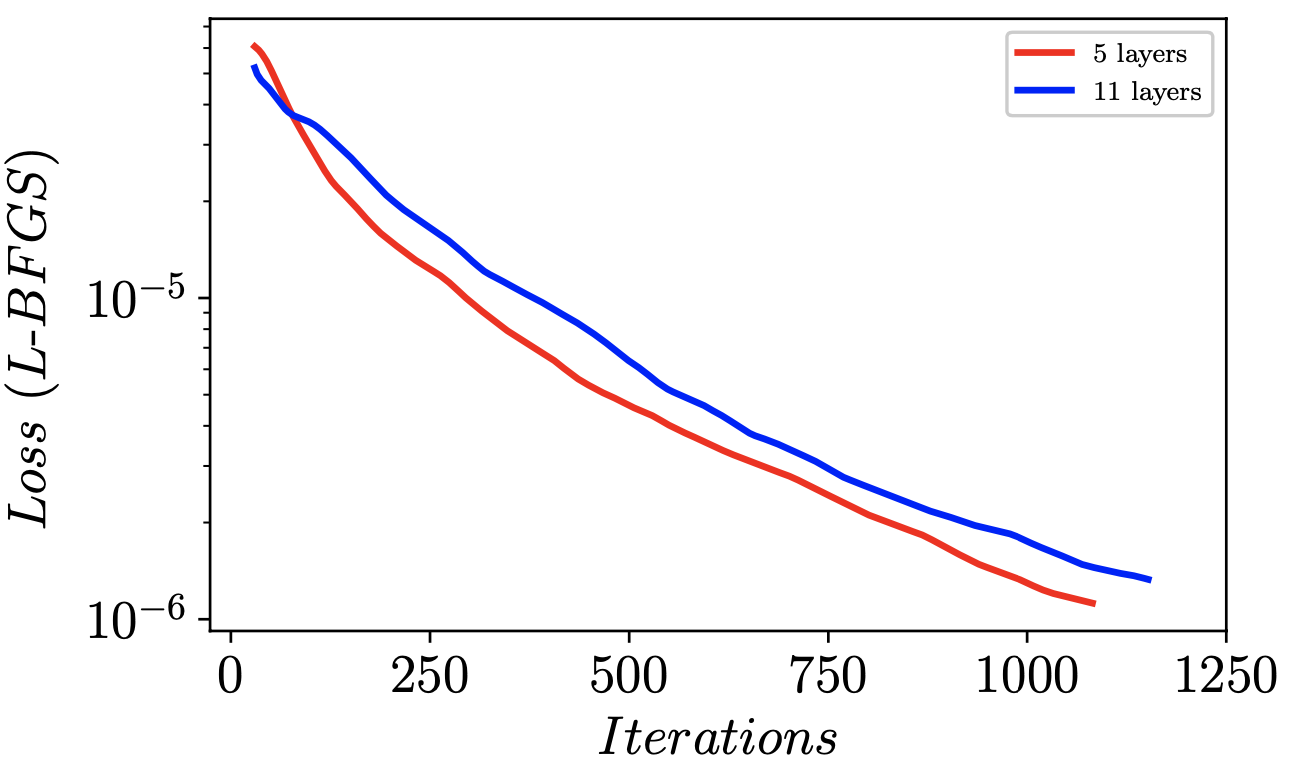}
\caption*{(a)}
\end{subfigure}
\begin{subfigure}[b]{0.45\textwidth}
\includegraphics[width=8cm,height=5.5cm]{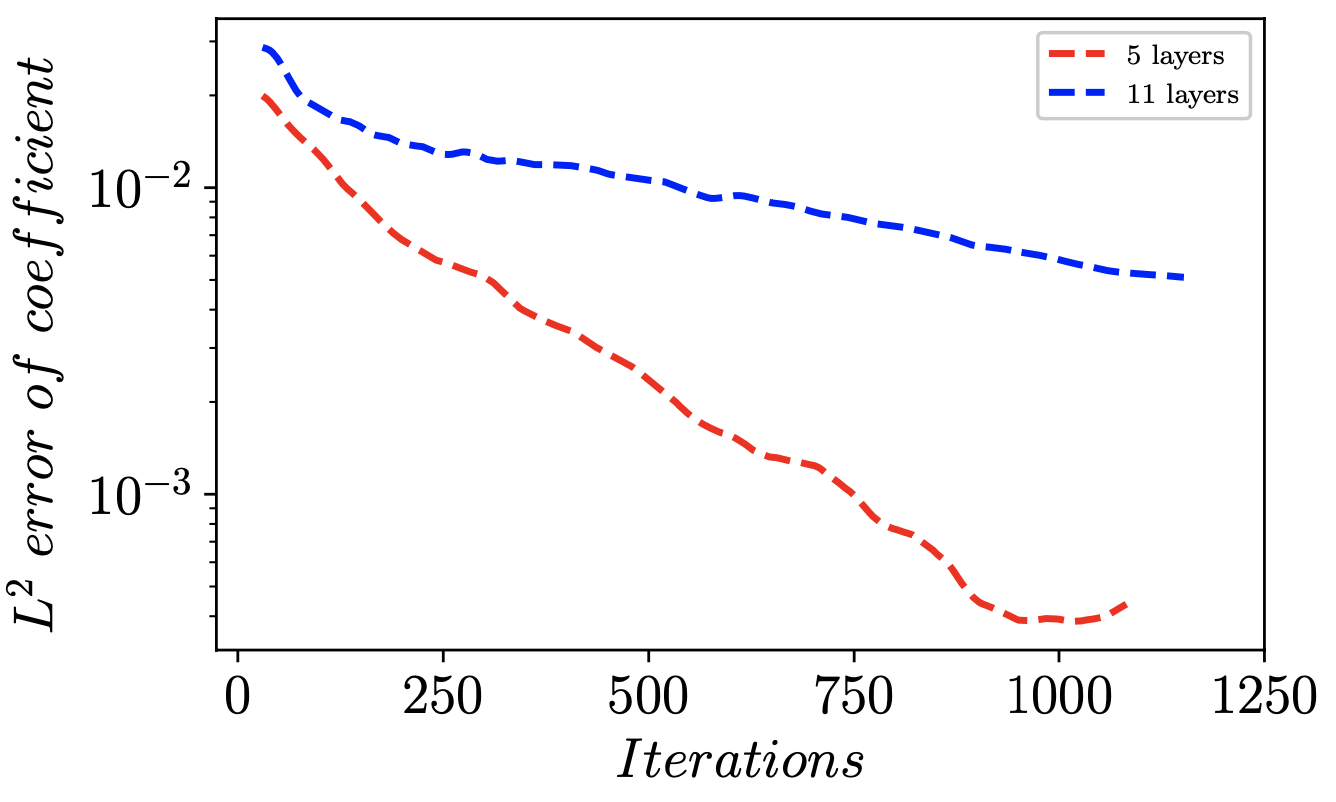}
\caption*{(b)}
\end{subfigure}\\
\caption{(Color online) The change curve of the loss and error of the deep network and the shallow network during the training process (only the $L$-$BFGS$ optimization part is intercepted, and the data is smoothed, but it does not affect the result). The data source is the same as Fig. \ref{heat_map}. (a) The curve of the change of the loss function. (b) Error curves for the coefficients.}
\label{loss_error}
\end{figure}
In addition, comparing the loss and error of the shallow network and the deep network (Fig. \ref{loss_error}), it can be shown that the abnormal results of the linear coefficients are not caused by overfitting. Because the loss and error of the deep network are higher than those of the shallow network. This phenomenon is more in line with the network degradation problem described in \cite{he2016identity, he2016deep}. Theoretically, a deep network should not perform worse (whether in terms of accuracy or loss) than a shallow network, since the former is the hyperspace of the latter. And a deeper network can be obtained by adding a network layer of identity mapping on the shallow network. The network degradation proposed in A shows the difficulty of approximating identity maps with multiple nonlinear layers, so that the disappointing performance of deep networks on linear coefficients can be understood here.

The structure of ResNet can effectively solve the degradation problem of the network. (c) and (d) in Fig. \ref{heat_map} are respectively the heat maps of the errors under the linear and nonlinear coefficients changing with the number of network layers after using the ResNet structure. (c) and (d) in Fig. \ref{heat_map} show that, no matter whether the coefficient is linear or not, as the number of branch network layers increases, the error tends to become smaller. In addition, it is worth noting that after comparing the scales of the color bars of the four images in Fig. \ref{heat_map}, it is easy to find that the ResNet structure has improved the accuracy to a certain extent while solving the degradation problem. Table \ref{ResNet_NoResNet} shows the comparison between using ResNet and not using ResNet.

\begin{table}[H]
\caption{Comparison of errors with and without ResNet}
\label{ResNet_NoResNet}
\begin{center}
\begin{tabular}{c||c|c|c}
\hline
\hline
 & No ResNet &  ResNet & Improve $(\%)$ \\
\hline
\hline
Linear & 2.40$\times 10^{-3}$ & 6.89$\times 10^{-4}$ & 71.36$\%$\\
Nonlinear  & 1.70$\times 10^{-3}$ & 1.25$\times 10^{-3}$ & 26.59$\%$\\
\hline
\hline
\end{tabular}
\end{center}
\end{table}

In addition, the following points need special explanation:
\begin{itemize}
	\item An obvious consequence is that, in the case of linear coefficients, using a linear activation function yields higher accuracy. But such a choice does not seem to be desirable in practical applications. On the one hand, it is impossible to choose a linear activation function in advance. We can only set up a special program for monitoring, or perform secondary training, but this undoubtedly increases the workload. On the other hand, in the case of multiple variable coefficients, when linear and nonlinear coefficients coexist, we can only make sacrifices and approximate them with different networks, because we cannot use both linear and nonlinear activation functions in the same network.
	\item Our analysis is mainly based on the inverse problem of the KdV equation. For the forward problem, because the variable coefficient is known in the discrete sense, the trunk network usually represents a nonlinear wave, so the degradation problem does not exist. In addition, in our experiments, a double ResNet structure is adopted (trunk network and branch network are both ResNet structures). Although ResNet seems to work mainly on the branch network, the results of the test show that the double ResNet structure unexpectedly has a better effect than the single ResNet structure.
\end{itemize}

In general, using the ResNet structure can improve accuracy on the basis of unifying linearity and nonlinearity, and it does not bring additional network parameters.

\subsection{The relationship between convexity of variable coefficients and learning}\label{sec_learning}
\quad

Understanding the learning process of neural networks is a huge challenge, but there are still some efforts in this area, such as deep neural networks usually fit the target function from low frequency to high frequency --``frequency principle" \cite{xu2019frequency}; in the early stage of network training, the input weights (including biases and weights) of hidden neurons will condense into isolated directions -- ``parameter condensation phenomenon" \cite{zhou2022towards}. However, in the context of variable coefficients, the learning of neural networks may lead to disappointing results or even outright failure when faced with more difficult problems (high dimensionality, coexistence of multiple variable coefficients, higher order derivatives). Although such a conclusion is regrettable, it seems reasonable in our cognition. In the numerical experiments of VC-PINN, we found that in addition to the above situations, some special variable coefficients will also make the learning of neural networks extremely difficult.

\begin{figure}[htpb]
\centering
\includegraphics[width=17cm,height=15.45cm]{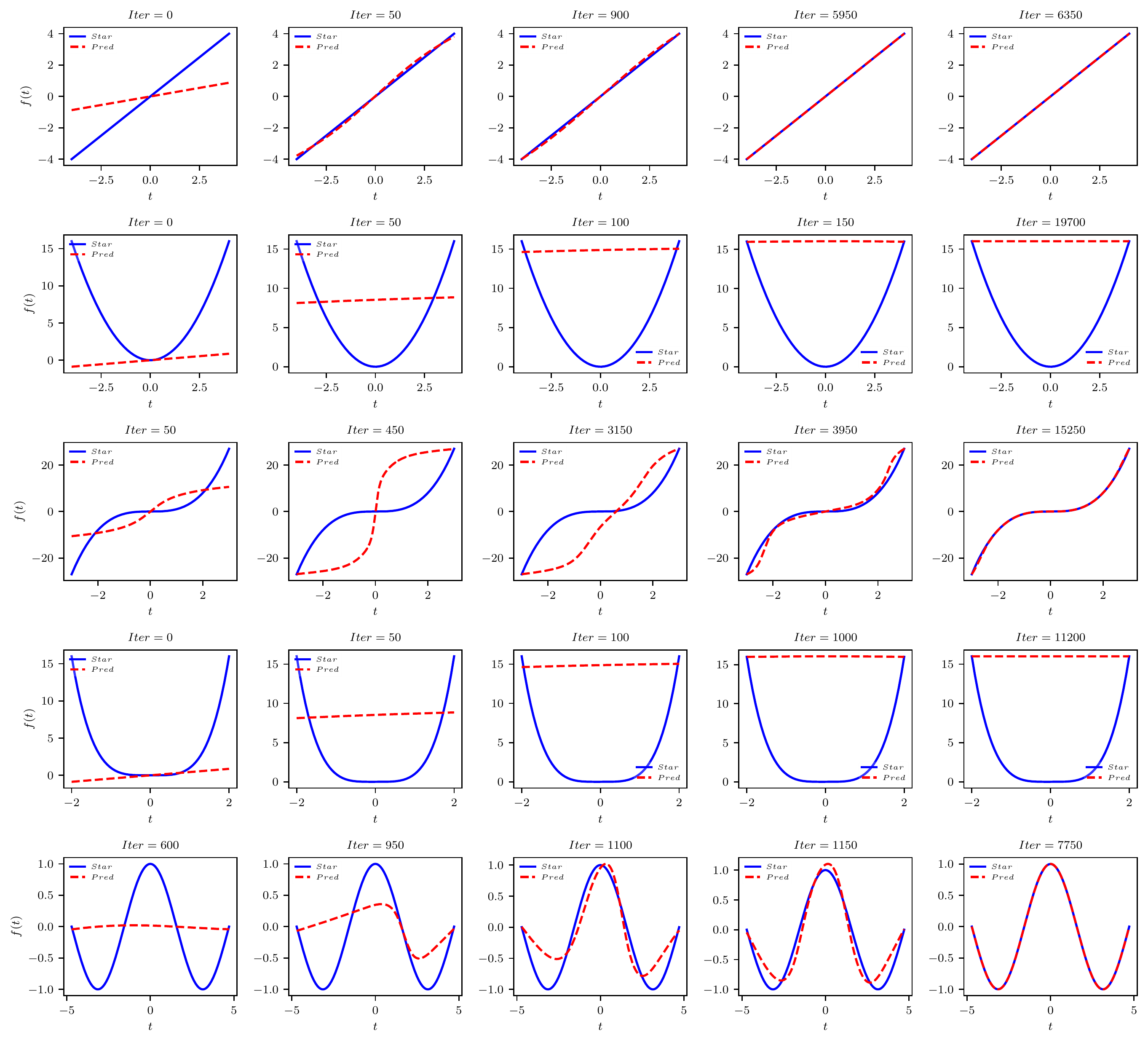}

\caption{(Color online) The learning process of the vKdV equation under five different variable coefficients. The coefficients from the first line to the fifth line are (1) $f(t)=t$; (2) $f(t)=t^2$; (3) $f(t)=t^3$; (4) $f(t)=t^4$; (5) $f(t)=\text{cos}(t)$.}
\label{NN_learn_1}
\end{figure}

In order to analyze the reasons for the failure of neural network learning, we consider observing the changes of the prediction results of variable coefficients during the training process. Fig. \ref{NN_learn_1} displays the learning process of the vKdV equation under five different variable coefficients (both success and failure in the example).

The learning process of the five examples in Fig. \ref{NN_learn_1} has a common law, that is, the neural network first learns the two endpoints of the variable coefficients, and then gradually learns the middle area. The soft constraints on the bounds of the variable coefficients we add to the loss function can reasonably explain this learning behavior. Furthermore, learning under quadratic and quartic polynomials fails, and the neural network stagnates after pre-learning (learning for bound constraints on coefficients in the loss function). Large gradients do not seem to be a plausible explanation for learning failures, since learning with cubic polynomial coefficients is completely successful. This seems to imply that the learning of neural networks is hindered in the face of strongly convex targets.

\begin{figure}[htpb]
\centering
\includegraphics[width=17cm,height=18.54cm]{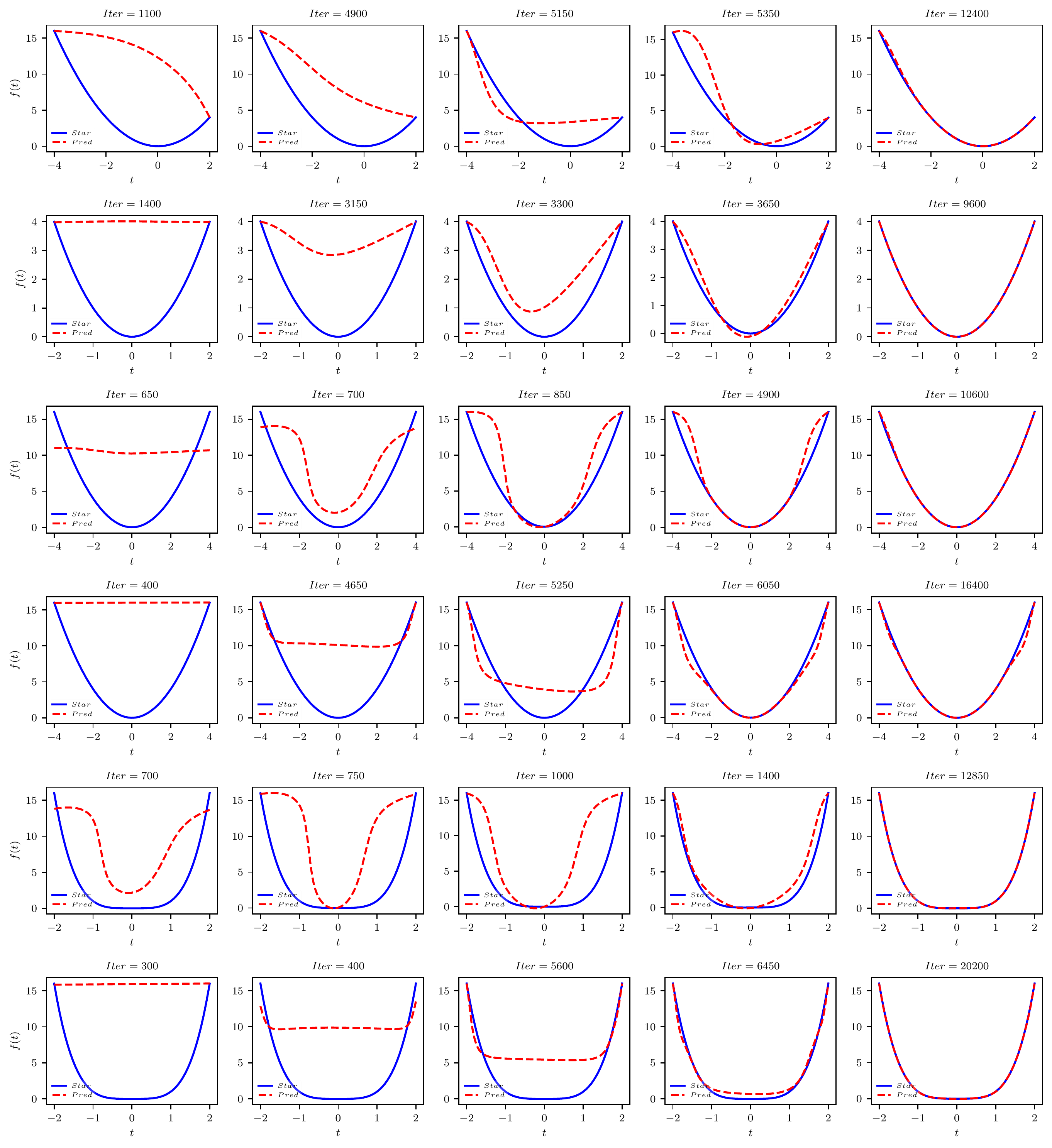}

\caption{(Color online) The learning process of the vKdV equation under quadratic and quartic polynomials. The middle coefficients from the first row to the sixth row are (1)-(4) $f(t)=t^2$; (5)-(6) $f(t)=t^4$. In (3) and (5), $Loss_c$ adds intermediate point information ($(0,0)$). (4) and (6) add the first derivative information of the boundary (condition \eqref{hdc}).}
\label{NN_learn_2}
\end{figure}

Curvature is a quantitative description of the convexity of the curve, and the curvature formula of the curve $f(t)$ is

\begin{equation}
	K|_{t=t_0} = \frac{|f^{''}|}{(1+{f^{'}}^{2})^{\frac{3}{2}}}\bigg |_{t=t_0}
\end{equation}
where $K|_{t=t_0}$ is the curvature of curve $f(t)$ at point $t=t_0$. According to the curvature formula, the curvature of the first-degree polynomial coefficient in Fig. \ref{NN_learn_1} is 0, and the curvature of the third-degree polynomial coefficient changes sign at point $t=0$ (meaning that the concave-convexity changes). The curvature of the quadratic and quartic polynomials at $t=0$ is the largest, respectively $K|_{t=0}=2$ and $K|_{t=0}=12$, and their curvature does not change the sign in the interval (concave-convexity does not change). This illustrates that the conclusion that convexity hinders learning seems reasonable.

In order to further explore the reasons for the learning failure of the neural network, we make different adjustment strategies in the previous examples of quadratic and quartic polynomial coefficients. The details are as follows: (1) Reduce the time interval on one side (from $[-4,4]$ to $[-4,2]$); (2) Reduce the time interval on both sides (from $[-4,4]$ to $[-2,2]$); (3) and (5) $Loss_c$ adds intermediate point information ($(0,0)$); (4) and (6) add the first-order derivative information of the boundary of the coefficient (condition \eqref{hdc}).

Both (1) and (2) in Fig. \ref{NN_learn_2} do not change the curve, but only shorten the time range, and $t=0$ with the largest curvature is still included in the interval. But such an adjustment enables the coefficients to be successfully learned, which shows that what affects the learning is the convexity of the entire interval rather than the convexity at a certain point. The accumulation of convexity on the entire interval makes the coefficient curve after pre-learning far from the real coefficient curve, which may lead to learning failure.

The two adjustment strategies involved in (3)-(6) in Fig. \ref{NN_learn_2} provide variable coefficient information, so that the pre-learning curve can be closer to the real coefficient curve, thereby avoiding the problems caused by the accumulation of convexity. The information provided by the two strategies is only a small amount. For example, the strategy at the middle point only adds one point of information to make learning from failure to success. Of course, this strategy is not mathematically tenable, and it is more suitable for the more common intermediate state problems in industrial applications described in Section \ref{forward_inverse_unity} later. The reason we mention such a strategy here is to show that the gap between the pre-learning curve and the true coefficient curve is important and seems to play a decisive role in the success or failure of learning. Learning with quadratic and quartic polynomial coefficients was successful with all tuning strategies. (Although the learning in Fig. \ref{NN_learn_2} (4) is still a little flawed)

By analyzing the learning process under different coefficients and several adjustment strategies, the following empirical conclusions about learning can be drawn:
\begin{itemize}
	\item The convexity of the variable coefficients hinders the learning of the neural network, and the convexity here refers to the cumulative effect on the entire interval. The cumulative effect of convexity makes the pre-learning curve have a large gap with the real coefficient curve, which leads to the failure of learning.
	\item Coefficients with concave-convex changes such as cubic polynomials and cosine functions are relatively easier to learn because their pre-learning curves are distributed on both sides of the real coefficient curve, rather than on one side (such as quadratic and quartic polynomial coefficients).
	\item Strategies such as narrowing the range, increasing internal data points, and adding boundary derivative information can promote the learning of neural networks by bringing the pre-training curve and the real coefficient curve closer together.
\end{itemize}

Although the discussion of the learning process of neural networks is not theoretical enough, it can explain the inseparable relationship between convexity and learning. I believe that a more in-depth discussion will become our future work. Finally, the detailed model setup and results of the numerical experiments in this section are presented in Appendix \ref{A_Learning}.

\subsection{Anti-noise analysis}\label{sec_noise}
\quad

Data resources are generally expensive, and obtaining clean (noise-free) data is impossible in the real world because errors in measurements are always inevitable. Therefore, the proposed method must have a certain anti-noise ability before it can be really applied to practical mathematical physics problems. The main purpose of this section is to test the anti-noise ability of VC-PINN. We still take the vKdV equation in Section \ref{KdV} as an example for testing. In the experiment, the $\alpha\%$ noise added to the clean data is a Gaussian distribution with zero mean, and its standard deviation is determined by the standard deviation of the training data and $\alpha$. We used $8$ sets of data with noise ranging from $0\%$ to $5\%$ to conduct experiments and tested the forward and inverse problems (double ResNet structure) using the VC-PINN framework under linear and nonlinear coefficients.

\begin{figure}[htpb]
\centering
\begin{subfigure}[b]{0.45\textwidth}
\includegraphics[width=8cm,height=5.5cm]{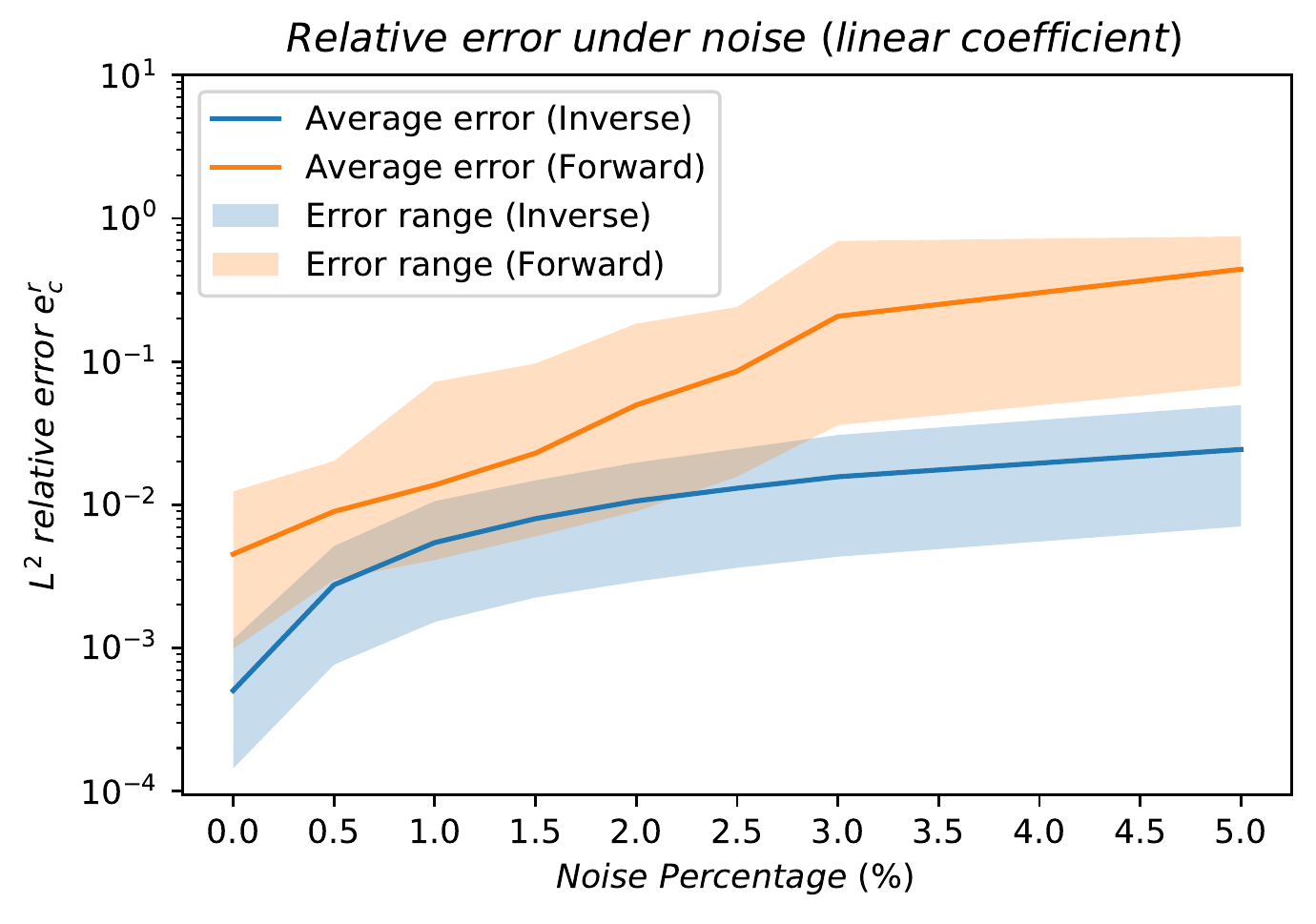}
\caption*{(a)}
\end{subfigure}
\begin{subfigure}[b]{0.45\textwidth}
\includegraphics[width=8cm,height=5.5cm]{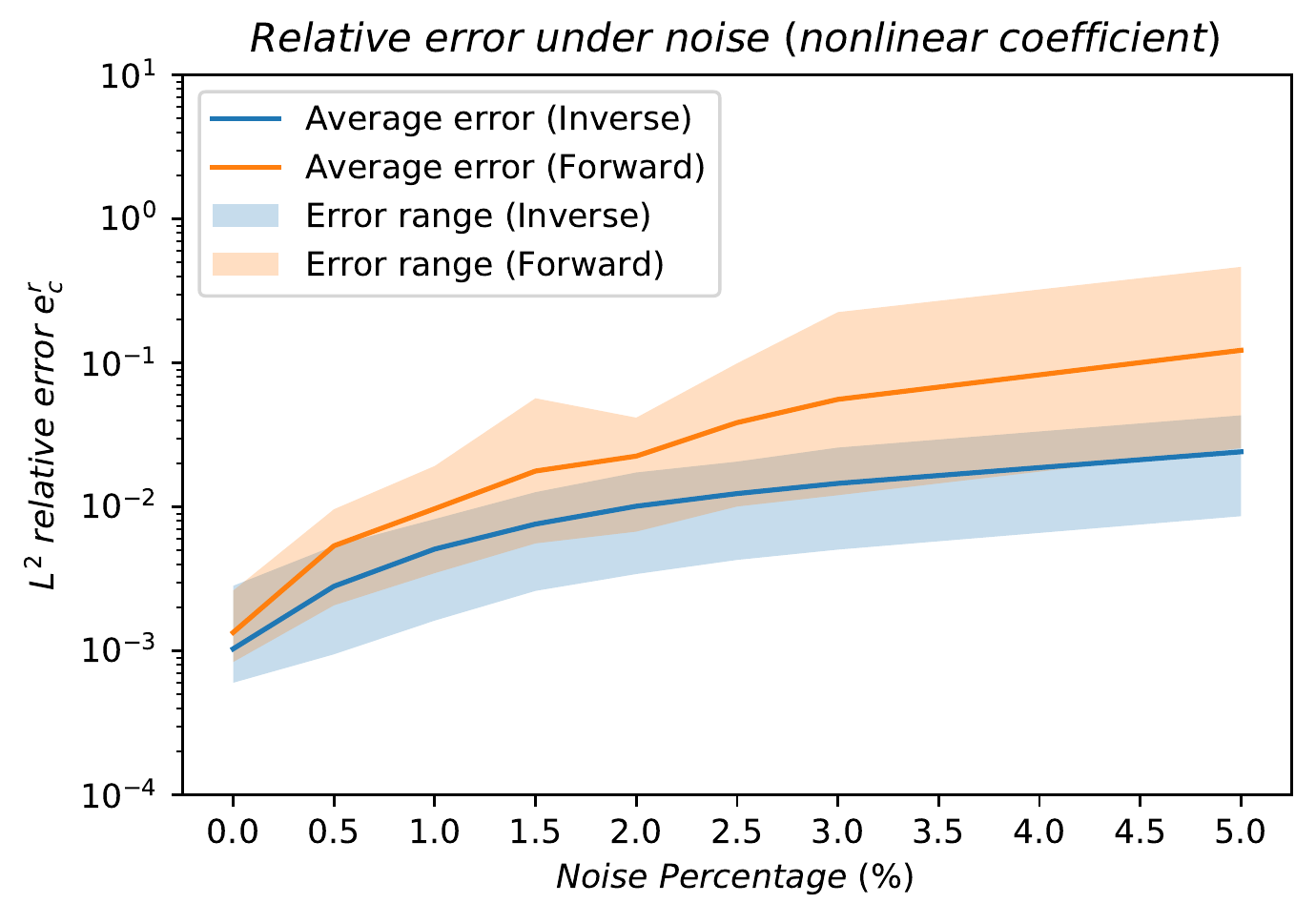}
\caption*{(b)}
\end{subfigure}\\
\caption{(Color online) Under different proportions of noise, the performance of VC-PINN with double ResNet structure (including forward and inverse problems). The solid line represents the mean error, and the semi-transparent area represents the error range of repeated experiments. (a) Error comparison under linear coefficients. (b) Error comparison under nonlinear coefficients.}
\label{Noise}
\end{figure}

The results in Fig. \ref{Noise} display that as the signal-to-noise ratio decreases, the relative error of the results increases. In addition, it can be found that the anti-noise ability of the inverse problem is significantly better than that of the forward problem. Regardless of whether it is a linear or nonlinear coefficient, even at a noise level of $5\%$, the relative error can still be maintained above the $10^{-2}$ level. The forward problem has not reached this error level under the noise percentage of $1.5\%$-$2\%$. And for the forward problem, the influence of noise on the linear coefficient seems to be greater than that of the nonlinear coefficient.

In general, the proposed method has a certain ability to resist noise, but the performance on the forward problem needs to be improved. Finally, it should be noted that the above-mentioned anti-noise test is carried out under the repeated experiment of selecting 10 groups of random seeds. More detailed model settings and experimental results are presented in Appendix \ref{A_noise}.

\subsection{The unity of forward and inverse problems/relationship with standard PINN}
\quad

This section mainly discusses our proposed VC-PINN from a framework perspective, including the unity of forward and inverse problems and the relationship with standard PINN.

\subsubsection{The unity of forward and inverse problems}\label{forward_inverse_unity}
\quad

In numerical experiments testing the performance of VC-PINN, we find that the proposed framework is uniform in solving forward and inverse problems with variable coefficients. And this kind of unity is not possessed by the constant coefficient equation.

Analyze from the perspective of the loss function. First, reviewing the method introduction in Section \ref{Forward_pro} and Section \ref{Inverse_pro}, the loss functions under the forward problem and the inverse problem are 
\begin{align}
	 &Forward:\ \ \ \  Loss(\theta) = Loss_{I}(\theta)+Loss_b(\theta)+Loss_f(\theta)+Loss_c(\theta),\\
	 &Inverse:\quad \ \ Loss(\theta) = Loss_{s}(\theta)+Loss_f(\theta)+Loss_c(\theta).
\end{align}

For the initial value loss $Loss_I(\theta)$ and boundary loss $Loss_b(\theta)$ in the forward problem, if we ignore their actual mathematical meaning and only consider them from the perspective of numerical calculation, they can be combined into a loss function, namely
\begin{align}
	Loss_I(\theta)+Loss_b(\theta)=Loss_{Ib}(\theta),
\end{align}
where the involved $I$-$type$ points and $b$-$type$ points are also merged, and mark them with $Ib$-$type$ points. Thus, both $Loss_{Ib}$ in the forward problem and $Loss_{s}$ in the inverse problem represent the mean square error loss between the network output and the true solution $u$. Therefore, the loss functions of the forward problem and the inverse problem under the variable coefficient version are completely consistent in form. They differ only in the distribution of points involved in $Loss_{Ib}$/$Loss_{s}$ and $Loss_c$ ($Loss_f$ is exactly the same in both forward and inverse problems). In the forward problem, the $Ib$-$type$ points involved in $Loss_{Ib}$ are distributed in the initial boundary value area of $u$, and the $c$-$type$ points involved in $Loss_c$ are distributed on the entire interval $[T_0,T_1]$. For the inverse problem, the $s$-$type$ points involved in $Loss_s$ are distributed in the entire area of $u$, and the $c$-$type$ points involved in $Loss_c$ are distributed on the two endpoints of the interval $[T_0,T_1]$. A general summary is that the inverse problem provides more information about $u$ but less information about the coefficient $c_1$ than the forward problem. Therefore, from the data level, the forward problem and the inverse problem under the variable coefficient problem are unified. And there is no essential difference in the implementation of the code.

However, such unity does not exist in the constant coefficient equations that PINN usually deals with. Because the coefficients at this time are a set of fixed constants rather than functions. For constants, there are only two states: fully known and completely unknown, so there is not even a loss item about the mean square error of the coefficient in the standard PINN, let alone the unity of the forward and inverse problems.

\begin{figure}[htpb]
\centering
\includegraphics[width=12cm,height=5.38cm]{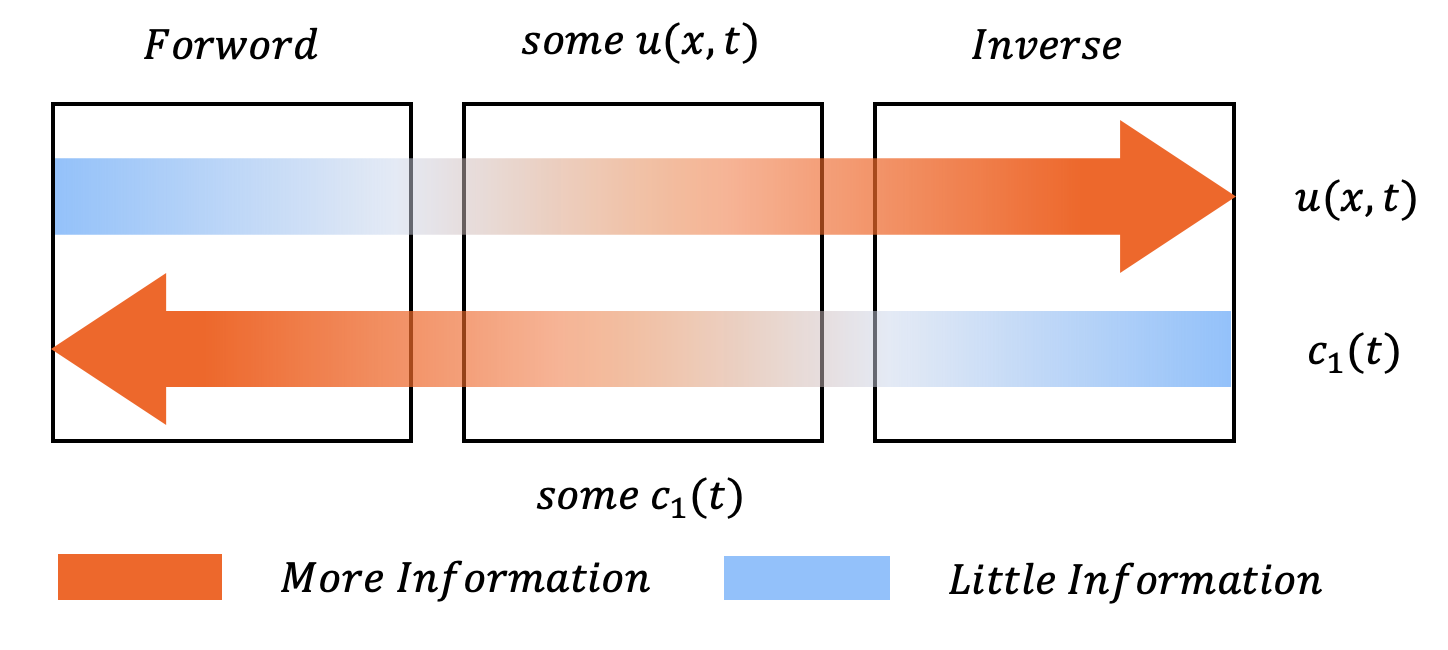}

\caption{(Color online) The relationship between the type of problem and the information content of $u(x,t)$ and $c_1(t)$.}
\label{unity}
\end{figure}

Referring to the graphical representation in \cite{karniadakis2021physics}, Fig. \ref{unity} shows the relationship between the type of problem and the information content of $u(x,t)$ and $c_1(t)$. Finally, it is worth mentioning that if we do not discuss the strict definition of forward or inverse problems in the mathematical sense, the scenario represented by the middle area in Fig. \ref{unity} may be the problem faced in industrial applications. That is, there are some information about $u$ and some information about the variable coefficient $c_1$, which is an intermediate state between the forward problem and the inverse problem.

\subsubsection{Relationship with standard PINN}\label{sec_relation}
\quad

Before analyzing the relationship between VC-PINN and standard PINN, we first discuss the example where the independent variable of the coefficient is two-dimensional.

If the variable coefficient in equation \eqref{vceq1} is also related to $x$, that is, ${\bm C}[t]$ is the function of $x$ and $t$. In this case, we test the performance of VC-PINN. Consider the vSG equation discussed in Section \ref{vSG_forward} and rewrite it as
\begin{equation}\label{2DSG}
	V_{xt}=M(x,t)\text{sin}(V),
\end{equation}
Where $V=V(x,t)$, and $M(x,t)$ is a variable coefficient. The solution of equation \eqref{2DSG} is given in \cite{yang2014analytical} using the self-similar method. Specifically, suppose the solution of equation \eqref{2DSG} satisfies
\begin{equation}
	V(x,t)=u(X,T),
\end{equation}
where $X=X(x)$ and $T=T(t)$ are two coordinate transformations respectively. If you consider some special coefficients such as $M(x,t)=\frac{dX}{dx}\frac{dT}{dt}$. Then there is 
\begin{equation}
	u_{X,T}=\text{sin}(u),
\end{equation}
which means that $u(X,T)$ satisfies the SG equation with constant coefficient, and the solution of equation \eqref{2DSG} can be derived naturally by using the solution of SG equation with constant coefficient, as follows:
\begin{equation}
	V(x,t)=4\text{arctan}\left(\lambda e^{aX(x)+\frac{T(t)}{a}} \right)
\end{equation}
where $a\neq 0$ and $\lambda$ are free parameters. In \cite{yang2014analytical}, the variable coefficients are taken as Chebyshev polynomials and a series of exact solutions are obtained, but here we only consider two simple cases and use them to test VC-PINN on the inverse problem when the coefficients are two-dimensional performance. Specifically ($\lambda = a=1$)
\begin{itemize}
	\item \textbf{Case 1:} Select $X(x)=\frac{x^2}{2}$, $T(t)=\frac{t^2}{2}$, then the corresponding $M(x,t)=xt$.
	\item \textbf{Case 2:} Select $X(x)=\text{sin}(2x)$, $T(t)=\text{sin}(2t)$, then the corresponding $M(x,t)=4\text{cos}(2x)\text{cos}(2t)$.
\end{itemize}

\begin{figure}[htpb]
\centering
\begin{subfigure}[b]{0.90\textwidth}
\includegraphics[width=16cm,height=5.5cm]{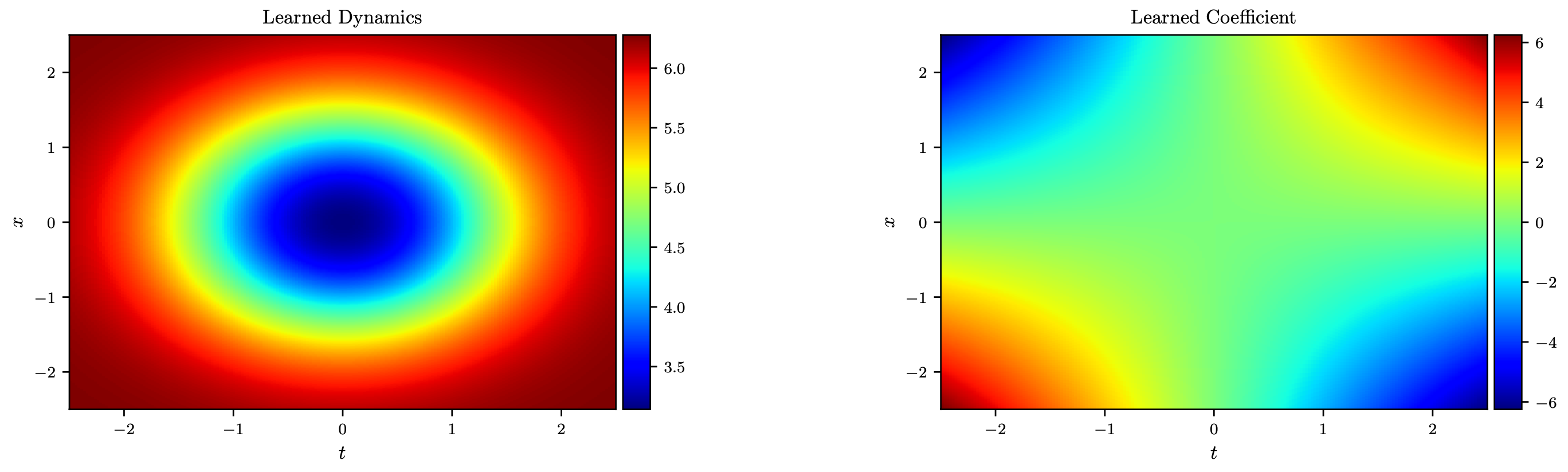}
\caption*{(a)}
\end{subfigure}\\

\begin{subfigure}[b]{0.90\textwidth}
\includegraphics[width=16cm,height=5.5cm]{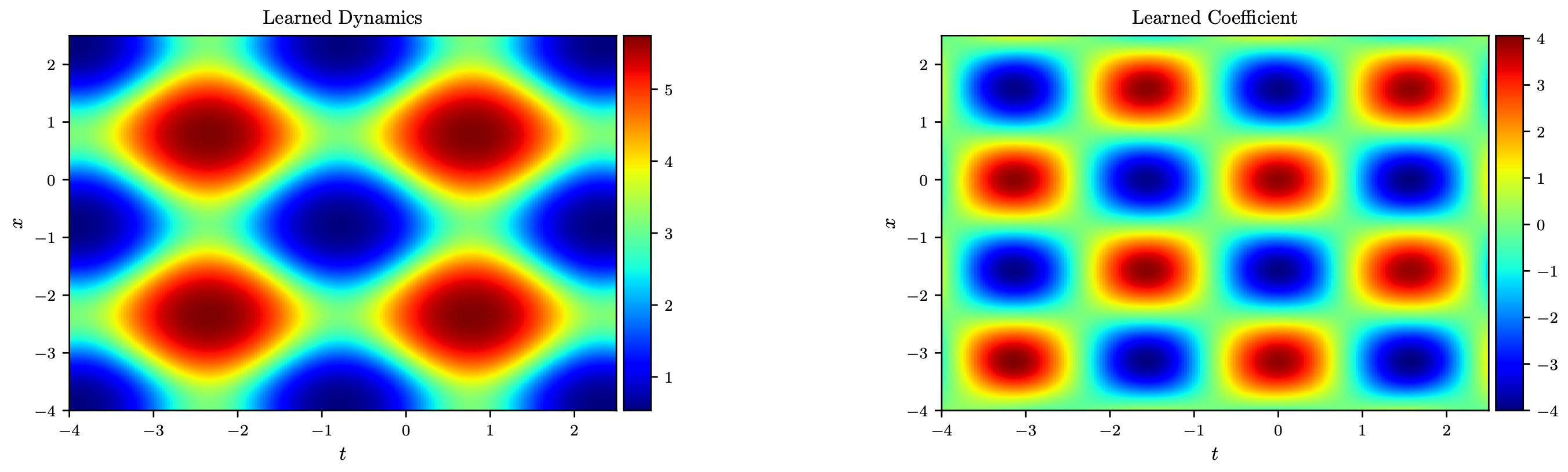}
\caption*{(b)}
\end{subfigure}
\caption{Prediction results of $V(x,t)$ and coefficient $M(x,t)$ under Case 1 and Case 2. (The first line is Case 1, and the second line is Case 2)}
\label{2DSG_fig}
\end{figure}

Fig. \ref{2DSG_fig} shows the dynamic behavior of $V(x,t)$ and coefficient $M(x,t)$ predicted by the network in these two cases, respectively. The $L^2$ relative errors $e^{r}_c$ of the coefficients in the two cases are $6.86\times 10^{-3}$ and $2.32\times 10^{-2}$, respectively. Therefore, the VC-PINN method is also applicable to the case of high-dimensional coefficients. The specific settings are shown in Appendix \ref{A_2D_SG}.

Next, back to the topic of this section. In addition to the ResNet structure, the difference between VC-PINN and the standard PINN is the addition of a branch network responsible for approximating variable coefficients. But for the above situation where the coefficient dimension and the equation dimension are equal, it is obviously feasible to use a standard PINN with only one network. It is only necessary to make the output of the network two-dimensional, representing $V(x,t)$ and $M(x,t)$ respectively. Therefore, the relationship between VC-PINN and standard PINN can be summarized as follows:
\begin{itemize}
	\item The constant coefficient can be regarded as the dimensionality reduction of the variable coefficient, which is what the standard PINN is good at. Of course, VC-PINN can also handle constant coefficients well, such as the inverse problem of Case 1 in Section \ref{gKP_inverse_1}.
	\item When encountering the situation where the variable coefficient dimension is inconsistent with the equation dimension (such as $u(x,t)$ and $c_1(t)$), VC-PINN can show its talents. Because the added branch network is equivalent to adding a hard constraint to the coefficient $c_1(t)$, which ensures that the coefficients at different spatial positions are equal at the same time. But it is difficult for the standard PINN to handle such situations, unless the unsatisfactory strategy of soft constraints is used.
	\item If the dimension of the variable coefficient continues to increase until it is equal to the dimension of the equation (such as $u(x,t)$ and $c_1(x,t)$), then VC-PINN is not the only choice. Then the problem to be solved can also be regarded as a coupled system composed of $u(x,t)$ and $c_1(x,t)$, and it is reasonable to use a standard PINN. It's just that the use of ResNet in standard PINN is not as free as in VC-PINN (trunk network and branch network can use different ResNets).
\end{itemize}

In general, VC-PINN is closely related to standard PINN, but the proposal of VC-PINN is necessary, and it can make up for the deficiency of PINN when the dimension of the variable coefficient and the dimension of the equation are different.

\section{Conclusion}\label{conclusion}

Compared with the constant coefficient model, the variable coefficient model is a more realistic description of nature, because it can describe phenomena such as inhomogeneous media, non-constant physical quantities, and variable external forces. However, in the context of variable coefficients, it is difficult for the standard PINN to handle the case where the independent variable dimension of the coefficient function is different from the equation dimension. In view of the importance of the variable coefficient model, this paper proposes a deep learning method -- VC-PINN, which specifically deals with the forward and inverse problem of variable coefficient PDEs. It adds a branch network responsible for approximating variable coefficients on the basis of standard PINN, which adds hard constraints to variable coefficients and avoids the problem of different dimensions between coefficients and equations. In addition, the ResNet structure without additional parameters is introduced into VC-PINN, which unifies linear and nonlinear coefficients while alleviating gradient vanishing.

In this paper, the proposed VC-PINN is applied to four equations including vSG, gvKP, vKdV, and gvSK. The exact solutions of these four integrable variable coefficient equations can be obtained by generalizing the classical integrable method. This provides exact samples (rather than high-precision numerical samples) for testing the performance of VC-PINN. VC-PINN has achieved success in forward and inverse problems with different forms of variable coefficients (polynomials, trigonometric functions, fractions, oscillation attenuation coefficients), high dimensions, and coexistence of multiple variable coefficients. It learns the dynamic behavior of the solution $u$ as well as the full variation of the variable coefficients with satisfactory accuracy. In the numerical experiments, we also draw some empirical conclusions: 1. In the forward problem, the large gradient and high wave height of the solution $u$ has a strong correlation with the error; 2. In the inverse problem, the error mainly comes from the region with large coefficient and the region with drastic change of coefficient; 3. In the inverse problem, the high-frequency oscillation of the coefficient will also cause the high-frequency oscillation of the error.

In the analysis and discussion, we conducted an in-depth analysis of VC-PINN in a combination of theory and numerical experiments, including four aspects: the necessity of ResNet; the relationship between the convexity of variable coefficients and learning; anti-noise analysis; the unity of forward and inverse problems/relationship with standard PINN.

When discussing the necessity of ResNet, we derived the backpropagation formula of the gradient in VC-PINN with ResNet structure, explaining how shortcut connections can alleviate gradient vanishing. In addition, we found that in the case of linear coefficients, the accuracy of ordinary FNN decreases with the increase of network depth (network degradation problem), which is completely opposite to the phenomenon in the case of nonlinear coefficients. Comparative experiments tell us that the ResNet structure reverses this phenomenon, thus unifying the linear and nonlinear coefficients. When analyzing the learning process, we found that the cumulative convexity of variable coefficients can seriously hinder the learning of neural networks. We also found that strategies that can make the pre-learning curve and the real coefficient curve closer have a certain promotion effect on the learning of the neural network. Anti-noise experiments show that the proposed method has a certain anti-noise ability, but the performance on the forward problem needs to be improved. Then, from the perspective of the loss function and data level, we found that the forward and inverse problems of VC-PINN under the background of variable coefficients are unified, and the difference is only in the data volume of $u$ and $c_1$. Finally, the close relationship between VC-PINN and standard PINN is explored through a two-dimensional variable coefficient example (vSG).

In general, the proposal of VC-PINN is very necessary, and it fills the gap of PINN in variable coefficient problems. Of course, the examples involved in the numerical experiments in this article are relatively standard models and simple cases, and the performance of VC-PINN on more complex coefficients and more complex equations (coupling equations, equations on the complex field) is expected. The capabilities of VC-PINN demonstrated in this paper are just the tip of the iceberg, and some transferable modular technologies can be armed to VC-PINN in the future. In addition, it is very meaningful to use neural operator networks to learn the mapping between solutions and variable coefficients, which will be our future work.

\section*{Acknowledgements}
The project is supported by the National Natural Science Foundation of China (No. 12175069 and No. 12235007), Science and Technology Commission of Shanghai Municipality (No. 21JC1402500 and No. 22DZ2229014), and Natural Science Foundation of Shanghai (No. 23ZR1418100).

\appendix
\section{Symbol Description}
\quad

The following table gives a description of the symbols involved in the appendix.

\begin{table}[H]
\caption{Symbol description}
\begin{center}
\begin{tabular}{c||c}
\hline
\hline
 & Description\\
\hline
\hline
$Seed$ & A setting that implements pseudo-randomness in code.\\
$N_{Adam}$ & The number of iterations of the Adam optimizer.\\
$\Omega$ & Spatial domain.\\
$n_x(n_y/n_t)$ & The division size in the $x(y/t)$-axis direction when meshing.\\
$nd_c$ & The division size of the standard coefficients.\\
$N_d^u/N_d^c$ & The number of hidden layer nodes of the trunk/branch network.\\
$N_B^u/N_B^c$ & The number of residual blocks contained in the trunk/ branch network.\\
$N_h^u/N_h^c$ & The number of hidden layers contained in each residual block of the trunk/branch network.\\
$n_I+n_b$ & The number of initial boundary value condition data points.\\
$n_s$ & The number of internal data points of $u$.\\
$n_c$ & The number of variable coefficients data points.\\
$n_f$ & The number of interior collocation points.\\
$e^r_u$ & The $L^2$ relative error of $u$.\\
$e^r_c$ & The $L^2$ relative error of variable coefficients.\\
$e^a_c$ & The absolute error of variable coefficients.\\
$T_{train}$ & The total training time of the model.\\ 
$Iter_L$ & The number of iterations for $L$-$BFGS$ optimization.\\
$Loss$ & The value of the loss function when training stops.\\
\hline
\hline
\end{tabular}
\end{center}
\end{table}

\section{Model parameters and results of numerical experiments}
\quad

\subsection{Model parameters and results of the forward problem of the vSG equation}\label{Appendix_vSG_forward}
\quad

The symbols $p1$, $p2$, and $cos$ in the following table represent the cases of first-degree polynomials, quadratic polynomial, and trigonometric function coefficients in Section \ref{vSG_forward}, respectively.

\begin{table}[H]
\caption{Space-time interval for the forward problem of the vSG equation}
\begin{center}
\begin{tabular}{c||c|c}
\hline
\hline
 & $[T_0, T_1]$ & $\Omega$ \\
\hline
\hline
$p1$ ($k_1=1$)& $[-5,5]$ & $[-5,5]$\\
$p1$ ($k_1=-1$)& $[-5,5]$ & $[-5,5]$\\
$p2$  & $[-3,3]$ & $[-5,5]$\\
$cos$ & $[-3,5]$ & $[-8,8]$\\
\hline
\hline
\end{tabular}
\end{center}
\end{table}

\begin{table}[H]
\caption{Model parameters for the forward problem of the vSG equation}
\begin{center}
\begin{tabular}{c||c|c|c|c|c|c|c|c|c|c|c|c|c}
\hline
\hline
 & $Seed$ & $N_{Adam}$ & $n_x$ & $n_t$ &$N_d^u$ & $N_B^u$ & $N_h^u$ & $N_d^c$ & $N_B^c$ & $N_h^c$  & $n_I+n_b$ & $n_c$ & $n_f$\\
\hline
\hline
$p1$ ($k_1=1$)& 6886 & 5000 & 512 & 200 & 40 & 0 & 0 & 30 & 2& 2 & 800 & 60 & 20000\\
$p1$ ($k_1=-1$)& 6886 & 5000 & 512 & 200 & 40 & 4 & 2 & 30 & 2& 2 & 800 & 60 & 20000\\
$p3$\ /\ $cos$ & 6886 & 5000 & 512 & 200 & 40 & 4 & 2 & 30 & 2& 2 & 800 & 60 & 20000\\
\hline
\hline
\end{tabular}
\end{center}
\end{table}

\begin{table}[H]
\caption{Model results for the forward problem of the vSG equation}
\begin{center}
\begin{tabular}{c||c|c|c|c|c|c}
\hline
\hline
 & $e^r_u$ & $e^r_c$ & $e^a_c$ & $T_{train}$ & $Iter_L$ & $Loss$ \\
\hline
\hline
$p1$ ($k_1=1$)& $5.23\times 10^{-4}$ & $7.66\times 10^{-5}$ & $1.92\times 10^{-4}$ & 260.06s & 1446 & $1.61\times 10^{-6}$\\
$p1$ ($k_1=-1$) & $1.52\times 10^{-4}$ & $1.66\times 10^{-4}$ & $4.11\times 10^{-4}$ & 327.59s & 2311 & $5.37\times 10^{-7}$ \\
$p3$  & $9.33\times 10^{-5}$ & $3.98\times 10^{-5}$ & $9.55\times 10^{-5}$ & 455.13s & 4993 & $5.28\times 10^{-7}$\\
$cos$  & $6.20\times 10^{-4}$ & $2.06\times 10^{-3}$ & $1.73\times 10^{-3}$ & 548.05s & 6999 & $9.84\times 10^{-7}$\\
\hline
\hline
\end{tabular}
\end{center}
\end{table}

\subsection{Model parameters and results of the forward problem of the gvKP equation}\label{Appendix_gvKP_forward}
\quad

The notations Case 1 to Case 4 here correspond to the four situations in Section \ref{KP}.

\begin{table}[H]
\caption{Space-time interval for the forward problem of the gvKP equation}
\begin{center}
\begin{tabular}{c||c|c|c}
\hline
\hline
  & $x$ & $y$ &$t$\\
\hline
\hline
$Case$ 1& [-6,6] & [-6,6]& [-6,6]\\
$Case$ 2& [-8,8] & [-8,8]& [-8,8]\\
$Case$ 3  & [-10,2] & [-6,6]& [-6,6]\\
$Case$ 4 & [-8,4] & [-6,6]& [-4,4]\\
\hline
\hline
\end{tabular}
\end{center}
\end{table}

\begin{table}[H]
\caption{Model parameters for the forward problem of the gvKP equation}
\begin{center}
\begin{tabular}{c||c|c|c|c|c|c|c|c|c|c|c|c|c|c}
\hline
\hline
 & $Seed$ & $N_{Adam}$ & $n_x$ & $n_y$ &$n_t$ &$N_d^u$ & $N_B^u$ & $N_h^u$ & $N_d^c$ & $N_B^c$ & $N_h^c$  & $n_I+n_b$ & $n_c$ & $n_f$\\
\hline
\hline
All Case& 8888 & 5000 & 100 & 100 & 100 & 40 & 4 & 2 & 30& 3 & 2 & 6000 & 100 & 50000\\
\hline
\hline
\end{tabular}
\end{center}
\end{table}

\begin{table}[H]
\caption{Model results for the forward problem of the gvKP equation}
\begin{center}
\begin{tabular}{c||c|c|c|c|c|c|c}
\hline
\hline
 & $e^r_u$ & $e^r_c(f(t))$ & $e^r_c(g(t))$ & $e^r_c(l(t))$ & $T_{train}$ & $Iter_L$ & $Loss$ \\
\hline
\hline
$Case$ 1 & $3.48\times 10^{-4}$ & $3.46\times 10^{-4}$ & $4.22\times 10^{-4}$ & $1.87\times 10^{-3}$ & 5415.26s & 7683 & $8.95\times 10^{-7}$\\
$Case$ 2 & $4.83\times 10^{-4}$ & $1.49\times 10^{-3}$ & $1.96\times 10^{-3}$ & $1.43\times 10^{-3}$ & 10122.47s & 18316 & $1.87\times 10^{-6}$\\
$Case$ 3 & $2.23\times 10^{-3}$ & $1.06\times 10^{-4}$ & $4.12\times 10^{-3}$ & $1.70\times 10^{-4}$ & 2746.84s & 1277 & $1.50\times 10^{-6}$\\
$Case$ 4  & $1.30\times 10^{-3}$ & $6.11\times 10^{-5}$ & $9.89\times 10^{-4}$ & $1.62\times 10^{-4}$ & 3606.26s & 3427 & $1.44\times 10^{-6}$\\
\hline
\hline
\end{tabular}
\end{center}
\end{table}

\subsection{Model parameters and results for the inverse problem of vKdV equation (single coefficient)}\label{Appendix_vKdV_Inverse_1}
\quad

The symbols $p1$, $p3$, and $cos$ in the following table represent the cases of first-degree polynomials, cubic polynomials, and trigonometric function coefficients in Section \ref{KdV-1v}, respectively.

\begin{table}[H]
\caption{Space-time interval for the inverse problem the vKdV equation (single coefficient)}
\begin{center}
\begin{tabular}{c||c|c}
\hline
\hline
 & $[T_0, T_1]$ & $\Omega$ \\
\hline
\hline
$p1$ & [-5,5] & [-5,5]\\
$p3$  & [-2,2] & [-5,5]\\
$cos$ & [-5,5] & [-5,5]\\
\hline
\hline
\end{tabular}
\end{center}
\end{table}

\begin{table}[H]
\caption{Model parameters for the inverse problem the vKdV equation (single coefficient)}
\begin{center}
\begin{tabular}{c||c|c|c|c|c|c|c|c|c|c|c}
\hline
\hline
 & $Seed$ & $N_{Adam}$ & $nd_c$ &$N_d^u$ & $N_B^u$ & $N_h^u$ & $N_d^c$ & $N_B^c$ & $N_h^c$  & $n_s$ & $n_f$\\
\hline
\hline
All & 6666 & 5000 & 500 & 40 & 3 & 2 & 30 & 2& 2 & 2000 & 20000\\
\hline
\hline
\end{tabular}
\end{center}
\end{table}

\begin{table}[H]
\caption{Model results for the inverse problem the vKdV equation (single coefficient)}
\begin{center}
\begin{tabular}{c||c|c|c|c|c|c}
\hline
\hline
 & $e^r_u$ & $e^r_c$ & $e^a_c$ & $T_{train}$ & $Iter_L$ & $Loss$ \\
\hline
\hline
$p1$ & $2.82\times 10^{-4}$ & $4.97\times 10^{-4}$ & $9.66\times 10^{-4}$ & 531.17s & 1844 & $1.09\times 10^{-6}$\\
$p3$ & $2.56\times 10^{-4}$ & $1.70\times 10^{-3}$ & $3.41\times 10^{-3}$ & 780.92s & 4494 & $6.67\times 10^{-7}$\\
$cos$ & $2.26\times 10^{-4}$ & $2.46\times 10^{-3}$ & $1.24\times 10^{-3}$ & 581.75s & 1837 & $9.38\times 10^{-7}$\\
\hline
\hline
\end{tabular}
\end{center}
\end{table}

\subsection{Model parameters and results for the inverse problem of vKdV equation (multiple coefficients)}\label{Appendix_vKdV_Inverse_2}
\quad

The symbols $p1$, $p3$ and $cos$ in the following table represent the cases of first-degree polynomials, cubic polynomials, and trigonometric function coefficients in Section \ref{KdV_2v_ex}, respectively. However, Case 1 and Case 2 are the corresponding situations in Section \ref{KdV_2v_ex}

\begin{table}[H]
\caption{Space-time interval for the inverse problem the vKdV equation (multiple coefficients)}
\begin{center}
\begin{tabular}{c||c|c}
\hline
\hline
 & $[T_0, T_1]$ & $\Omega$ \\
\hline
\hline
$p1$ & [-5,5] & [-5,5]\\
$p3$  & [-2,2] & [-5,5]\\
$cos$ & $[-3\pi /2, 3\pi /2]$ & [-5,5]\\
$Case$ 1 & $[1/2, 2]$ & [-4,4]\\
$Case$ 2 & $[-\pi /2 , 5\pi /2]$ & [-4,4]\\
\hline
\hline
\end{tabular}
\end{center}
\end{table}

\begin{table}[H]
\caption{Model parameters for the inverse problem the vKdV equation (multiple coefficients)}
\begin{center}
\begin{tabular}{c||c|c|c|c|c|c|c|c|c|c|c}
\hline
\hline
 & $Seed$ & $N_{Adam}$ & $nd_c$ &$N_d^u$ & $N_B^u$ & $N_h^u$ & $N_d^c$ & $N_B^c$ & $N_h^c$  & $n_s$ & $n_f$\\
\hline
\hline
All & 6666 & 5000 & 500 & 40 & 3 & 3 & 30 & 3& 2 & 5000 & 20000\\
\hline
\hline
\end{tabular}
\end{center}
\end{table}

\begin{table}[H]
\caption{Model results for the inverse problem the vKdV equation (multiple coefficients)}
\begin{center}
\begin{tabular}{c||c|c|c|c|c|c|c|c}
\hline
\hline
 & $e^r_u$ & $e^r_c(f(t))$ & $e^a_c(f(t))$ & $e^r_c(g(t))$ & $e^a_c(g(t))$ & $T_{train}$ & $Iter_L$ & $Loss$ \\
\hline
\hline
$p1$ & $1.88\times 10^{-4}$ & $2.95\times 10^{-3}$ & $5.75\times 10^{-3}$ & $2.56\times 10^{-3}$ & $9.92\times 10^{-3}$ & 1091.52s & 4062 & $7.73\times 10^{-7}$\\
$p3$  & $3.07\times 10^{-4}$ & $2.42\times 10^{-3}$ & $3.89\times 10^{-3}$ & $3.90\times 10^{-3}$ & $1.19\times 10^{-2}$ & 1409.46s & 6951 & $7.06\times 10^{-7}$\\
$cos$ & $8.79\times 10^{-5}$ & $4.53\times 10^{-4}$ & $2.25\times 10^{-4}$ & $7.88\times 10^{-4}$ & $7.87\times 10^{-4}$ & 1464.19s & 7506 & $4.61\times 10^{-7}$\\
$Case$ 1& $1.66\times 10^{-5}$ & $1.35\times 10^{-3}$ & $1.53\times 10^{-3}$ & $2.86\times 10^{-3}$ & $6.59\times 10^{-3}$ & 1084.69s & 3886 & $8.12\times 10^{-7}$\\
$Case$ 2& $1.76\times 10^{-4}$ & $1.85\times 10^{-3}$ & $6.25\times 10^{-4}$ & $2.77\times 10^{-3}$ & $1.83\times 10^{-3}$ & 1708.11s & 9281 & $1.06\times 10^{-6}$\\
\hline
\hline
\end{tabular}
\end{center}
\end{table}

\subsection{Model parameters and results for the inverse problem of gvSK equation (multiple coefficients)}\label{Appendix_gvSK_Inverse}
\quad

The symbols $p1$, $p2$ and $p3$ in the following table represent the cases of first-degree polynomials, quadratic polynomial and cubic polynomials coefficients in Section \ref{vSK}, respectively.

\begin{table}[H]
\caption{Space-time interval for the inverse problem the gvSK equation (multiple coefficients)}
\begin{center}
\begin{tabular}{c||c|c}
\hline
\hline
 & $[T_0, T_1]$ & $\Omega$ \\
\hline
\hline
$p1$ & [-2,2] & [-4,4]\\
$p2$  & [-2,2] & [-2,2] \\
$p3$ & [-2,2] & [-3,3]\\
\hline
\hline
\end{tabular}
\end{center}
\end{table}

\begin{table}[H]
\caption{Model parameters for the inverse problem the gvSK equation (multiple coefficients)}
\begin{center}
\begin{tabular}{c||c|c|c|c|c|c|c|c|c|c|c}
\hline
\hline
 & $Seed$ & $N_{Adam}$ & $nd_c$ &$N_d^u$ & $N_B^u$ & $N_h^u$ & $N_d^c$ & $N_B^c$ & $N_h^c$  & $n_s$ & $n_f$\\
\hline
\hline
All & 6666 & 5000 & 500 & 40 & 3 & 3 & 30 & 3& 2 & 2000 & 20000\\
\hline
\hline
\end{tabular}
\end{center}
\end{table}

\begin{table}[H]
\caption{Model results for the inverse problem the gvSK equation (multiple coefficients)}
\begin{center}
\begin{tabular}{c||c|c|c|c|c|c|c}
\hline
\hline
 & $e^r_u$ & $e^r_c(\alpha(t))$ & $e^r_c(\beta(t))$ & $e^r_c(\gamma(t))$ & $T_{train}$ & $Iter_L$ & $Loss$ \\
\hline
\hline
$p1$ & $5.04\times 10^{-4}$ & $3.05\times 10^{-2}$ & $4.18\times 10^{-2}$ & $2.94\times 10^{-2}$ & 1314.41s & 5105 & $2.13\times 10^{-6}$\\
$p2$  & $1.00\times 10^{-3}$ & $1.68\times 10^{-2}$ & $1.49\times 10^{-2}$ & $1.63\times 10^{-2}$ & 1417.54s & 6013 & $5.37\times 10^{-7}$\\
$p3$ & $6.69\times 10^{-4}$ & $1.54\times 10^{-2}$ & $1.93\times 10^{-2}$ & $1.46\times 10^{-2}$ & 1675.60s & 7410 & $3.12\times 10^{-6}$\\
\hline
\hline
\end{tabular}
\end{center}
\end{table}

\subsection{Model parameters and results for the inverse problem of gvKP equation (multiple coefficients)}\label{Appendix_gvKP_Inverse}
\quad

The notations Case 1 to Case 4 here correspond to the four situations in Section \ref{KP}.

\begin{table}[H]
\caption{Space-time interval for the inverse problem the gvKP equation (multiple coefficients)}
\begin{center}
\begin{tabular}{c||c|c|c}
\hline
\hline
  & $x$ & $y$ &$t$\\
\hline
\hline
$Case$ 1& [-6,6] & [-6,6]& $[-2\pi,2\pi]$\\
$Case$ 2& [-8,8] & [-8,8]& $[-\pi,\pi]$\\
$Case$ 3  & [-10,2] & [-6,6]& [-3,3]\\
$Case$ 4 & [-8,4] & [-6,6]& [-2,2]\\
\hline
\hline
\end{tabular}
\end{center}
\end{table}

\begin{table}[H]
\caption{Model parameters for the inverse problem the gvKP equation (multiple coefficients)}
\begin{center}
\begin{tabular}{c||c|c|c|c|c|c|c|c|c|c|c|c|c|c}
\hline
\hline
 & $Seed$ & $N_{Adam}$ & $nd_c$ & $n_x$ & $n_y$ &$n_t$ & $N_d^u$ & $N_B^u$ & $N_h^u$ & $N_d^c$ & $N_B^c$ & $N_h^c$  & $n_s$ & $n_f$\\
\hline
\hline
All & 6666 & 5000 & 500 & 100 & 100 & 100 & 40 & 3 & 3 & 30 & 3& 2 & 20000 & 50000\\
\hline
\hline
\end{tabular}
\end{center}
\end{table}

\begin{table}[H]
\caption{Model results for the inverse problem the gvKP equation (multiple coefficients)}
\begin{center}
\begin{tabular}{c||c|c|c|c|c|c|c}
\hline
\hline
 & $e^r_u$ & $e^r_c(f(t))$ & $e^r_c(g(t))$ & $e^r_c(l(t))$ & $T_{train}$ & $Iter_L$ & $Loss$ \\
\hline
\hline
$Case$ 1& $3.96\times 10^{-4}$ & $2.98\times 10^{-3}$ & $2.06\times 10^{-3}$ & $5.94\times 10^{-3}$ & 7037.44s & 9254 & $9.93\times 10^{-7}$\\
$Case$ 2 & $3.47\times 10^{-4}$ & $1.95\times 10^{-3}$ & $4.01\times 10^{-3}$ & $2.10\times 10^{-3}$ & 8504.51s & 12091 & $9.13\times 10^{-7}$\\
$Case$ 3& $1.05\times 10^{-3}$ & $2.45\times 10^{-2}$ & $1.10\times 10^{-2}$ & $6.98\times 10^{-3}$ & 3314.94s & 1700 & $1.25\times 10^{-6}$\\
$Case$ 4& $7.84\times 10^{-4}$ & $1.29\times 10^{-2}$ & $3.98\times 10^{-3}$ & $4.42\times 10^{-3}$ & 3335.84s & 1737 & $9.56\times 10^{-7}$\\
\hline
\hline
\end{tabular}
\end{center}
\end{table}

\section{Supplement to analysis and discussion}

\subsection{Derivation of gradient propagation}\label{A_grad}
\quad

First, some obviously established formulas needed in the derivation process are given, as follows:
\begin{align*}
	&R^{[i+1]}=\mathcal{L}_i(R^{[i]})+\mathcal{K}R^{[i]},\ i=0,1,...,N_B,\\
	&Loss_s(\theta)=\frac{1}{n_s}\sum_{i=1}^{n_s}|\tilde{u}({\bm x}_s^i,t_s^i;\theta_u)-u_s^i|^2,\\
	&Loss_f(\theta)=\frac{1}{n_f}\sum_{i=1}^{n_f}|f({\bm x}_f^i,t_f^i;\tilde{u}({\bm x},t;\theta_u),\tilde{c}(t;\theta_c))|^2.
\end{align*}

The gradient of loss function $Loss_s(\theta)$ to $X^{[j]}, j\geq 1$: \footnote{\ The gradient calculation in this section is based on a single sample.}
\begin{itemize}
    \item If $X^{[j]}$ is the output inside a residual block, let $[j]$ be between $[iN_h+1]$ and $[(i+1)N_h+1]$, i.e. $iN_h+1\le j< [(i+1)N_h+1],i=0,1,...,N_B$, then we have
\begin{align*}
	\frac{\partial Loss_s}{\partial X^{[j]}}&=\frac{\partial Loss_s}{\partial R^{[i+1]}}\cdot \frac{\partial R^{[i+1]}}{\partial X^{[j]}},\\
	&=\frac{\partial Loss_s}{\partial R^{[N_B]}}\cdot \left[\prod_{k=i+1}^{N_B-1}\frac{\partial R^{[k+1]}}{\partial R^{[k]}}\right]\cdot \frac{\partial R^{[i+1]}}{\partial X^{[j]}},\\
	&=\frac{\partial Loss_s}{\partial R^{[N_B]}}\cdot\prod_{k=i+1}^{N_B-1}\left[\mathcal{K}+\frac{1}{\partial R^{[k]}}\left(\mathcal{L}_k(R^{[k]})\right)\right]\cdot\frac{\partial X^{[(i+1)N_h+1]}}{\partial X^{[j]}},\\
	&=\frac{\partial Loss_s}{\partial R^{[N_B]}}\cdot \prod_{k=j}^{(i+1)N_h}\frac{\partial X^{[k+1]}}{\partial X^{[k]}}\cdot \prod_{k=i+1}^{N_B-1}\left[\mathcal{K}+\frac{1}{\partial R^{[k]}}\left(\mathcal{L}_k(R^{[k]})\right)\right].
\end{align*}
\item In particular, $X^{[j]}$ happens to be the output/input of a residual block, assuming $j=iN_h+1,i=0,1,...,N_B$, then the above formula is simplified to
\begin{align*}
	\frac{\partial Loss_s}{\partial X^{[j]}}=\frac{\partial Loss_s}{\partial R^{[N_B]}}\cdot\prod_{k=i}^{N_B-1}\left[\mathcal{K}+\frac{1}{\partial R^{[k]}}\left(\mathcal{L}_k(R^{[k]})\right)\right].	
\end{align*}
\end{itemize}

The gradient calculation formulas of the loss items $Loss_I(\theta)$, $Loss_b(\theta)$ and $Loss_c(\theta)$ involved in the forward and inverse problems are similar to the above, while the loss item $Loss_f(\theta)$ is different. The gradient of the above loss function only involves the trunk network, but for $Loss_f(\theta)$, it involves both the trunk network and the branch network. Therefore, in the subsequent derivation, $X^{[i]}_{[\cdot]}$ and $R^{[i]}_{[\cdot]}$ are subscripted with $u$ or $c$ to distinguish them. In addition, $D^{[\cdot]}$, $N_B^{[\cdot]}$ $N_h^{[\cdot]}$ and $\mathcal{K}^{[\cdot]}$ are all superscripted accordingly. Suppose the equation is in a simplified form, as
\begin{equation*}
	u_{t}=c_1(t)\mathcal{N}[u]
\end{equation*}

The gradient of the $Loss_f(\theta)$ to the weights $W^{[j]}_u$ and $W^{[j]}_c$ of the trunk network and the branch network:
\begin{itemize}
	\item Take the gradient contribution of item $(\tilde{u})_t$ in $f$ to weight $W^{[j]}_u$ as an example to calculate (assuming $iN_h^u+1\le j<[(i+1)N_h^u+1],i=0,1,...,N_B^u$):
\begin{align*}
	\frac{\partial Loss_f}{\partial W^{[j]}_u}&=\frac{\partial Loss_f}{\partial f}\cdot\left[\frac{\partial f}{\partial [(\tilde{u})_t]}\cdot \frac{\partial [(\tilde{u})_t]}{\partial W^{[j]}_u}+c_1\frac{\partial f}{\partial \mathcal{N}[\tilde{u}]}\cdot \frac{\partial \mathcal{N}[\tilde{u}]}{\partial W^{[j]}_u}\right],\\
	\frac{\partial [(\tilde{u})_t]}{\partial W^{[j]}_u}&=\frac{1}{\partial W^{[j]}_u}\cdot\left(\frac{\partial X^{[D^u]}_u}{\partial X^{[D^u-1]}_u}\cdot \prod_{k=0}^{N_B^u-1} \frac{\partial R^{[k+1]}_u}{\partial R^{[k]}_u}\cdot \frac{\partial R^{[0]}_u}{\partial t}\right),\\
	&=\frac{\partial X^{[D^u]}_u}{\partial X^{[D^u-1]}_u}\cdot \prod_{\substack{0\le k \le N^u_B-1\\k\neq i}} \left[\frac{\partial R^{[k+1]}_u}{\partial R^{[k]}_u}\right]\cdot\frac{\partial R^{[0]}_u}{\partial t}\cdot \frac{1}{\partial W^{[j]}_u}\left(\frac{\partial R^{[i+1]}_u}{\partial R^{[i]}_u}\right),\\
	&=W^{[D^u-1]}_u\cdot\frac{\partial R^{[0]}_u}{\partial t}\cdot \prod_{\substack{0\le k \le N^u_B-1\\k\neq i}} \left[\mathcal{K}^u+\frac{1}{\partial R^{[k]}_u}\left(\mathcal{L}_k(R^{[k]}_u)\right)\right]\cdot \frac{1}{\partial W^{[j]}_u}\left[\mathcal{K}^u+\frac{1}{\partial R^{[i]}_u}\left(\mathcal{L}_i(R^{[i]}_u)\right)\right]
\end{align*}
    \item The gradient of the loss term $Loss_f(\theta)$ to the weight $W^{[j]}_c$ of the branch network (assuming $iN_h^c+1\le j+1<[(i+1)N_h^c+1],i=0,1,...,N_B^c$):
\begin{align*}
	\frac{\partial Loss_f}{\partial W^{[j]}_c}&=\frac{\partial Loss_f}{\partial f}\cdot \frac{\partial f}{\partial(c_1)}\cdot\frac{\partial c_1}{\partial W^{[j]}_c},\\
	&=\frac{\partial Loss_f}{\partial f}\cdot\mathcal{N}[\tilde{u}]\cdot \frac{\partial X^{[D^c]}_c}{\partial R^{[N_B^c]}_c}\cdot \frac{\partial R^{[N_B^c]}_c}{\partial W^{[j]}_c},\\
	&=\frac{\partial Loss_f}{\partial f}\cdot\mathcal{N}[\tilde{u}]\cdot W^{[D^c-1]}_c\cdot\frac{\partial R^{[N_B^c]}_c}{\partial X^{[j+1]}_c}\cdot\frac{\partial X^{[j+1]}_c}{\partial W^{[j]}_c},\\ 
	&=\frac{\partial Loss_f}{\partial f}\cdot\frac{\partial X^{[j+1]}_c}{\partial W^{[j]}_c}\cdot\mathcal{N}[\tilde{u}]\cdot W^{[D^c-1]}_c\cdot \prod_{k=j+1}^{(i+1)N_h^c}\frac{\partial X^{[k+1]}_c}{\partial X^{[k]}_c}\cdot \prod_{k=i+1}^{N_B^c-1}\left[\mathcal{K}^c+\frac{1}{\partial R^{[k]}_c}\left(\mathcal{L}_k(R^{[k]}_c)\right)\right].
\end{align*}

\end{itemize}

\subsection{Repeated experiments in the unity of linearity and nonlinearity}\label{A_linear_nonlinear}
\quad

For the solution $u^{(vKdV)}_1$ (linear/$p1$) and solution $u^{(vKdV)}_2$ (nonlinear/$p3$) of the vKdV equation in Section \ref{KdV}, the inverse problem test is performed under the two situations of using the ResNet structure and not using the ResNet structure. It contains the results under different layers of trunk network and branch network. Except for the different random seeds, other model settings are the same. The model settings are shown below, along with the average results for 5 different random seeds. The results of this appendix mainly provide data support for the discussion in Section \ref{sec_linear_nonlinear}.

\begin{table}[H]
\caption{Space-time interval for the inverse problem the vKdV equation (single coefficient)}
\begin{center}
\begin{tabular}{c||c|c}
\hline
\hline
 & $[T_0, T_1]$ & $\Omega$ \\
\hline
\hline
$p1$ & [-5,5] & [-5,5]\\
$p3$  & [-2,2] & [-5,5]\\
\hline
\hline
\end{tabular}
\end{center}
\end{table}

\begin{table}[H]
\caption{Model parameters for the inverse problem the vKdV equation (single coefficient)}
\begin{center}
\begin{tabular}{c||c|c|c|c|c|c|c|c}
\hline
\hline
 & $N_{Adam}$ & $nd_c$ &$N_d^u$ & $N_h^u$ & $N_d^c$ & $N_h^c$  & $n_s$ & $n_f$\\
\hline
\hline
All & 5000 & 500 & 40 & 2 & 30 & 2 & 2000 & 20000\\
\hline
\hline
\end{tabular}
\end{center}
\end{table}

\begin{table}[H]
\caption{The $L^2$ relative error of linear coefficient (without ResNet)}
\begin{center}
\begin{tabular}{c||c|c|c|c|c|c}
\hline
\hline
 \diagbox{trunk}{branch}& 5 & 6 & 7 & 8 & 9 & 10\\
\hline
\hline
7& $6.11\times 10^{-4}$ & $1.32\times 10^{-3}$ & $1.57\times 10^{-3}$ & $2.18\times 10^{-3}$ & $3.10\times 10^{-3}$ & $2.09\times 10^{-3}$\\

8& $8.04\times 10^{-4}$ & $5.26\times 10^{-4}$ & $1.99\times 10^{-3}$ & $3.60\times 10^{-3}$ & $3.33\times 10^{-3}$ & $2.88\times 10^{-3}$\\

9& $1.01\times 10^{-3}$ & $9.67\times 10^{-4}$ & $2.37\times 10^{-3}$ & $3.80\times 10^{-3}$ & $3.39\times 10^{-3}$ & $2.95\times 10^{-3}$\\

10& $8.22\times 10^{-4}$ & $6.55\times 10^{-4}$ & $3.57\times 10^{-3}$ & $3.26\times 10^{-3}$ & $3.69\times 10^{-3}$ & $3.48\times 10^{-3}$\\

11& $1.22\times 10^{-3}$ & $1.15\times 10^{-3}$ & $2.64\times 10^{-3}$ & $3.42\times 10^{-3}$ & $4.04\times 10^{-3}$ & $3.36\times 10^{-3}$\\

12& $6.60\times 10^{-4}$ & $7.29\times 10^{-4}$ & $3.34\times 10^{-3}$ & $3.81\times 10^{-3}$ & $4.21\times 10^{-3}$ & $4.03\times 10^{-3}$\\
\hline
\hline
\end{tabular}
\end{center}
\end{table}

\begin{table}[H]
\caption{The $L^2$ relative error of nonlinear coefficient (without ResNet)}
\begin{center}
\begin{tabular}{c||c|c|c|c|c|c}
\hline
\hline
 \diagbox{trunk}{branch}& 5 & 6 & 7 & 8 & 9 & 10\\
\hline
\hline
7& $3.07\times 10^{-3}$ & $2.16\times 10^{-3}$ & $2.04\times 10^{-3}$ & $1.44\times 10^{-3}$ & $9.31\times 10^{-4}$ & $1.34\times 10^{-3}$\\

8& $3.96\times 10^{-3}$ & $2.32\times 10^{-3}$ & $1.52\times 10^{-3}$ & $1.72\times 10^{-3}$ & $1.23\times 10^{-3}$ & $1.35\times 10^{-3}$\\

9& $3.73\times 10^{-3}$ & $1.97\times 10^{-3}$ & $1.62\times 10^{-3}$ & $1.34\times 10^{-3}$ & $1.23\times 10^{-3}$ & $7.67\times 10^{-4}$\\

10& $3.56\times 10^{-3}$ & $1.57\times 10^{-3}$ & $1.38\times 10^{-3}$ & $1.05\times 10^{-3}$ & $1.14\times 10^{-3}$ & $8.95\times 10^{-4}$\\

11& $3.29\times 10^{-3}$ & $2.28\times 10^{-3}$ & $1.23\times 10^{-3}$ & $1.03\times 10^{-3}$ & $7.14\times 10^{-4}$ & $9.57\times 10^{-4}$\\

12& $3.37\times 10^{-3}$ & $1.46\times 10^{-3}$ & $1.23\times 10^{-3}$ & $9.22\times 10^{-4}$ & $9.34\times 10^{-4}$ & $6.42\times 10^{-4}$\\
\hline
\hline
\end{tabular}
\end{center}
\end{table}

\begin{table}[H]
\caption{The $L^2$ relative error of linear coefficient (with ResNet)}
\begin{center}
\begin{tabular}{c||c|c|c|c|c|c}
\hline
\hline
 \diagbox{trunk}{branch}& 5 & 6 & 7 & 8 & 9 & 10\\
\hline
\hline
7& $1.10\times 10^{-3}$ & $6.38\times 10^{-4}$ & $5.39\times 10^{-4}$ & $4.57\times 10^{-4}$ & $3.79\times 10^{-4}$ & $1.12\times 10^{-3}$\\

8& $7.87\times 10^{-4}$ & $5.57\times 10^{-4}$ & $4.44\times 10^{-4}$ & $4.84\times 10^{-4}$ & $3.33\times 10^{-4}$ & $4.14\times 10^{-4}$\\

9& $9.70\times 10^{-4}$ & $9.40\times 10^{-4}$ & $4.48\times 10^{-4}$ & $4.23\times 10^{-4}$ & $6.00\times 10^{-4}$ & $4.53\times 10^{-4}$\\

10& $1.02\times 10^{-3}$ & $8.77\times 10^{-4}$ & $4.98\times 10^{-4}$ & $5.19\times 10^{-4}$ & $3.72\times 10^{-4}$ & $5.79\times 10^{-4}$\\

11& $1.04\times 10^{-3}$ & $1.15\times 10^{-3}$ & $9.16\times 10^{-4}$ & $8.95\times 10^{-4}$ & $4.86\times 10^{-4}$ & $3.34\times 10^{-4}$\\

12& $1.16\times 10^{-3}$ & $1.12\times 10^{-3}$ & $8.94\times 10^{-4}$ & $7.11\times 10^{-4}$ & $5.97\times 10^{-4}$ & $5.37\times 10^{-4}$\\
\hline
\hline
\end{tabular}
\end{center}
\end{table}

\begin{table}[H]
\caption{The $L^2$ relative error of nonlinear coefficient (with ResNet)}
\begin{center}
\begin{tabular}{c||c|c|c|c|c|c}
\hline
\hline
 \diagbox{trunk}{branch}& 5 & 6 & 7 & 8 & 9 & 10\\
\hline
\hline
7& $1.62 \times 10^{-3}$ & $1.20 \times 10^{-3}$ & $1.17 \times 10^{-3}$ & $8.65 \times 10^{-4}$ & $1.63 \times 10^{-3}$ & $1.80 \times 10^{-3}$ \\

8 & $1.57 \times 10^{-3}$ & $1.36 \times 10^{-3}$ & $1.12 \times 10^{-3}$ & $1.07 \times 10^{-3}$ & $1.29 \times 10^{-3}$ & $1.17 \times 10^{-3}$ \\

9 & $1.71 \times 10^{-3}$ & $1.33 \times 10^{-3}$ & $1.23 \times 10^{-3}$ & $8.84 \times 10^{-4}$ & $9.80 \times 10^{-4}$ & $1.16 \times 10^{-3}$ \\

10 & $1.57 \times 10^{-3}$ & $1.32 \times 10^{-3}$ & $9.54 \times 10^{-4}$ & $7.57 \times 10^{-4}$ & $1.24 \times 10^{-3}$ & $1.23 \times 10^{-3}$ \\

11 & $1.68 \times 10^{-3}$ & $1.43 \times 10^{-3}$ & $1.01 \times 10^{-3}$ & $1.01 \times 10^{-3}$ & $1.42 \times 10^{-3}$ & $1.05 \times 10^{-3}$ \\

12 & $2.12 \times 10^{-3}$ & $1.69 \times 10^{-3}$ & $8.20 \times 10^{-4}$ & $7.26 \times 10^{-4}$ & $9.55 \times 10^{-4}$ & $9.75 \times 10^{-4}$ \\
\hline
\hline
\end{tabular}
\end{center}
\end{table}

\subsection{Data support for the relationship between convexity and learning of neural networks}\label{A_Learning}
\quad

Notations (1)-(5) or (1)-(6) correspond to the corresponding cases in Fig. \ref{NN_learn_1} and Fig. \ref{NN_learn_2} in Section \ref{sec_learning}, respectively.

\begin{table}[H]
\caption{Space-time interval for the inverse problem the vKdV equation (Fig. \ref{NN_learn_1})}
\begin{center}
\begin{tabular}{c||c|c}
\hline
\hline
 & $[T_0, T_1]$ & $\Omega$ \\
\hline
\hline
(1) & [-4,4] & [-4,4]\\
(2)  & [-4,4] & [-3,3]\\
(3)  & [-3,3] & [-5,5]\\
(4)  & [-2,2] & [-4,4]\\
(5)  & $[-3\pi/2,3\pi/2]$ & [-5,5]\\
\hline
\hline
\end{tabular}
\end{center}
\end{table}

\begin{table}[H]
\caption{Space-time interval for the inverse problem the vKdV equation (Fig. \ref{NN_learn_2})}
\begin{center}
\begin{tabular}{c||c|c}
\hline
\hline
 & $[T_0, T_1]$ & $\Omega$ \\
\hline
\hline
(1) & [-4,2] & [-3,3]\\
(2)  & [-2,2] & [-3,3]\\
(3)/(4)  & [-4,4] & [-3,3]\\
(5)/(6)  & [-2,2] & [-4,4]\\
\hline
\hline
\end{tabular}
\end{center}
\end{table}

\begin{table}[H]
\caption{Model parameters for the inverse problem of the vKdV equation (Fig. \ref{NN_learn_1} and Fig. \ref{NN_learn_2})}
\begin{center}
\begin{tabular}{c||c|c|c|c|c|c|c|c|c|c|c}
\hline
\hline
 & $Seed$ & $N_{Adam}$ & $nd_c$ &$N_d^u$ & $N_B^u$ & $N_h^u$ & $N_d^c$ & $N_B^c$ & $N_h^c$  & $n_s$ & $n_f$\\
\hline
\hline
All & 8888 & 5000 & 500 & 40 & 2 & 3 & 30 & 2& 2 & 5000 & 20000\\
\hline
\hline
\end{tabular}
\end{center}
\end{table}

\begin{table}[H]
\caption{Model results for the inverse problem the vKdV equation (Fig. \ref{NN_learn_1})}
\begin{center}
\begin{tabular}{c||c|c|c|c|c|c}
\hline
\hline
 & $e^r_u$ & $e^r_c$ & $e^a_c$ & $T_{train}$ & $Iter_L$ & $Loss$ \\
\hline
\hline
(1) & $3.78\times 10^{-4}$ & $1.47\times 10^{-3}$ & $2.71\times 10^{-3}$ & 494.16s & 1382 & $9.89\times 10^{-7}$\\
(2) & $2.10\times 10^{-1}$ & $1.62\times 10^{0}$ & $1.06\times 10^{0}$ & 1549.39s & 14744 & $8.03\times 10^{-2}$\\
(3) & $3.52\times 10^{-4}$ & $2.77\times 10^{-2}$ & $1.20\times 10^{-1}$ & 1235.97s & 10268 & $1.12\times 10^{-6}$\\
(4) & $3.20\times 10^{-1}$ & $2.51\times 10^{0}$ & $1.28\times 10^{1}$ & 887.64s & 6203 & $2.66\times 10^{-1}$\\
(5) & $2.28\times 10^{-4}$ & $1.44\times 10^{-3}$ & $8.48\times 10^{-4}$ & 597.61s & 2750 & $5.72\times 10^{-7}$\\
\hline
\hline
\end{tabular}
\end{center}
\end{table}

\begin{table}[H]
\caption{Model results for the inverse problem the vKdV equation (Fig. \ref{NN_learn_2})}
\begin{center}
\begin{tabular}{c||c|c|c|c|c|c}
\hline
\hline
 & $e^r_u$ & $e^r_c$ & $e^a_c$ & $T_{train}$ & $Iter_L$ & $Loss$ \\
\hline
\hline
(1) & $2.59\times 10^{-4}$ & $3.80\times 10^{-2}$ & $1.01\times 10^{-1}$ & 999.60s & 7418 & $4.67\times 10^{-7}$\\
(2) & $1.92\times 10^{-4}$ & $9.67\times 10^{-4}$ & $1.46\times 10^{-3}$ & 767.36s & 4646 & $5.02\times 10^{-7}$\\
(3) & $2.28\times 10^{-4}$ & $1.55\times 10^{-2}$ & $4.45\times 10^{-2}$ & 828.71s & 5605 & $5.82\times 10^{-6}$\\
(4) & $7.07\times 10^{-4}$ & $6.32\times 10^{-2}$ & $2.05\times 10^{-1}$ & 1393.76s & 11443 & $4.51\times 10^{-6}$\\
(5) & $5.43\times 10^{-4}$ & $4.71\times 10^{-3}$ & $1.30\times 10^{-2}$ & 1027.98s & 7895 & $2.04\times 10^{-6}$\\
(6) & $1.26\times 10^{-3}$ & $6.01\times 10^{-3}$ & $2.19\times 10^{-2}$ & 1675.01s & 15209 & $1.05\times 10^{-5}$\\

\hline
\hline
\end{tabular}
\end{center}
\end{table}

\subsection{Data support in anti-noise test}\label{A_noise}
\quad

The solution $u^{(vKdV)}_1$ (linear/$p1$) and solution $u^{(vKdV)}_2$ (nonlinear/$p3$) of the vKdV equation in Section \ref{KdV} are tested for forward and inverse problems in the case of using the ResNet structure. The results of this appendix mainly provide data support for the discussion in Section \ref{sec_noise}.

\begin{table}[H]
\caption{Space-time interval for the vKdV equation (Fig. \ref{Noise})}
\begin{center}
\begin{tabular}{c||c|c}
\hline
\hline
 & $[T_0, T_1]$ & $\Omega$ \\
\hline
\hline
$p1$ & [-5,5] & [-5,5]\\
$p3$  & [-2,2] & [-5,5]\\
\hline
\hline
\end{tabular}
\end{center}
\end{table}

\begin{table}[H]
\caption{Model parameters for the vKdV equation (Fig. \ref{Noise})}
\begin{center}
\begin{tabular}{c||c|c|c|c|c|c|c|c|c|c|c|c|c}
\hline
\hline
&$N_{Adam}$ & $nd_c$ & $n_x$ & $n_t$ &$N_d^u$ & $N_B^u$ & $N_h^u$ & $N_d^c$ & $N_B^c$ & $N_h^c$  & $n_s/n_I+n_b$ & $n_c$ &$n_f$\\
\hline
\hline
$Forward$ & 5000 & 500 & 512 & 200 & 40 & 3 & 2 & 30 & 3& 2 & 800 & 50 &20000\\
$Inverse$ & 5000 & 500 & 512 & 200 & 40 & 3 & 2 & 30 & 3& 2 & 2000 & None &20000\\
\hline
\hline
\end{tabular}
\end{center}
\end{table}

\begin{table}[H]
\caption{Anti-noise test of inverse problem with linear coefficients ($L^2$ relative error)}
\begin{center}
\begin{tabular}{c||c|c|c|c|c|c|c|c}
\hline
\hline
 \diagbox{seed}{noise}& $0\%$ & $0.5\%$ & $1.0\%$ & $1.5\%$ & $2.0\%$ & $2.5\%$ & $3.0\%$ & $5.0\%$\\
\hline
\hline
1000 & 1.15 $\times 10^{-3}$ & 2.64 $\times 10^{-3}$ & 4.30 $\times 10^{-3}$ & 6.41 $\times 10^{-3}$ & 8.56 $\times 10^{-3}$ & 1.04 $\times 10^{-2}$ & 1.23 $\times 10^{-2}$ & 2.11 $\times 10^{-2}$ \\

2000 & 2.50 $\times 10^{-4}$ & 1.39 $\times 10^{-3}$ & 2.65 $\times 10^{-3}$ & 4.08 $\times 10^{-3}$ & 5.25 $\times 10^{-3}$ & 6.63 $\times 10^{-3}$ & 7.70 $\times 10^{-3}$ & 1.24 $\times 10^{-2}$ \\

3000 & 4.66 $\times 10^{-4}$ & 3.95 $\times 10^{-3}$ & 7.87 $\times 10^{-3}$ & 1.17 $\times 10^{-2}$ & 1.56 $\times 10^{-2}$ & 1.89 $\times 10^{-2}$ & 2.29 $\times 10^{-2}$ & 3.20 $\times 10^{-2}$ \\

4000 & 7.39 $\times 10^{-4}$ & 2.68 $\times 10^{-3}$ & 5.41 $\times 10^{-3}$ & 7.95 $\times 10^{-3}$ & 1.05 $\times 10^{-2}$ & 1.31 $\times 10^{-2}$ & 1.58 $\times 10^{-2}$ & 2.51 $\times 10^{-2}$ \\

5000 & 3.65 $\times 10^{-4}$ & 1.21 $\times 10^{-3}$ & 2.25 $\times 10^{-3}$ & 3.45 $\times 10^{-3}$ & 4.41 $\times 10^{-3}$ & 5.59 $\times 10^{-3}$ & 6.66 $\times 10^{-3}$ & 1.05 $\times 10^{-2}$ \\

6000 & 6.69 $\times 10^{-4}$ & 1.03 $\times 10^{-3}$ & 2.41 $\times 10^{-3}$ & 3.82 $\times 10^{-3}$ & 5.14 $\times 10^{-3}$ & 6.45 $\times 10^{-3}$ & 7.79 $\times 10^{-3}$ & 1.31 $\times 10^{-2}$ \\

7000 &  4.25 $\times 10^{-4}$ & 7.64 $\times 10^{-4}$ & 1.52 $\times 10^{-3}$ & 2.25 $\times 10^{-3}$ & 2.91 $\times 10^{-3}$ & 3.63 $\times 10^{-3}$ & 4.34 $\times 10^{-3}$ & 7.06 $\times 10^{-3}$ \\

8000 & 3.36 $\times 10^{-4}$ & 5.14 $\times 10^{-3}$ & 9.99 $\times 10^{-3}$ & 1.46 $\times 10^{-2}$ & 1.97 $\times 10^{-2}$ & 2.44 $\times 10^{-2}$ & 2.94 $\times 10^{-2}$ & 4.26 $\times 10^{-2}$ \\

9000 & 5.00 $\times 10^{-4}$ & 3.68 $\times 10^{-3}$ & 7.41 $\times 10^{-3}$ & 1.09 $\times 10^{-2}$ & 1.45 $\times 10^{-2}$ & 1.67 $\times 10^{-2}$ & 1.92 $\times 10^{-2}$ & 2.98 $\times 10^{-2}$ \\

10000& 1.44 $\times 10^{-4}$ & 5.09 $\times 10^{-3}$ & 1.06 $\times 10^{-2}$ & 1.48 $\times 10^{-2}$ & 1.97 $\times 10^{-2}$ & 2.46 $\times 10^{-2}$ & 3.07 $\times 10^{-2}$ & 4.97 $\times 10^{-2}$ \\
\hline
\hline
\end{tabular}
\end{center}
\end{table}

\begin{table}[H]
\caption{Anti-noise test of inverse problem with nonlinear coefficients ($L^2$ relative error)}
\begin{center}
\begin{tabular}{c||c|c|c|c|c|c|c|c}
\hline
\hline
 \diagbox{seed}{noise}& $0\%$ & $0.5\%$ & $1.0\%$ & $1.5\%$ & $2.0\%$ & $2.5\%$ & $3.0\%$ & $5.0\%$\\
\hline
\hline
1000 & 6.64 $\times 10^{-4}$ & 2.37 $\times 10^{-3}$ & 4.55 $\times 10^{-3}$ & 6.61 $\times 10^{-3}$ & 8.78 $\times 10^{-3}$ & 1.13 $\times 10^{-2}$ & 1.29 $\times 10^{-2}$ & 2.13 $\times 10^{-2}$ \\

2000 & 1.14 $\times 10^{-3}$ & 1.68 $\times 10^{-3}$ & 3.03 $\times 10^{-3}$ & 3.95 $\times 10^{-3}$ & 4.99 $\times 10^{-3}$ & 6.20 $\times 10^{-3}$ & 7.51 $\times 10^{-3}$ & 1.19 $\times 10^{-2}$ \\

3000 &2.84 $\times 10^{-3}$ & 5.38 $\times 10^{-3}$ & 7.49 $\times 10^{-3}$ & 1.08 $\times 10^{-2}$ & 1.44 $\times 10^{-2}$ & 1.67 $\times 10^{-2}$ & 2.02 $\times 10^{-2}$ & 3.36 $\times 10^{-2}$ \\

4000 &6.03 $\times 10^{-4}$ & 2.77 $\times 10^{-3}$ & 5.37 $\times 10^{-3}$ & 8.13 $\times 10^{-3}$ & 1.15 $\times 10^{-2}$ & 1.40 $\times 10^{-2}$ & 1.65 $\times 10^{-2}$ & 2.67 $\times 10^{-2}$ \\

5000 &1.02 $\times 10^{-3}$ & 1.58 $\times 10^{-3}$ & 2.70 $\times 10^{-3}$ & 3.92 $\times 10^{-3}$ & 5.26 $\times 10^{-3}$ & 6.40 $\times 10^{-3}$ & 6.61 $\times 10^{-3}$ & 1.21 $\times 10^{-2}$ \\

6000 &6.88 $\times 10^{-4}$ & 9.50 $\times 10^{-4}$ & 1.63 $\times 10^{-3}$ & 2.62 $\times 10^{-3}$ & 3.43 $\times 10^{-3}$ & 4.30 $\times 10^{-3}$ & 5.07 $\times 10^{-3}$ & 8.65 $\times 10^{-3}$ \\

7000 &6.91 $\times 10^{-4}$ & 1.78 $\times 10^{-3}$ & 2.74 $\times 10^{-3}$ & 4.51 $\times 10^{-3}$ & 5.51 $\times 10^{-3}$ & 6.66 $\times 10^{-3}$ & 7.56 $\times 10^{-3}$ & 1.30 $\times 10^{-2}$ \\

8000 &1.01 $\times 10^{-3}$ & 4.11 $\times 10^{-3}$ & 8.26 $\times 10^{-3}$ & 1.23 $\times 10^{-2}$ & 1.65 $\times 10^{-2}$ & 2.04 $\times 10^{-2}$ & 2.42 $\times 10^{-2}$ & 3.87 $\times 10^{-2}$ \\

9000 &9.02 $\times 10^{-4}$ & 3.62 $\times 10^{-3}$ & 7.14 $\times 10^{-3}$ & 1.07 $\times 10^{-2}$ & 1.37 $\times 10^{-2}$ & 1.75 $\times 10^{-2}$ & 1.97 $\times 10^{-2}$ & 3.28 $\times 10^{-2}$ \\
10000 &8.11 $\times 10^{-4}$ & 3.96 $\times 10^{-3}$ & 8.24 $\times 10^{-3}$ & 1.27 $\times 10^{-2}$ & 1.74 $\times 10^{-2}$ & 2.07 $\times 10^{-2}$ & 2.59 $\times 10^{-2}$ & 4.34 $\times 10^{-2}$ \\
\hline
\hline
\end{tabular}
\end{center}
\end{table}

\begin{table}[H]
\caption{Anti-noise test of forward problem with linear coefficients ($L^2$ relative error)}
\begin{center}
\begin{tabular}{c||c|c|c|c|c|c|c|c}
\hline
\hline
 \diagbox{seed}{noise}& $0\%$ & $0.5\%$ & $1.0\%$ & $1.5\%$ & $2.0\%$ & $2.5\%$ & $3.0\%$ & $5.0\%$\\
\hline
\hline
1000 & 2.49 $\times 10^{-3}$ & 1.09 $\times 10^{-2}$ & 7.21 $\times 10^{-2}$ & 9.68 $\times 10^{-2}$ & 1.20 $\times 10^{-1}$ & 1.29 $\times 10^{-1}$ & 2.03 $\times 10^{-1}$ & 5.37 $\times 10^{-1}$ \\

2000 & 3.43 $\times 10^{-3}$ & 3.00 $\times 10^{-3}$ & 9.47 $\times 10^{-3}$ & 1.46 $\times 10^{-2}$ & 1.95 $\times 10^{-2}$ & 2.36 $\times 10^{-2}$ & 6.95 $\times 10^{-1}$ & 7.48 $\times 10^{-1}$ \\

3000 & 1.63 $\times 10^{-3}$ & 6.72 $\times 10^{-3}$ & 9.00 $\times 10^{-3}$ & 1.42 $\times 10^{-2}$ & 1.84 $\times 10^{-1}$ & 2.40 $\times 10^{-1}$ & 1.31 $\times 10^{-1}$ & 4.91 $\times 10^{-1}$ \\

4000 & 8.62 $\times 10^{-3}$ & 2.02 $\times 10^{-2}$ & 8.06 $\times 10^{-3}$ & 7.69 $\times 10^{-3}$ & 8.99 $\times 10^{-3}$ & 1.57 $\times 10^{-2}$ & 3.60 $\times 10^{-2}$ & 6.67 $\times 10^{-1}$ \\

5000 & 9.85 $\times 10^{-4}$ & 6.69 $\times 10^{-3}$ & 1.13 $\times 10^{-2}$ & 1.70 $\times 10^{-2}$ & 4.28 $\times 10^{-2}$ & 1.23 $\times 10^{-1}$ & 1.77 $\times 10^{-1}$ & 3.17 $\times 10^{-1}$ \\

6000 & 1.54 $\times 10^{-3}$ & 3.91 $\times 10^{-3}$ & 4.12 $\times 10^{-3}$ & 6.02 $\times 10^{-3}$ & 1.51 $\times 10^{-2}$ & 1.86 $\times 10^{-1}$ & 3.62 $\times 10^{-1}$ & 6.83 $\times 10^{-1}$ \\

7000 & 8.66 $\times 10^{-3}$ & 1.58 $\times 10^{-2}$ & 6.07 $\times 10^{-3}$ & 1.17 $\times 10^{-2}$ & 3.21 $\times 10^{-2}$ & 3.26 $\times 10^{-2}$ & 4.88 $\times 10^{-2}$ & 1.85 $\times 10^{-1}$ \\

8000 & 1.24 $\times 10^{-2}$ & 8.73 $\times 10^{-3}$ & 5.32 $\times 10^{-3}$ & 2.09 $\times 10^{-2}$ & 2.89 $\times 10^{-2}$ & 4.01 $\times 10^{-2}$ & 6.28 $\times 10^{-2}$ & 3.46 $\times 10^{-1}$ \\

9000 & 3.85 $\times 10^{-3}$ & 9.29 $\times 10^{-3}$ & 5.68 $\times 10^{-3}$ & 2.96 $\times 10^{-2}$ & 3.54 $\times 10^{-2}$ & 4.47 $\times 10^{-2}$ & 1.44 $\times 10^{-1}$ & 6.79 $\times 10^{-2}$ \\

10000 & 1.61 $\times 10^{-3}$ & 4.46 $\times 10^{-3}$ & 6.25 $\times 10^{-3}$ & 1.07 $\times 10^{-2}$ & 1.07 $\times 10^{-2}$ & 2.15 $\times 10^{-2}$ & 2.13 $\times 10^{-1}$ & 3.70 $\times 10^{-1}$ \\
\hline\hline
\hline
\end{tabular}
\end{center}
\end{table}

\begin{table}[H]
\caption{Anti-noise test of forward problem with nonlinear coefficients ($L^2$ relative error)}
\begin{center}
\begin{tabular}{c||c|c|c|c|c|c|c|c}
\hline
\hline
 \diagbox{seed}{noise}& $0\%$ & $0.5\%$ & $1.0\%$ & $1.5\%$ & $2.0\%$ & $2.5\%$ & $3.0\%$ & $5.0\%$\\
\hline
\hline
1000 & 1.04 $\times 10^{-3}$ & 6.62 $\times 10^{-3}$ & 1.21 $\times 10^{-2}$ & 5.70 $\times 10^{-2}$ & 4.18 $\times 10^{-2}$ & 9.96 $\times 10^{-2}$ & 5.17 $\times 10^{-2}$ & 8.11 $\times 10^{-2}$ \\

2000 & 1.01 $\times 10^{-3}$ & 2.59 $\times 10^{-3}$ & 5.20 $\times 10^{-3}$ & 1.55 $\times 10^{-2}$ & 2.16 $\times 10^{-2}$ & 2.53 $\times 10^{-2}$ & 4.17 $\times 10^{-2}$ & 1.01 $\times 10^{-1}$ \\

3000 & 1.04 $\times 10^{-3}$ & 7.48 $\times 10^{-3}$ & 1.11 $\times 10^{-2}$ & 1.74 $\times 10^{-2}$ & 2.06 $\times 10^{-2}$ & 3.64 $\times 10^{-2}$ & 4.14 $\times 10^{-2}$ & 1.19 $\times 10^{-1}$ \\

4000 & 2.63 $\times 10^{-3}$ & 5.14 $\times 10^{-3}$ & 1.77 $\times 10^{-2}$ & 7.69 $\times 10^{-3}$ & 3.05 $\times 10^{-2}$ & 6.72 $\times 10^{-2}$ & 2.26 $\times 10^{-1}$ & 4.66 $\times 10^{-1}$ \\
5000 & 9.73 $\times 10^{-4}$ & 4.72 $\times 10^{-3}$ & 1.04 $\times 10^{-2}$ & 2.00 $\times 10^{-2}$ & 2.74 $\times 10^{-2}$ & 4.03 $\times 10^{-2}$ & 4.51 $\times 10^{-2}$ & 8.43 $\times 10^{-2}$ \\

6000 & 1.37 $\times 10^{-3}$ & 7.13 $\times 10^{-3}$ & 6.62 $\times 10^{-3}$ & 1.08 $\times 10^{-2}$ & 1.81 $\times 10^{-2}$ & 2.07 $\times 10^{-2}$ & 4.21 $\times 10^{-2}$ & 8.68 $\times 10^{-2}$ \\

7000 & 1.84 $\times 10^{-3}$ & 3.28 $\times 10^{-3}$ & 3.48 $\times 10^{-3}$ & 5.60 $\times 10^{-3}$ & 7.08 $\times 10^{-3}$ & 1.01 $\times 10^{-2}$ & 1.21 $\times 10^{-2}$ & 2.88 $\times 10^{-2}$ \\

8000 & 8.40 $\times 10^{-4}$ & 2.08 $\times 10^{-3}$ & 5.92 $\times 10^{-3}$ & 1.03 $\times 10^{-2}$ & 1.36 $\times 10^{-2}$ & 1.87 $\times 10^{-2}$ & 1.76 $\times 10^{-2}$ & 2.57 $\times 10^{-2}$ \\
9000 & 1.16 $\times 10^{-3}$ & 9.66 $\times 10^{-3}$ & 1.93 $\times 10^{-2}$ & 2.66 $\times 10^{-2}$ & 3.86 $\times 10^{-2}$ & 5.31 $\times 10^{-2}$ & 6.14 $\times 10^{-2}$ & 1.33 $\times 10^{-1}$ \\
10000 & 1.57 $\times 10^{-3}$ & 5.16 $\times 10^{-3}$ & 5.67 $\times 10^{-3}$ & 7.18 $\times 10^{-3}$ & 6.76 $\times 10^{-3}$ & 1.49 $\times 10^{-2}$ & 2.11 $\times 10^{-2}$ & 1.02 $\times 10^{-1}$ \\
\hline\hline
\hline
\end{tabular}
\end{center}
\end{table}

\subsection{Model parameters and results for the inverse problem of the vSG equation (2D-coefficients)}\label{A_2D_SG}
\quad

Notations Case 1 and Case 2 correspond to the two situations in Section \ref{sec_relation}.

\begin{table}[H]
\caption{Space-time interval for the vSG equation (two-dimensional coefficients)}
\begin{center}
\begin{tabular}{c||c|c}
\hline
\hline
 & $[T_0, T_1]$ & $\Omega$ \\
\hline
\hline
$Case$ 1 & [-2.5,2.5] & [-2.5,2.5]\\
$Case$ 2 & [-4,2.5] & [-4,2.5]\\
\hline
\hline
\end{tabular}
\end{center}
\end{table}

\begin{table}[H]
\caption{Model parameters for the inverse problem of the vSG equation (two-dimensional coefficients)}
\begin{center}
\begin{tabular}{c||c|c|c|c|c|c|c|c|c|c|c|c|c}
\hline
\hline
 & $Seed$ & $N_{Adam}$ & $n_x$ & $n_t$ & $N_d^u$ & $N_B^u$ & $N_h^u$ & $N_d^c$ & $N_B^c$ & $N_h^c$  & $n_s$ & $n_c$ & $n_f$\\
\hline
\hline
All & 6666 & 5000 & 512 & 200 & 40 & 4 & 2 & 30 & 3 & 2 & 2000 &1000 & 20000\\
\hline
\hline
\end{tabular}
\end{center}
\end{table}

\begin{table}[H]
\caption{Model results for the inverse problem the vKdV equation (single coefficient)}
\begin{center}
\begin{tabular}{c||c|c|c|c|c|c}
\hline
\hline
 & $e^r_u$ & $e^r_c$ & $e^a_c$ & $T_{train}$ & $Iter_L$ & $Loss$ \\
\hline
\hline
$Case$ 1& $6.92\times 10^{-5}$ & $6.86\times 10^{-3}$ & $6.81\times 10^{-3}$ & 465.72s & 3426 & $5.97\times 10^{-7}$\\
$Case$ 2& $3.69\times 10^{-4}$ & $2.32\times 10^{-2}$ & $2.66\times 10^{-2}$ & 1199.90s & 16324 & $2.85\times 10^{-6}$\\
\hline
\hline
\end{tabular}
\end{center}
\end{table}

\bibliographystyle{unsrt} 
\bibliography{VC-PINN.bib} 

\end{document}